\pdfoutput=1
\documentclass[11pt,a4paper,american]{extarticle}

\usepackage{a4wide}
\usepackage[active]{srcltx}
\usepackage[export]{adjustbox}
\usepackage{amssymb}
\usepackage{amsmath}
\usepackage{amsfonts}
\usepackage{amsthm}
\usepackage{array,multirow,bigdelim}
\usepackage{authblk}
\usepackage{babel}
\usepackage{bbm}
\usepackage{bm}
\usepackage{braket}
\usepackage{cite}
\usepackage{color}
\usepackage{comment}
\usepackage{enumerate}
\usepackage{esint}
\usepackage{etoolbox}
\usepackage{float}
\usepackage[T1]{fontenc}
\usepackage{graphicx}
\usepackage{hyperref}
\usepackage[latin1]{inputenc}
\usepackage{latexsym}
\usepackage{mathrsfs}
\usepackage{mathtools}
\usepackage{relsize}
\usepackage{setspace}
\usepackage{slashed}
\usepackage{soul}
\usepackage{subcaption}
\usepackage{todonotes}
\usepackage{xcolor}

\newcommand{\xdownarrow}[1]{%
	{\left\downarrow\vbox to #1{}\right.\kern-\nulldelimiterspace}
}

\newcommand{\<}{\langle}
\renewcommand{\>}{\rangle}

\apptocmd{\sloppy}{\hbadness 10000\relax}{}{}

\def\a{\alpha}
\def\dif{\text{d}}

\def\be{\begin{equation}}
\def\ee{\end{equation}}
\def\ba{\begin{eqnarray}}
\def\ea{\end{eqnarray}}

\def\CP1{\mathbb{CP}^1}
\def\SL2C{\mathrm{SL}(2,\mathbb{C})}

\def\Z2{\mathbb{Z}_2}

\def\su2{{SU(2)}}

\def\a{{\alpha}}

\def\L{\Lambda}

\def\s{\sigma}
\def\a{\alpha}

\def\({\left(}
\def\){\right)}

\def\<{\langle}
\def\>{\rangle}

\def\i2{\frac{i}{2}}

\def\2F1{\,_2{\rm F}_1}

\date{}

\makeatletter

\allowdisplaybreaks
\numberwithin{equation}{section}
\author[1,2]{Humberto Gomez\thanks{humberto.gomez@durham.ac.uk}}
\author[3]{Renann Lipinski Jusinskas\thanks{renannlj@fzu.cz}}
\author[1]{Arthur Lipstein\thanks{arthur.lipstein@durham.ac.uk}}

\affil[1]{Department of Mathematical Sciences, Durham University,
\authorcr  Stockton Road, DH1 3LE, Durham, United Kingdom}
\affil[2]{Facultad de Ciencias Basicas,  Universidad Santiago de Cali,
\authorcr  Calle 5 $N^\circ$  62-00 Barrio Pampalinda, Cali, Valle, Colombia}
\affil[3]{Institute of Physics of the Czech Academy of Sciences \& CEICO,
\authorcr  Na Slovance 2, 182 21, Prague, Czech Republic}

\makeatother

\begin{document}

\title{Cosmological Scattering Equations at Tree-level and One-loop}

\maketitle

\begin{abstract}
We recently proposed a formula for tree-level $n$-point correlators of massive $\phi^4$ theory in de Sitter momentum space which consists of an integral over $n$ punctures on the Riemann sphere and differential operators in the future boundary dubbed the cosmological scattering equations. This formula was explicitly checked up to six points via a map to Witten diagrams using the global residue theorem. In this work we provide further details of these calculations and present an alternative formulation based on a double cover of the Riemann sphere. This framework can be used to derive simple graphical rules for evaluating the integrals more efficiently. Using these rules, we check the validity of our formula up to eight points and sketch the derivation of $n$-point correlators. Finally, we propose a similar formula for 1-loop $n$-point correlators in terms of an integral over $(n+2)$ punctures on the Riemann sphere, which we verify at four points.  The 1-loop formula holds for small masses in de Sitter space and arbitrary masses satisfying the Breitenlohner-Freedman bound after Wick-rotating to Anti-de Sitter space. 
\end{abstract}

\pagebreak

\tableofcontents

\section{Introduction}
The scattering equations of Cachazo, He, and Yuan (CHY) \cite{Cachazo:2013hca,Cachazo:2013iea,Mason:2013sva} revolutionized the study of scattering amplitudes, revealing new perturbative dualities \cite{Cachazo:2014xea,Casali:2015vta,Cheung:2017ems}, and new techniques for computing loop amplitudes \cite{Adamo:2013tsa,Geyer:2015bja,Baadsgaard:2015twa,Gomez:2017lhy,Farrow:2020voh} and soft limits \cite{Schwab:2014xua,Adamo:2014yya,Geyer:2014lca,Cachazo:2016njl,Nandan:2016ohb}. These equations  take a very compact and universal form, describing a broad range of quantum field theories in flat space.  In a nutshell, this approach is used to map scattering amplitudes to an integration over punctures on a Riemann sphere. The role of the scattering equations is to constrain the locations of punctures corresponding to the external legs of the scattering amplitude. The specific details of the interactions are encoded by the integrand which can be constructed from a simple set of building blocks. It is fair to say that the CHY formulae were the culmination of decades of research on perturbative scattering amplitudes that led to a wealth of many other powerful techniques such as twistor string theory \cite{Nair:1988bq,Witten:2003nn,Berkovits:2004hg,Roiban:2004yf,Boels:2006ir,Skinner:2013xp}, recursion relations \cite{Cachazo:2004kj,Britto:2005fq,Arkani-Hamed:2010zjl}, unitarity methods \cite{Bern:1994zx,Bern:1994cg}, the double copy relating guage theory to gravity \cite{Bern:2008qj,Bern:2010ue}, and new geometric formulations \cite{Hodges:2009hk,Arkani-Hamed:2012zlh,Arkani-Hamed:2013jha}.

In contrast, correlators in de Sitter (dS) space are far less understood despite their relevance to inflationary cosmology \cite{Guth:1980zm,Linde:1981mu,Albrecht:1982wi,Mukhanov:1981xt}. These correlators can be computed using the in-in formalism \cite{Maldacena:2002vr,Weinberg:2005vy}, or alternatively from expectation values involving a cosmological wavefunction \cite{Hartle:1983ai}. The latter can be computed perturbatively by Witten diagrams ending on the future boundary, Wick-rotating the corresponding diagrams in Anti-de Sitter (AdS) space \cite{Maldacena:2002vr,McFadden:2009fg,McFadden:2010vh,Maldacena:2011nz,Raju:2011mp,Ghosh:2014kba,Sleight:2019hfp,Sleight:2021plv}. The wavefunction coefficients are constrained by conformal Ward identities (CWI) associated with the spacetime isometries, and can be treated like correlation functions of a conformal field theory (CFT) living at the boundary \cite{Strominger:2001gp,Bzowski:2013sza,Bzowski:2015pba}. We will refer to them as cosmological correlators.

When Fourier transformed to momentum space, cosmological correlators develop singularities as the energy (defined as the sum of the magnitudes of the boundary momenta) goes to zero. This is interpreted as a flat space limit, such that the coefficients of these singularities correspond to scattering amplitudes in one higher dimension \cite{Raju:2012zr}. The relation between scattering amplitudes and conformal correlators opens up the possibilty of importing amplitude methods to cosmology. Recent progress along these lines includes geometric approaches \cite{Arkani-Hamed:2017fdk,Bzowski:2020kfw}, reconstruction from symmetries and factorization \cite{Arkani-Hamed:2015bza,Arkani-Hamed:2018kmz,Baumann:2019oyu,Baumann:2020dch,Baumann:2021fxj,Meltzer:2021zin,Hillman:2021bnk}, unitarity methods \cite{Alday:2017vkk,Meltzer:2020qbr,Goodhew:2020hob,Melville:2021lst,Jazayeri:2021fvk}, color/kinematics duality \cite{Armstrong:2020woi,Albayrak:2020fyp,Alday:2021odx,Diwakar:2021juk}, and the double copy \cite{Farrow:2018yni,Lipstein:2019mpu,Zhou:2021gnu,Sivaramakrishnan:2021srm}. There has also been recent progress in the non-perturbative calculation of non-gravitational correlators in rigid dS background \cite{Hogervorst:2021uvp,DiPietro:2021sjt}. 

In a recent publication \cite{Gomez:2021qfd}, we conjectured a worldsheet formula for tree-level cosmological correlators of massive $\phi^4$ theory, which is a toy model for inflation \cite{Martin:2013tda}. This formula was inspired by previous proposals for massless bi-adjoint scalar theories in AdS embedding space \cite{Roehrig:2020kck,Eberhardt:2020ewh}. One of the key insights of these papers was to promote the flat space scattering equations to differential operators built out of conformal generators acting on the boundary. In our work, the scattering equations are formulated in de Sitter momentum space, which is the natural language for cosmology and makes the flat space limit much more transparent, so they are referred to as the cosmological scattering equations (CSE). The CSE do not trivially follow from the scattering equations in AdS embedding space, and our construction introduces an operatorial Pfaffian which is the core ingredient for describing more general interactions in de Sitter space. Potential ambiguities that could arise in curved background were shown to be absent and we explicitly checked the formula up to six points by using the global residue theorem (GRT) \cite{Harris} to map it to a sum of Witten diagrams.

In this paper we will provide detailed derivations of the above results and extend them in several directions. In particular, we will present an alternative worldsheet formula based on the double cover formulation developed in \cite{Gomez:2016bmv} for flat space. The basic idea is to consider the Riemann sphere as a quadratic curve embedded in a two-dimensional complex projective space, promoting the branch-cut parameter to a new variable. By integrating this variable, we can access different factorization channels. This formulation streamlines the derivation of simple graphical rules for a more efficient evaluation of the worldsheet integrals, which can also be applied to the single-cover approach. Using these rules, we then verify our proposal for $\phi^4$ correlators up to eight-points, which include diagrams with non-ladder topology. While ladder diagrams can be computed recursively, non-ladder ones have a more complicated structure and we will present a systematic way to evaluate them. With the insights we gain at eight points, we then sketch the calculation of $n$-point correlators. 

Finally, we also propose a formula for 1-loop correlators of massive $\phi^4$ theory in de Sitter space. The basic idea is to consider a tree-level correlator with two auxiliary punctures. They encode the loop momentum and have unfixed scaling dimensions, which must be integrated over. As we show explicitly at 4-points, the bulk-to-boundary propagators for the auxiliary punctures get glued together to form a bulk-to-bulk propagator after integrating over their scaling dimensions, giving rise to 1-loop Witten diagrams. This gluing relies on a split representation for the bulk-to-bulk propagator first proposed in AdS$_{d+1}$ \cite{Meltzer:2020qbr,Penedones:2010ue}, which we Wick-rotate to dS$_{d+1}$. The resulting dS propagator is only valid for masses $0 \leq m \leq d/2$, or equivalently scaling dimensions in the complementary series. On the other hand, after Wick-rotating back to AdS, our 1-loop formula is valid for any mass satisfying the Breitenlohner-Freedman bound $m^2 \geq -d^2/4$ \cite{Breitenlohner:1982jf}. 

The structure of this paper is as follows. In section \ref{review} we review some basic facts about the cosmological wavefunction and its perturbative calculation using Witten diagrams. We also review the worldsheet formula for tree-level cosmological correlators proposed in \cite{Gomez:2021qfd}. In section \ref{DoubleCover} we describe a new formula in terms of the double cover and use it to derive simple graphical ruels for evaluating worldsheet integrals. In section \ref{examples4pt6} we show that the formula reproduces the expected Witten diagram expansion at four and six points, and in section \ref{sec:eightpoint} we extend this to eight-points and comment on $n$ points. In section \ref{1loop} we propose a new worldsheet formula for 1-loop correlators and verify it at 4-points. Finally in section \ref{conclusion} we present the concluding remarks. We also have a number of Appendices giving more details about dS isometries in momentum space, ${\rm SL}(2,\mathbb{C})$ symmetry of the worldsheet formula, the double cover formalism, and 6-point correlators.

{\bf{Note added:}} While completing this manucript, a proposal for constructing 1-loop Witten diagrams from tree-level ones in AdS embedding space appeared in \cite{Herderschee:2021jbi} which has some overlap with the results in section \ref{1loop} of this paper. Note that there are some important differences in the two approaches. For example, our 1-loop formula is for correlators in de Sitter momentum space and makes use of the CSE. Moreover, while \cite{Herderschee:2021jbi} focuses on $\phi^3$ theory, we focus on $\phi^4$ theory, although we believe both approaches can be extended to more general theories in (A)dS. 

\section{Review} \label{review}

We will work in the Poincar\'e patch of $(d+1)$-dimensional de
Sitter with radius $R$ and metric ${\rm d}s^{2}=(R/\eta)^{2}(-{\rm d}\eta^{2}+{\rm d}\vec{x}^{2})$,
where $-\infty<\eta<\eta_0$ is the conformal
time (the limit $\eta_0 \to 0^-$ is implicit), and $\vec{x}$ denotes the Euclidean boundary directions, with
individual components $x^{i}$. For simplicity, we will set $R=1$. In this section we will define the correlators we wish to compute, 
 and describe in detail how to do it perturbatively using Witten diagrams.

\subsection{Cosmological Correlators}

In-in correlators in de Sitter can be computed from a cosmological wavefunction as follows:
\begin{equation}
\left\langle \phi(\vec{k}_{1})...\phi(\vec{k}_{n})\right\rangle =\frac{\int\mathcal{D}\phi \, \phi(\vec{k}_{1})...\phi(\vec{k}_{n})\left|\Psi\left[\phi\right]\right|^{2}}{\int\mathcal{D}\phi\left|\Psi\left[\phi\right]\right|^{2}},
\end{equation}
where $\phi$ are values of the fields in the future boundary Fourier transformed to momentum space, and $\Psi\left[\phi\right]$ is the cosmological wavefuntion, which is a functional of $\phi$. In principle, we should integrate over the boundary values of all the fields, including the metric, but perturatively we can restrict to matter fields and for simplicity we will only consider scalar fields. In more detail, the wavefunction can be perturbatively expanded as follows:
\begin{equation}
\ln\Psi\left[\phi\right]=-\sum_{n=2}^{\infty}\frac{1}{n!}\int\prod_{i=1}^{n}\frac{{\rm d}^{d}k_{i}}{(2\pi)^{d}}\Psi_{n}\left(\vec{k}_{1},...\vec{k}_{n}\right)\phi(\vec{k}_{1})...\phi(\vec{k}_{n}),
\end{equation}
where the wavefunction coefficients $\Psi_n$ can be treated as an $n$-point CFT correlators in the future boundary, our main focus here. We will refer to them as cosmological correlators.

The $n$-point cosmological correlator can be expressed in momentum space as
\begin{equation}
\Psi_{n}=\delta^{d}(\vec{k}_T)\left\langle \left\langle \mathcal{O}\left(\vec{p}_{1}\right)...\mathcal{O}\left(\vec{p}_{n}\right)\right\rangle \right\rangle ,
\label{psid}
\end{equation}
where $\vec{k}_T=\vec{k}_{1}+...+\vec{k}_{n}$ and the object in double brackets can be treated as a CFT correlator in the future boundary. We will work with scalar operators $\mathcal{O}$ of scaling dimension $\Delta$, dual to massive scalar fields $\phi$ in the bulk satisfying
\begin{equation}
\eta^{2}\tfrac{\partial^{2}}{\partial\eta^{2}}\phi+(1-d)\eta\tfrac{\partial}{\partial\eta}\phi-\delta^{ij}\eta^{2}\tfrac{\partial^{2}}{\partial x^{i}\partial x^{j}}\phi=-m^{2}\phi.\label{eq:massive-dS}
\end{equation}
The mass is related to the scaling dimension by
\begin{equation}
m^{2}=\Delta(d-\Delta).
\label{eq:massdelta}
\end{equation}
For a given $m^2$, the two linearly independent solutions of \eqref{eq:massive-dS} have conformal dimensions
\begin{equation}
\Delta_{\pm}=\frac{d}{2}\pm\nu,\,\,\,\nu=\sqrt{\frac{d^{2}}{4}-m^{2}},
\end{equation}
which are related by a shadow symmetry $\Delta_{\pm} = (d-\Delta_{\mp})$. In practice we choose the scaling dimension with relative plus sign. The light solutions ($m^2 \leq d^2/4$) are parametrized by $\nu \in \mathbb{R}$, while for heavy solutions ($m^2>d^2/4$) $\nu$ is imaginary. The former is known as the complementary series, while the latter is the principal series. Note that $\Delta=d$ and $\Delta=(d+1)/2$ correspond to bulk scalars which are minimally or conformally coupled, respectively. 

The conformal Ward identities (CWI) for the cosmological correlator can be expressed as
\begin{equation}
\sum_{a=1}^{n}P^i_a\Psi_{n}=\sum_{a=1}^{n}D_{a}\Psi_{n}=\sum_{a=1}^{n}K^i_a\Psi_{n}=0,
\label{cwi}
\end{equation}
where $a,b,...$ are particle labels and the conformal generators in momentum space are
\begin{eqnarray}
P^{i} & = & k^{i}, \nonumber\\
D& = & k^{i}\partial_{i}+(d-\Delta), \label{eq:CGG-boundary}\\
K_{i} & = & k_{i}\partial^{j}\partial_{j}-2k^{j}\partial_{j}\partial_{i}-2(d-\Delta)\partial_{i}, \nonumber
\end{eqnarray}
with $\partial_i=\tfrac{\partial}{\partial k^i}$. A derivation of these generators can be found in Appendices \ref{props} and \ref{sl2c}. 
We will not need to consider rotation generators $L_{ij}$ since we focus on correlators of scalar operators.

\subsection{de Sitter propagators in momentum space} \label{dspropsmomentum}

It is convenient to consider Fourier modes of the scalar field $\phi$ along the boundary,
\begin{equation}
\phi=\mathcal{K}_{\nu}(k,\eta)e^{i\vec{k}\cdot\vec{x}},
\end{equation}
where $k=|\vec{k}|$, $\nu=\Delta-d/2$. Plugging this into \eqref{eq:massive-dS} then implies
\begin{eqnarray}
\mathcal{D}_{k}^{2}\mathcal{K}_{\nu} & = & \Delta(\Delta-d)\mathcal{K}_{\nu},\label{eq:eigenstates}\\
\mathcal{D}_{k}^{2} & \equiv & \eta^{2}\partial_{\eta}^{2}+(1-d)\eta\partial_{\eta}+\eta^{2}k^{2}. \label{eq:D2def}
\end{eqnarray}

The solutions $\mathcal{K}_{\nu}(k,\eta)$ are related to the Hankel functions of order $\nu$. Here we are going to work with the bulk-to-boundary propagators in the following form:
\begin{eqnarray}
\mathcal{K}_{\nu}(k,\eta)&=&\frac{\left(-i\right)^{d/2+1}}{2^{\nu}\Gamma(1+\nu)}\eta^{d/2}k^{\nu}K_{\nu}\left(-ik\eta\right), \label{kprop}\\
&=&\frac{\pi\left(-i\right)^{\Delta}}{2^{\nu+1}\Gamma(\nu+1)}k^{\nu}\eta^{d/2}H_{\nu}^{(2)}(-k\eta),
\end{eqnarray}
where $K_{\nu}$ is the modified Bessel function of the second kind and $H_{\nu}^{(2)}$ denotes the Hankel function of the second kind. Note that the first line was obtained by Wick rotating the standard bulk-to-boundary propagator in AdS \cite{Meltzer:2020qbr} according to $z\rightarrow -i \eta$, and the second line was obtained using the identity
\begin{equation}
K_{\nu}\left(-ix\right)=\frac{\pi}{2}\left(-i\right)^{\nu+1}H_{\nu}^{(2)}(-x),
\end{equation}
for $x\in\mathbb{R}^{-}$. The asymptotic behaviour of $\mathcal{K}_{\nu}$ when $\eta \to -\infty$ has a positive-frequency Minkowski mode\footnote{More details on the boundary conditions involved can be found in e.g. \cite{Goodhew:2021oqg}.}. 
Equation \eqref{eq:massive-dS} is invariant under global scaling transformations $(\eta,x^{i})\to(\lambda\eta,\lambda x^{i})$,
with parameter $\lambda$. In momentum space, this is translated to $(\eta,k^{i})\to(\lambda\eta,k^{i}/\lambda)$, and $\mathcal{K}_{\nu}$ has scaling dimension $d-\Delta$:
\begin{equation}
\mathcal{K}_{\nu}(k/\lambda,\lambda\eta)  =  \lambda^{d-\Delta}\mathcal{K}_{\nu}(k,\eta). \label{eq:scaling-prop}
\end{equation}
The other linearly independent solution of \eqref{eq:eigenstates} is denoted by $\mathcal{P}_{\nu}(k,\eta)$, and will be specified below.

The bulk-to-bulk propagators $G_{\nu}(\eta, \bar{\eta} ; k) $ satisfy
\begin{equation}
\left(\mathcal{D}_{k}^{2}+m^{2}\right)G_{\nu}(k,\eta,\bar{\eta})=\eta^{d+1}\delta(\eta-\bar{\eta}), \label{eq:eom-bulkprop}
\end{equation}
with boundary conditions
\begin{equation}
\lim_{\eta \to -\infty} G_{\nu}(k,\eta, \bar{\eta}) = G_{\nu}(k,\eta_0, \bar{\eta}) =0,
\end{equation}
where we take the limit $\eta_0 \rightarrow 0$ as explained in the beginning of the section. The solution is given by \cite{Mueck:1998wkz}
\begin{eqnarray}
G_{\nu}(k,\eta, \bar{\eta}) &\equiv&  \theta(\eta-\bar{\eta})\mathcal{K}_{\nu}(k, \bar{\eta}) \mathcal{P}_{\nu}(k, \eta) + \theta (\bar{\eta}-\eta) \mathcal{K}_{\nu}(k, \eta) \mathcal{P}_{\nu}(k, \bar{\eta}) \nonumber \\ &&-\frac{\mathcal{P}_{\nu}(k, \eta_{0})}{\mathcal{K}_{\nu}(k, \eta_{0})} \mathcal{K}_{\nu}(k, \bar{\eta}) \mathcal{K}_{\nu}(k, \eta) \label{eq:bulktobulkprop}
\end{eqnarray}
where $\theta(\eta-\bar{\eta})$ is the Heaviside step function. In order to show that this expression satisfies \eqref{eq:eom-bulkprop}, recall that the Wrosnkian of the two linearly independent solutions of \eqref{eq:eigenstates} satisfies
\begin{equation}
\mathcal{K}_{\nu} \partial_{\eta}\mathcal{P}_{\nu}-\mathcal{P}_{\nu} \partial_{\eta}\mathcal{K}_{\nu}=\eta^{d-1},
\end{equation}
for a convenient  normalization of $\mathcal{P}_{\nu}(k, \eta)$. More explicitly, the time-ordered bulk-to-bulk propagator in dS can be cast as
\begin{equation}
\label{dsprop}
G_{\nu}(k,\eta,\bar{\eta})=\frac{\pi}{4i}(\eta\bar{\eta})^{d/2}\left[H_{\nu}^{(1)}(-k\eta)H_{\nu}^{(2)}(-k\bar{\eta})-\frac{H_{\nu}^{(1)}(-k\eta_{0})}{H_{\nu}^{(2)}(-k\eta_{0})}H_{\nu}^{(2)}(-k\eta)H_{\nu}^{(2)}(-k\bar{\eta})\right],
\end{equation}
for $\eta>\bar{\eta}$. If $\eta<\bar{\eta}$, we simply exchange $\eta$ and $\bar{\eta}$. Here, $H_{\nu}^{(1)}$ is the Hankel function of the first kind, and the limit $\eta_0\rightarrow 0$ is implicit.

Let us briefly comment on the relation to the bulk-to-boundary propagator in AdS, which has the following split representation \cite{Penedones:2010ue,Meltzer:2020qbr}:
\begin{equation}
G_{\nu}^{AdS}\left(k,z,\tilde{z}\right)=\frac{1}{\pi}\left(z\tilde{z}\right)^{d/2}\int_{-\infty}^{\infty} {\rm d}\omega\frac{\omega^{2}}{\omega^{2}+\nu^{2}}\frac{K_{i\omega}\left(kz\right)K_{-i\omega}\left(k\tilde{z}\right)}{\Gamma\left(1+i\omega\right)\Gamma\left(1-i\omega\right)}. \label{eq:AdSbuktobulk}
\end{equation}
After Wick-rotating the dS propagator in \eqref{dsprop} according to $\eta\rightarrow iz$, we have numerically verified the agreement between \eqref{dsprop} and \eqref{eq:AdSbuktobulk} when the spectral parameter $\nu$ is real. On the other hand, they disagree when the spectral parameter is complex. In AdS$_{d+1}$, this corresponds to masses which violate the Breitenlohner-Freedman bound, $m^2 \geq -d^2/4$ \cite{Breitenlohner:1982jf}. In dS$_{d+1}$ this corresponds to heavy masses $m \geq d/2$. In section \ref{1loop} we will propose a worldsheet formula for 1-loop dS correlators which makes use of \eqref{eq:AdSbuktobulk} Wick-rotated via $z\rightarrow -i \eta$. We therefore expect this proposal to hold for $0\leq m \leq d/2$ in dS and $m^2 \geq -d^2/4$ after Wick-rotating back to AdS.

\subsection{Witten diagrams}

Cosmological correlators admit a perturbative expansion in terms of bulk Witten
diagrams ending on the future boundary. Here we take
the bulk theory to be a scalar with mass $m$ and quartic self-interaction. 

The bulk-to-boundary propagators are the building blocks of the contact diagrams $\mathcal{C}^{\Delta}_n$:
\begin{equation}
\mathcal{C}^{\Delta}_n  \equiv  \int  \frac{ {\rm d}\eta}{\eta^{d+1}}\prod_{a=1}^{n}\mathcal{K}_{\nu}(k_{a},\eta),
\end{equation}
where momentum conservation in the boundary directions and the integration over $\eta\in\left\{ -\infty,0\right\}$ are implicit. As we will see, all tree-level Witten diagrams can be obtained from contact diagrams by acting with certain differential operators.

A central object in our analysis is the action of the operator
\begin{equation}\label{eq:10R}
\mathcal{D}_{a}\cdot\mathcal{D}_{b}= \tfrac{1}{2}(P_{a}^{i}K_{bi}+K_{ai}P_{b}^{i})+D_{a}D_{b},
\end{equation}
on the product $\mathcal{K}_{\nu}(k_{a},\eta)\mathcal{K}_{\nu}(k_{b},\eta)\equiv\mathcal{K}_{\nu}^{a}\mathcal{K}_{\nu}^{b}$. When acting on $\mathcal{K}_{\nu}$, the boundary generators in \eqref{eq:CGG-boundary} can be written in terms of derivatives with respect to conformal time
\begin{equation}
\begin{array}{ccc}
D\mathcal{K}_{\nu} = \eta\partial_{\eta}\mathcal{K}_{\nu}, & P_{i}\mathcal{K}_{\nu}= k_{i}\mathcal{K}_{\nu}, & K^{i}\mathcal{K}_{\nu}  = \eta^{2}k^{i}\mathcal{K}_{\nu},
\end{array}
\label{eq:confgenboundary}
\end{equation}
This can be demonstrated by choosing a convenient integral representation for $\mathcal{K}_{\nu}$, e.g.
\begin{equation}
\mathcal{K}_{\nu}(k,\eta) = \eta^{\Delta}\int {\rm d}^{d}y\left(\eta^{2}-y^{2}\right)^{-\Delta}e^{i\vec{k}\cdot\vec{y}},
\end{equation}
which comes from a Fourier transformation of the bulk-to-boundary propagator in position space. Therefore, we obtain
\begin{equation}
(\mathcal{D}_{a}\cdot\mathcal{D}_{b})\mathcal{K}_{\nu}^{a}\mathcal{K}_{\nu}^{b}=\eta^{2}[\partial_{\eta}\mathcal{K}_{\nu}^{a}\partial_{\eta}\mathcal{K}_{\nu}^{b}+(\vec{k}_{a}\cdot\vec{k}_{b})\mathcal{K}_{\nu}^{a}\mathcal{K}_{\nu}^{b}].\label{eq:DaDb}
\end{equation}

Now let us consider the action of the operator $\mathcal{D}_{ab}^{2}\equiv \mathcal{D}_{k}^{2}$, with $k=|\vec{k}_a + \vec{k}_b|$, on the product $\mathcal{K}_{\nu}^{a}\mathcal{K}_{\nu}^{b}$:
\begin{eqnarray}
\mathcal{D}_{ab}^{2}(\mathcal{K}_{\nu}^{a}\mathcal{K}_{\nu}^{b})  =  (\mathcal{D}_{a}^{2}\mathcal{K}_{\nu}^{a})\mathcal{K}_{\nu}^{b}+\mathcal{K}_{\nu}^{a}(\mathcal{D}_{b}^{2}\mathcal{K}_{\nu}^{b})+2\eta^{2}[\partial_{\eta}\mathcal{K}_{\nu}^{a}\partial_{\eta}\mathcal{K}_{\nu}^{b}+(k_{a}\cdot k_{b})\mathcal{K}_{\nu}^{a}\mathcal{K}_{\nu}^{b}].
\end{eqnarray}
Since $\mathcal{D}_k^{2}\mathcal{K}_{\nu}=\Delta(\Delta-d)\mathcal{K}_{\nu}$,
a direct comparison with \eqref{eq:DaDb} yields
\begin{eqnarray}
\mathcal{D}_{ab}^{2}(\mathcal{K}_{\nu}^{a}\mathcal{K}_{\nu}^{b}) & = & 2[\Delta(\Delta-d)+(\mathcal{D}_{a}\cdot\mathcal{D}_{b})]\mathcal{K}_{\nu}^{a}\mathcal{K}_{\nu}^{b},\nonumber \\
 & = & (\mathcal{D}_{a}+\mathcal{D}_{b})^{2}\mathcal{K}_{\nu}^{a}\mathcal{K}_{\nu}^{b}.
\end{eqnarray}
The generalization to $n$-particles is straightforward. Let us define
\begin{eqnarray}
\mathcal{D}_{1...n}^{2} & \equiv & \eta^{2}\partial_{\eta}^{2}+(1-d)\eta\partial_{\eta}+\eta^{2}k_{1...n}^{2},\label{eq:BulkCasimir}\\
U_{m,n}(\eta) & \equiv & \prod_{a=m}^{n}\mathcal{K}_{\nu}^{a}(k_{a},\eta),
\end{eqnarray}
where $k_{1...n}=|\vec{k}_{1}+\ldots+\vec{k}_{n}|$. Then, using the results in (\ref{eq:confgenboundary}) and (\ref{eq:DaDb}),
we obtain
\begin{eqnarray}
(\mathcal{D}_{1...p}^{2}U_{1,p})U_{p+1,n} & = & (\mathcal{D}_{1}+\ldots+\mathcal{D}_{p})^{2}U_{1,n},\nonumber \\
 & = &[p\Delta(\Delta-d)+2\sum_{1\leq a<b \leq p}(\mathcal{D}_{a}\cdot\mathcal{D}_{b})]U_{1,n}, \label{eq:bulkvsboundary-props}
\end{eqnarray}
with $p<n$. The left-hand-side is the bulk Casimir in (\ref{eq:BulkCasimir})
and the right-hand-side is written in terms of the boundary conformal generators
in momentum space \eqref{eq:CGG-boundary}.

In practice we will encounter the inverse of boundary differential operators constructed from those in \eqref{eq:10R} acting on the product of bulk-to-boundary propagators. Using \eqref{eq:bulkvsboundary-props} we can then replace them with the inverse of the bulk differential operator in \eqref{eq:D2def} leading to bulk-to-bulk propagator insertions:
\begin{equation} \label{eq:invD-G}
[(\mathcal{D}_1+\ldots +\mathcal{D}_p)^2+m^2]^{-1} U_{1,p}(\eta) = \int \frac{ {\rm d}\bar{\eta}}{\bar{\eta}^{d+1}}  G_{\nu}(k_{1...p},\eta,\bar{\eta}) \, U_{1,p} (\bar{\eta}).
\end{equation}
Note that we are not explicitly inverting the differential operator on the left-hand side, a much more involved problem that is not being addressed here. Instead, the equation above is a formal equivalence that can be immediately verified by acting on both sides with $(\mathcal{D}_{1...p}^{2} + m^2)$. On the left-hand side we use the first line of \eqref{eq:bulkvsboundary-props}. As for the right-hand side we simply recall the equation of motion \eqref{eq:eom-bulkprop}. While the inverse operator is naively blind to boundary conditions, we promptly observe that the right-hand side of  \eqref{eq:invD-G} carries this information through the bulk-to-bulk propagator. This construction leads to a fundamental result expressing exchange Witten diagrams in terms of differential operators acting on contact diagrams:
\begin{equation}\label{eq:prop-insertion}
[(\mathcal{D}_1+\ldots +\mathcal{D}_p)^2+m^2]^{-1} \, \mathcal{C}^{\Delta}_n = \int  \frac{ {\rm d}\eta}{\eta^{d+1}}  \frac{ {\rm d}\bar{\eta}}{\bar{\eta}^{d+1}} \,   U_{p+1,n}(\eta) \, G_{\nu}(k_{1...p},\eta,\bar{\eta}) \, U_{1,p} (\bar{\eta}),
\end{equation}
which will be very useful later on. Explicit formulae for the bulk-to-bulk propagator are given in the previous subsection.


\subsection{Worldsheet Formula} \label{wsformulasc}

In \cite{Gomez:2021qfd} we proposed a new formula for computing tree-level correlators of massive $\phi^4$ theory in de Sitter momentum space based on a curved space analogue of the scattering equations. In more detail, the correlators are expressed as integrals over the Riemann sphere, mapping each external leg to a puncture, and the contour of integration is defined to encircle the points where the following differential operators vanish when acting on the rest of the integrand:
\begin{align}\label{eq:20R}
S_a = \sum_{b=1 \atop b\neq a}^{n} \frac{2\,  ({\cal D}_a \cdot {\cal D}_b )+\mu_{ab} }{\sigma_{ab}} \equiv \sum_{b=1 \atop b\neq a}^{n} \frac{\a_{ab} }{\sigma_{ab}} \, \, ,
\end{align}
where $\sigma_a$ is the holomorphic coordinate of the $a$'th puncture and $\mu_{ab}$ is a mass deformation equal to $-m^2$ when $a$ and $b$ are adjacent and zero otherwise. This mass deformation assumes canonical ordering of the external legs, {\it i.e.} $(1,2,\ldots ,n)\equiv \mathbb{I}_n $ \cite{Dolan:2013isa}. Different orderings are obtained by permutations. 

We refer to the equations which define the contour of integration as the cosmological scattering equations. In flat space, the scattering equations can be explicilty solved and the worldsheet integral can  be evaluated by summing over solutions to the scattering equuations. Due to the operatorial nature of \eqref{eq:20R}, it is not yet known how to do this in de Sitter space, so we instead deform the contour of integration in order to convert the worldsheet integral to a sum of Witten diagrams.  This approach was inspired by the ambistwistor string formulae in AdS position space first proposed in \cite{Eberhardt:2020ewh,Roehrig:2020kck}.  

In Appendix \ref{sl2c} we show that the  CSE enjoy an ${\rm SL}(2,\mathbb{C})$ symmetry:\footnote{The coefficients in \eqref{eq:26H} are the ${\rm SL}(2,\mathbb{C})$ generators, $L_{-1}=\sum_{a}\partial_{\sigma_a}, \, L_{0}=\sum_{a} \sigma_a\, \partial_{\sigma_a}, \, L_{1}=\sum_{a} \sigma_a^2\, \partial_{\sigma_a} $.} 
\begin{equation}\label{eq:26H}
\sum_{a=1}^n S_a = \sum_{a=1}^n \sigma_a\, S_a= \sum_{a=1}^n  \sigma_a^2\, S_a= 0.
\end{equation} 
This fact means there are only $(n-3)$ independent scattering equations.
It can then be used to fix the location of three punctures using a standard Fadeev-Popov procedure familiar from string theory. A generic worldsheet integral will then take the form
\begin{equation}\label{eq:22R}
{\cal A}=
\int_\gamma  \prod_{a=1 \atop a\neq b,c,d}^n  \dif\sigma_a \, (S_a)^{-1} \,  (\sigma_{bc}\sigma_{cd}\sigma_{db})^2 \, \mathcal{I}_n,
\end{equation}
where $\{\s_b,\s_c,\s_d\}$ denote the fixed punctures. The integration contour is defined by the intersection $\gamma= \bigcap_{a\neq b,c,d} \gamma_{S_a}$, where $\gamma_{S_a}$ encircles the pole where $S_a$ vanishes when acting on the theory-dependent integrand $\mathcal{I}_n$.
Following similar steps to \cite{Eberhardt:2020ewh,Roehrig:2020kck}, it is simple to check that the differential operators in \eqref{eq:20R} commute, 
\begin{equation}
\left[ S_a, S_b \right]=0, \qquad \forall\,\, a,b, 
\end{equation}
so the measure,
\begin{equation}
 {\rm d}\mu_n=\prod_{a=1 \atop a\neq b,c,d}^n  \dif\sigma_a \, (S_a)^{-1} \,  (\sigma_{bc}\sigma_{cd}\sigma_{db})^2 \, ,
\end{equation}
is well-defined and ${\rm SL}(2,\mathbb{C})$ invariant.

Using the CSE defined above, we can now define a worldsheet formula for $n$-point correlators of massive $\phi^4$ theory in dS momentum space \cite{Gomez:2021qfd}: 
\begin{equation}\label{eq:totalphi4}
\Psi_n=  \frac{\delta^{d}(\vec{k}_T)}{(3!)^{p-1}} \!\! \sum_{\rho\in {\rm S}_{n-1}} \!\!\!{\rm sgn}_\rho\,\,  {\cal A}(\rho(1,2,\ldots,n-1),n) \, {\cal C}^{\Delta}_n,
\end{equation}
where $n=2p \in \rm{even}$, ${\rm S}_{n-1}$ is the permutation group and
\begin{equation}\label{eq:7}
{\cal A}(\mathbb{I}_n ) \!= \int_\gamma  \prod^n_{a\neq b,c,d}  \dif \sigma_a\, (S_a)^{-1}   (\s_{bc} \s_{cd} \s_{db} )^2  \,\,  {\cal I}(\mathbb{I}_n) ,
\end{equation}
with
\begin{align}\label{}
{\cal I}(\mathbb{I}_n)= 
 {\rm PT}(\mathbb{I}_n) \,\,\,
{\rm Pf}^\prime A\, \times
\!\!\!\!\!\!\!\!\!\!
 \sum_{\{a,b\}\in cp(\mathbb{I}_n)} \frac{{\rm sgn}(\{a,b\})  }{  \s_{a_1b_1}\cdots \s_{a_{p} b_{p} } }   . 
\end{align}
Here  ${\rm PT}(\mathbb{I}_n) = \left(\sigma_{12}\sigma_{23}...\sigma_{n1}\right)^{-1}$ is the Parke-Taylor factor, $cp(\mathbb{I}_n)$ denotes all connected perfect matchings related to the ordering $(1,2,...,n)$ \cite{Cachazo:2014xea}, and the reduced Pfaffian ${\rm Pf}^\prime A$ is given by 
\begin{equation}
{\rm Pf}^\prime A=\frac{(-1)^{c+d}}{\sigma_{cd}}  {\rm Pf}A^{cd}_{cd},
\end{equation}
where
\begin{align}
{\rm Pf}A^{cd}_{cd}=
\frac{ \epsilon^{r_1 s_1\ldots r_{p-1} s_{p-1}} (A^{cd}_{cd})_{r_1 s_1} \cdots (A^{cd}_{cd})_{r_{p-1} s_{p-1}}   }{2^{p-1}(p-1)!}.
\end{align}
The matrix $A^{cd}_{cd}$ is obtained from the $n \times n$ matrix
\begin{align}
A_{rs} & =
\begin{cases} 
\displaystyle \frac{ \a_{rs} }{\sigma_{rs}},  & r\neq s,\\
\displaystyle  ~~ 0 , & r=s,
\end{cases}
\end{align}
by removing any pair of rows and columns $\{c,d\}$. From the commutation relations 
\begin{equation}
[\a_{rs}, \a_{pq}]=0, \qquad  \text{for} \,\,\, r\neq s\neq p \neq q,
\end{equation}
one can see that the reduced Pfaffian ${\rm Pf}^\prime A$ is well defined. 
In addition,  it is straightforward to demonstrate the commutators
\begin{equation}
\sum_{a=1}^n \left[ P_a^i,\a_{rs} \right]=\sum_{a=1}^n \left[ D_a,\a_{rs} \right]= \sum_{a=1}^n \left[ K_a^i,\a_{rs} \right]=0, \qquad \forall \,\, r,s,
\end{equation}
which imply that $\Psi_n$ satisfies the CWI. Note that ${\cal A}(\mathbb{I}_n)$ is independent of the choice of rows and columns one removes from the $A$-matrix. In Appendix \ref{sixpoint-16} we explicitly show this with a six-point example.

\begin{figure}[h]
\centering
\parbox[c]{11.1em}{\includegraphics[scale=0.25]{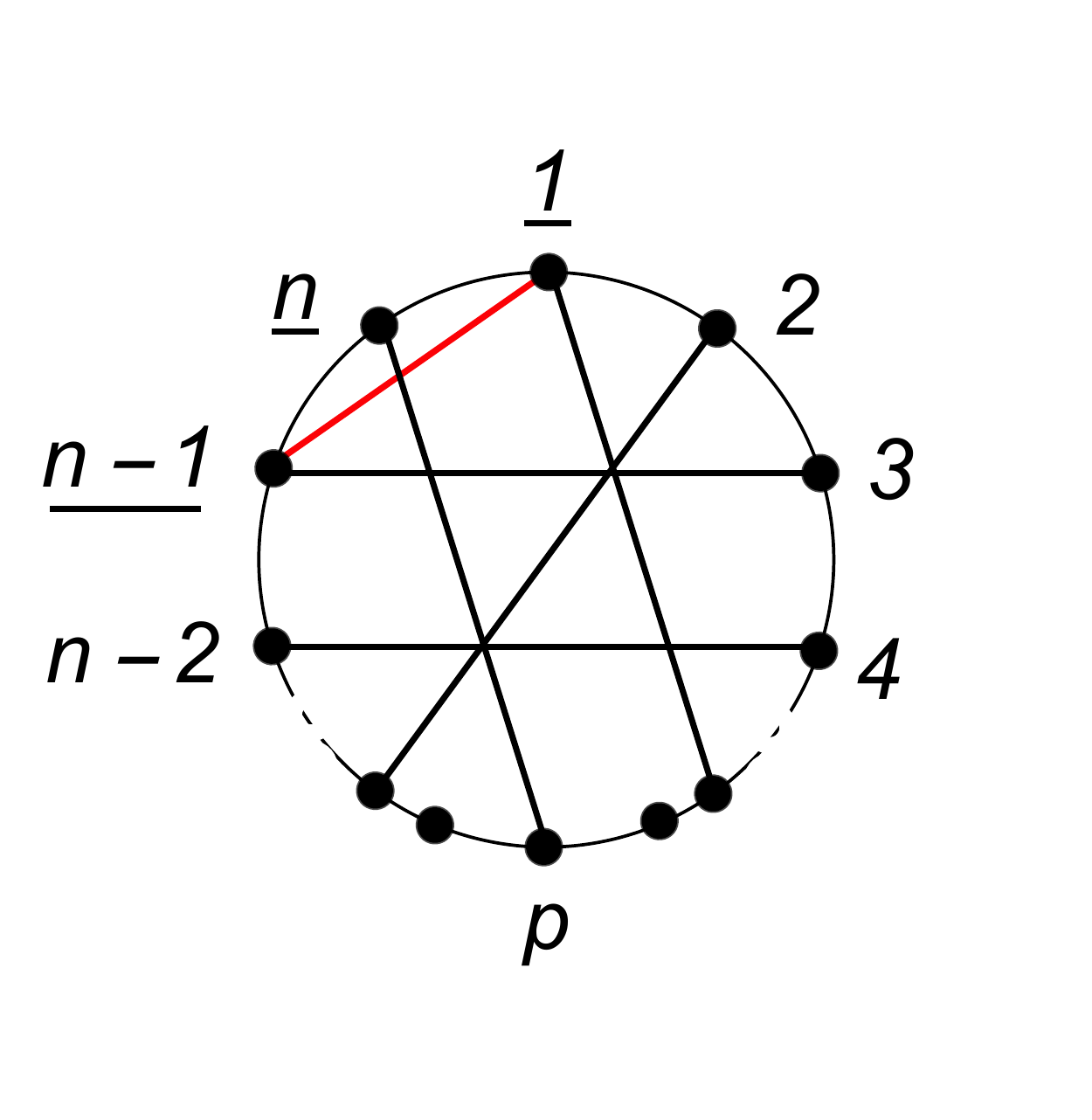}}
\vspace{-0.4cm}
\caption{Graph representation of ${\cal A}(\mathbb{I}_{n}:1\,p-2,2\,p+2,3\,n-1,\ldots,n\,p)$.}\label{Fig0} 
\end{figure}

Finally, notice that each connected perfect matching term in ${\cal A}(\mathbb{I}_n)$ has a  natural graph representation. For example, the graph correponding to $(\s_{1,p-2} \s_{2,p+2}\s_{3,n-1} \s_{4,n-2} \cdots \s_{n,p})^{-1}$ is given by
Fig. \ref{Fig0}. 
The external circle represents the Parke-Taylor factor, ${\rm PT}(\mathbb{I}_n)$, the
black lines depict the perfect matching, and the red line
indicates the rows and columns removed from the $A$-matrix.
The underlined labels $\{n-1, n, 1\}$ are the coordinates fixed
by the ${\rm SL}(2, \mathbb{C})$ symmetry, i.e. the punctures  $\{ \s_{n-1}, \s_n, \s_1 \}$ are not integrated. The notation we are going to use for this type of graph is ${\cal A}(\mathbb{I}_{n}:1\,p-2,2\,p+2,3\,n-1,\ldots,n\,p)$, where the argument following the double dots denotes the connected perfect matching.

The flat space limit of \eqref{eq:totalphi4} can accessed by taking $\eta\rightarrow-\infty$ in the conformal time integrals. Using the asymptotic form of the bulk-to-boundary propagators and  \eqref{eq:DaDb}, it is not difficult to show that our proposal has the expected flat space limit \cite{Gomez:2021qfd}.

\section{Double Cover}\label{DoubleCover}



In order to further test our construction, we will show that it reproduces the correct Witten diagram expansion for tree-level correlators up to eight points, with a sketch of the proof for $n$ points. In practice, we will evaluate the worldsheet integrals in \eqref{eq:7} using certain graphical rules which make calculations much more efficient. These rules can be conveniently derived by reformulating the worldsheet formula using a double cover formalism in which the worldsheet is represented by a two-sheeted Riemann surface connected by a cut which naturally encodes factorization \cite{Gomez:2016bmv}. A detailed review of this formalism can be found in Appendix \ref{double}, but we will state some basic definitions below. 

The idea of the double cover is  to consider  the Riemann sphere embedded in $\mathbb{CP}^2$ by the quadratic curve $y^2=z^2-\Lambda^2$, where $\Lambda\neq0$ is a constant and $(z,y)  \in \mathbb{C}^2$. The points $(\Lambda,0)$ and $(-\Lambda,0)$ are branch points. Since the curve is quadratic, there are only two-sheets. Thus, without loss of generality, we can say that ``$y$'' tells us on which of the two sheets the point is and ``$z$'' gives us the position on the sheet. For example, we state the puncture $(z_a,y_a)=(z_a,\sqrt{z_a^2-\Lambda^2})$ is on the upper sheet and $(z_b,y_b)=(z_b, -\sqrt{z_b^2-\Lambda^2})$ on the lower one. There are $2n-3$ parameters to be integrated in this double cover approach. This in contrast to the single-cover approach described in section \ref{wsformulasc}, which involves the integration over $n-3$ complex parameters. In particular, one starts with $2n+1$ integration variables, namely $(z_1,\ldots,z_n,y_1,\ldots,y_n)$ and $\Lambda$. The ${\rm SL}(2,\mathbb{C})$ symmetry and the additional scaling symmetry of the $\mathbb{CP}^2$ space can then be used to gauge-fix four of them. In practice, we will fix four of the $z$ punctures. 

Let us now sketch how the building blocks in the single cover approach can be adapted to the double cover language. The Parke-Taylor factor and the CSE become
\begin{align}
&{\rm PT}^\Lambda(\mathbb{I}_n) = \tau_{1:2}\tau_{2:3} \cdots \tau_{n:1}\, , \qquad S^{\Lambda}_a = \sum_{b=1 \atop b\neq a}^n \a_{ab} \, \tau_{a:b}\, , \qquad \text{where}\,\,\,\, \,
\tau_{a:b} = \frac{(yz)_a}{y_a} \, T_{ab} \,  ,  \nonumber  \\
&  T_{ab}\equiv \frac{1}{(yz)_a-(yz)_b}, \quad 
(yz)_i\equiv y_i+z_i,  \quad
\text{with} \,\, \, \, y_i^2= z^2_i-\Lambda^2. \label{}
\end{align}
Noting that  $T_{ab}$ is antisymmetric, the $A$-matrix and the reduced Pfaffian are then given by
\begin{align}
A_{rs} & =
\begin{cases} 
\displaystyle  \a_{rs} \, T_{rs},  & r\neq s\\
\displaystyle  ~~ 0 , & r=s
\end{cases}\,\, , \qquad \,
{\rm Pf}^\prime A = \left[\prod_{a=1}^n\frac{(yz)_a}{y_a}\right]  (-1)^{c+d} \, T_{cd} \, {\rm Pf}  A^{cd}_{cd} \, .
\end{align}
With these definitions, we can now explain how to adapt the worldsheet formulae in \eqref{eq:totalphi4} to the double cover approach and use it to derive some useful integration rules. More details are presented in Appendix \ref{double}.

\subsection{Alternative Worldsheet Formula}
  
Using the ingredients above, the corresponding version of the tree-level proposal in \eqref{eq:7} is given by
\begin{equation}
{\cal A}(\mathbb{I}_n) = \int_{\gamma} \dif \mu^\Lambda_n \, \Delta_{(pqr)} \, \Delta_{(pqr|m)} \prod_{a=1 \atop a\neq p,q,r}^n  (S_a^\Lambda)^{-1} \, {\cal I} (\mathbb{I}_n)\, ,
\end{equation}
where
\begin{align}\label{eq:Integrand}
{\cal I}(\mathbb{I}_n)= 
 {\rm PT}^\Lambda(\mathbb{I}_n) \,\,\,
{\rm Pf}^\prime A\, \times
\!\!\!\!\!\!\!\!\!\!
 \sum_{\{a,b\}\in cp(\mathbb{I}_n)} \!\!\!\!\!\!  {\rm sgn}(\{a,b\})  \,  T_{a_1 b_1}\cdots T_{a_{p}  b_{p} }    \, , 
\end{align}
and the measure $\dif \mu_n^\Lambda$ has the form
\begin{eqnarray}
\dif\mu_n^{\L}= 
\frac{1}{2^2}   \prod_{a=1}^n \frac{y_a\, \dif y_a}{{\rm C}_a}  
\times\hspace{-0.2cm} \prod_{a=1 \atop  a\neq p,q,r,m}^n
\hspace{-0.2cm}
\dif z_a \times \frac{\dif \L}{\L} \, .
\label{measureGF}
\end{eqnarray}

In the double cover approach $\Lambda$ is treated like an integration variable and we are able to gauge away four punctures. The Faddeev-Popov determinants for this gauge fixing are
\begin{eqnarray}
&&\Delta_{(pqr)} = (\tau_{p:q} \tau_{q:r} \tau_{r:p})^{-1}, \nonumber \\
&&\Delta_{(pqr|m)} =
\Delta_{(pqr)} \, \s_m -  \Delta_{(mpq)} \, \s_r +  \Delta_{(rmp)} \, \s_q -\Delta_{(qrm)} \, \s_p \, ,  \label{}
\end{eqnarray} 
where $\{z_p,z_q,z_r,z_m\}$ are the gauge fixed punctures. The integration contour $\gamma$ is determined by the $(n-3)$ scattering equations $(S_a^\Lambda)^{-1}$ and the hypersurfaces defined by the $n$ quadratic curves,
\begin{equation}
{\rm C}_a = y_a^2-z_a^a+\Lambda^2=0, \quad a=1,2,\ldots, n.
\end{equation}
Notice that the first part of the measure, $\prod_{a} \frac{y_a\, \dif y_a}{{\rm C}_a}  $, sums over all possible ways to place the punctures $(z_a,y_a)$ on the different sheets ($y_a$  is thought of as an independent variable), and $\frac{\dif \Lambda}{\Lambda}$ is the scale invariant measure over $\Lambda$.

Using the global residue theorem (GRT) \cite{Harris}, one of the scattering equations can be swapped by the contour $|\Lambda|=\varepsilon$ so, ${\cal A}(\mathbb{I}_n)$ can be written as
\begin{equation}
{\cal A}(\mathbb{I}_n) = \int_{\Gamma} \Big(\dif \mu^\Lambda_n \!\!\!\! \prod_{a=1 \atop a\neq p,q,r,m}^n  \!\!\!\! (S_a^\Lambda)^{-1} \Big)  \, \Big[(-1)\, \Delta_{(pqr)} \, \Delta_{(pqr|m)} \,  (S_m^\Lambda)^{-1} \Big] \, {\cal I} (\mathbb{I}_n).
\label{Prescription}
\end{equation}
where, without loss of generality, $\Gamma$ is defined by the $(n-4)$ scattering equations, $(S_a^\Lambda)^{-1},\, a\neq p,q,r,m,$ and the solution of the $(n+1)$ equations
\begin{equation}
\Lambda=0, \quad {\rm C}_a=0, \, \, \, \, \, \,  a=1,\ldots, n. 
\end{equation} 

\begin{figure}[h]
\centering
\parbox[c]{11.1em}{\includegraphics[scale=0.25]{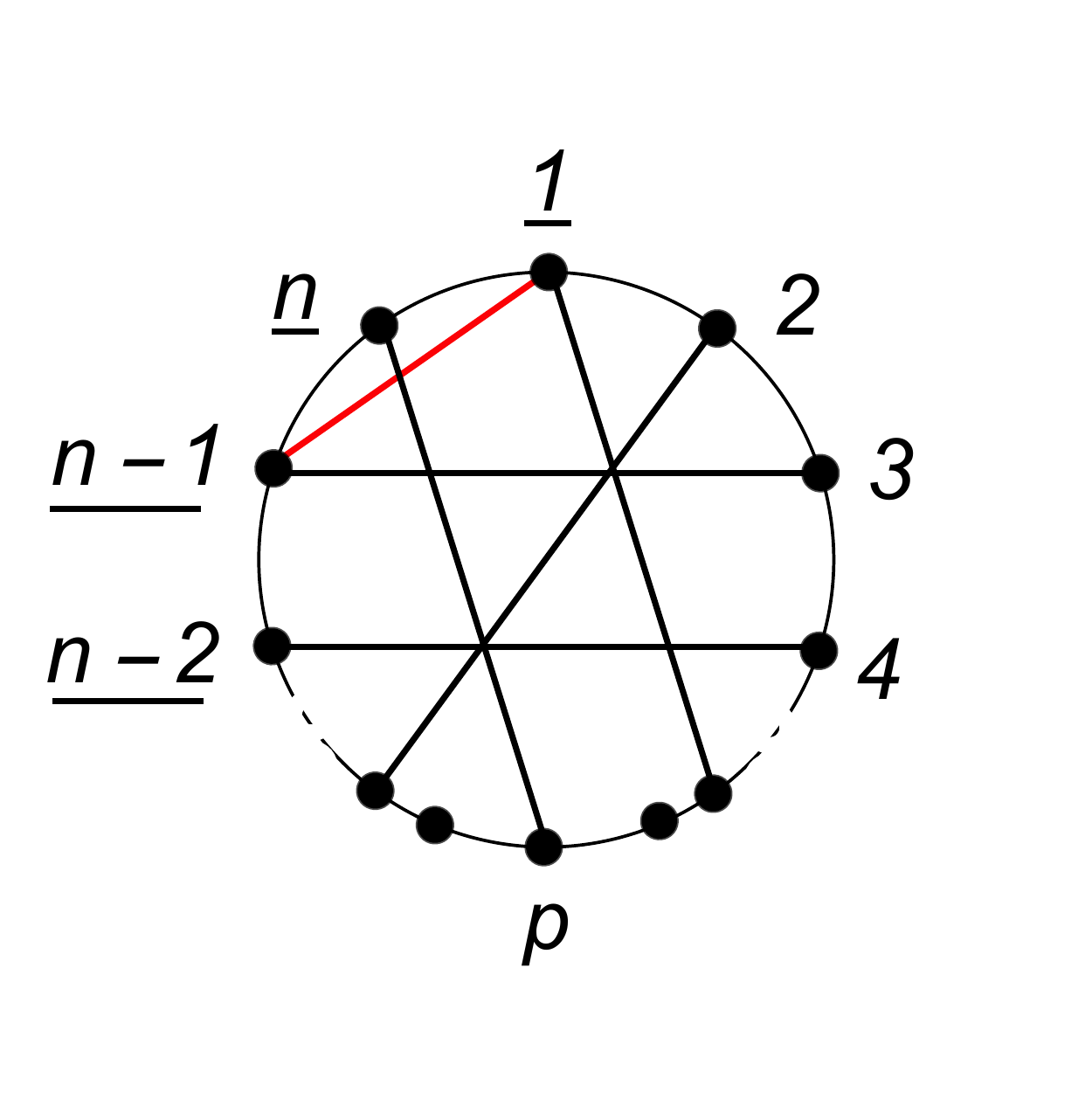}}
\vspace{-0.4cm}
\caption{Graph representation of ${\cal A}(\mathbb{I}_n: 1p-2, 2p+2, 3 n-1,\ldots, np)$ in the double cover language.}\label{FigDC} 
\end{figure}

As in the single cover approach, each connected perfect matching in ${\cal A}(\mathbb{I}_n)$
has a natural graph representation. The only difference with Fig. \ref{Fig0} is that there are four fixed punctures, {\it i.e.} four underlined labels. For example, if we consider the same perfect matching term as in Fig. \ref{Fig0}, $T_{1,p-2} T_{2,p+2} T_{3,n-1} T_{4,n-2} \cdots T_{n,p}$, then the graph related to this term is given by the Fig. \ref{FigDC},
\vspace{-0.0cm}
where we have chosen $\{z_{n-2},z_{n-1}, z_n, z_1  \}$ as the four fixed punctures.

\subsection{Integration rules}\label{IntegrationRules}  
 
In our previous work \cite{Gomez:2021qfd}, we proposed a couple of rules for evaluating worldsheet integrals for the $\phi^4$ model in a simple and straightforward way. They share a strong resemblance to the Yang-Mills and NLSM cases investigated in \cite{Gomez:2018cqg,Gomez:2019cik}. In this section we will prove these rules from the double cover point of view, which can then be applied to the single cover approach. They can be conveniently visualized using the graphs introduced in section \ref{wsformulasc} (see Figures \ref{Fig0} and \ref{FigDC} in the single and double cover formalisms, respectively). More specifically, the proposed rules are used to identify the vanishing graphs through factorization cuts, guiding a convenient choice of gauge (fixed punctures) and reduced Pfaffian.

\begin{figure}[h]
\centering
\parbox[c]{7.9em}{\includegraphics[scale=0.25]{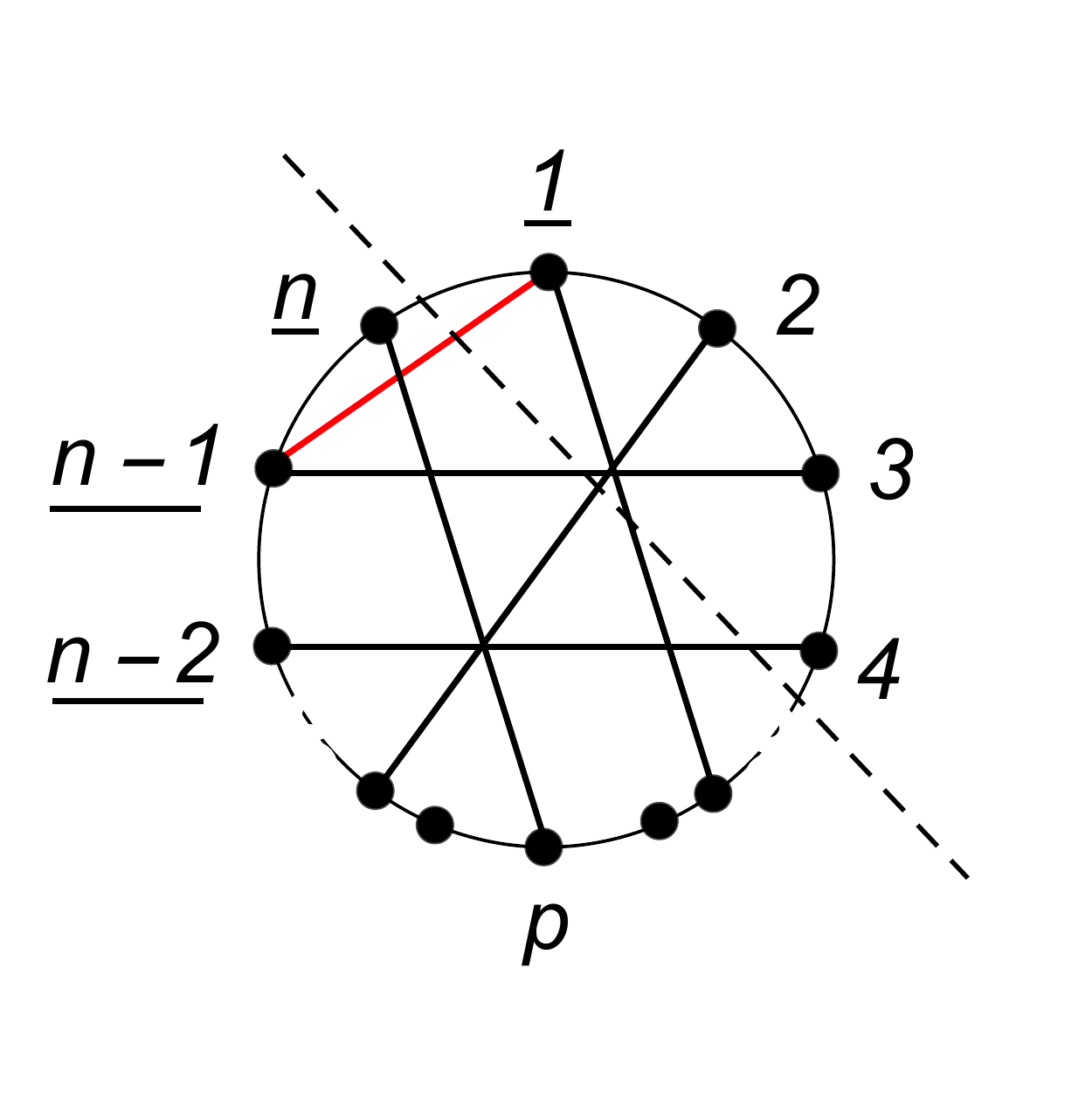}}= \,\,\, 0
\vspace{-0.4cm}
\caption{This particular factorization cut vanishes trivially since there is only one fixed puncture on the upper side of the  dashed black line (rule I).}\label{FigDcR1}
\end{figure}

Towards the first rule, we observe that the integration over the previously introduced $y_a$ variables localizes the integrand on the curves ${\rm C}_a=0$, with the solutions $y_a = \pm\sqrt{z_a^2 - \L^2}$. The punctures are distributed among the two sheets in all $2^n$ possible combinations. Due to the $\mathbb{Z}_2$ symmetry between the upper and lower sheets, only $2^{n-1}$ of them are inequivalent. After computing the integration over $|\L|=\varepsilon$, the two sheets factorize into two single-covers connected by a propagator corresponding to a differential operator such as in \eqref{eq:prop-insertion}. On each of the two single-cover sheets, three punctures must be fixed due to the ${\rm SL}(2,\mathbb{C})$ redundancy. When $\L =  0$, the branch-cut closes at a point giving two new punctures at the origin of each sheet. Each sheet must then have two more fixed punctures, which come from the fixed punctures in the original
double cover prescription, {\it i.e.} the points $(z_p,z_q,z_r,z_m)$. If there are not exactly two of these marked-points on each of the single covers, the configuration vanishes trivially. We summarize this in the rule:
\begin{itemize}
	\item {\bf Rule I.} {\it All configurations (or factorization cuts) with fewer  than two fixed
		marked-points vanish.}
\end{itemize}
In Fig. \ref{FigDcR1}, we give a simple example of this rule  for the term ${\cal A}(\mathbb{I}_n: 1p-2, 2p+2, 3 n-1,\ldots, np)$,
where the dashed black line denotes the factorization cut (the splitting between the upper and lower sheet\footnote{Since the prescription is invariant under the $\mathbb{Z}_2$ symmetry, $y_a\rightarrow -y_a$ so, either of the two areas next to the dashed black line can be considered as the upper (or lower) sheet.}). 
   
Let us now prove another useful rule. Without loss of generality, we will choose $(z_p,z_q,z_r,z_m)=(z_1,z_2,z_3,z_4)$. Following rule I, we will only be concerned with configurations where only two of these punctures are on the same sheet.
We would like to determine the behaviour of the different terms of the integrand around $\L=0$. 
In order to evaluate the integration measure, the Faddeev-Popov determinant, and the term $(S^\Lambda_4)^{-1}$, we consider a configuration where the punctures $\{z_{p+1},\dots,z_n,z_1,z_2\}$
are located on the upper sheet, {\it i.e.} $y_a =+\sqrt{z_a^2 - \L^2}$, and the punctures $\{z_3,z_4,\dots,z_p\}$ are located on the lower one, {\it i.e.} $y_a =-\sqrt{z_a^2 - \L^2}$ .  By expanding around $\L=0$, we obtain
\begin{align}
&
\Big(\dif \mu^\Lambda_n  \prod_{a=5}^n   (S_a^\Lambda)^{-1} \Big) \Big|^{p+1,\ldots,n,1,2}_{3,\, 4,\ldots ,p}  \nonumber\\
&
= \frac{d\L}{\L}    \Big[\dif z_{5}\cdots \dif z_{p}   \prod_{b=5}^p   (S_b)^{-1}  \Big] 
\Big[ \dif z_{p+1} \cdots   \dif z_{n}   \!\! \prod_{a=p+1}^n   (S_a)^{-1}   \Big]
+ {\cal O}(\L) \, , \label{eq:MeasureL0}\\
& 
\Big( (-1)\,\Delta_{(123)}  \Delta_{(123|4)} ( S^{\L}_4)^{-1} \Big)
\Big|^{p+1,\ldots,n,1,2}_{3,\, 4,\ldots ,p}  
\nonumber 
\\
&
=  \frac{2^5}{\L^4} (z_{12} z_{2L}   z_{L1}  )^2 
\left[ ( {\cal D}_4+\cdots + {\cal D}_p)^2 + m^2
\right]^{-1} 
(z_{R3}  z_{34}  z_{4R})^2 + {\cal O}(\L^{-2}) \, ,
\label{eq:fpL0}
\end{align}
where  $z_L=z_R=0$, $\alpha_{aL}= \a_{a3}+\a_{a4}+\cdots + \alpha_{ap}$ ($a=p+1,\ldots,n$) and $\alpha_{bR}= \a_{b\,p+1}+\cdots +\a_{bn}+ \alpha_{b1} + \alpha_{b2}$ ($b=5,6,\ldots,p$). 

Next, we need to determine the leading order expansion of the Parke-Taylor factor, the reduced Pfaffian, and the connected perfect matching terms when $\L\rightarrow 0$.
To do that,  we are going to use the graph representation explained in Fig. \ref{FigDC}. 
For a given configuration, with punctures distributed between the upper and the lower sheet, the lines that connect the vertices cross the banch-cut.  After expanding the terms in the integrand around $\L = 0$, it is straightforward to note that the leading order contribution is related to the number of lines cut by the branch-cut. In Table \ref{intruleTable} we have classified the $\L$-behaviour of these integrands.   
\begin{table}[ht]
	\large
	\centering
	\begin{tabular}{c|c||c|c|}
		\multicolumn{2}{c}{} & \multicolumn{2}{c}{{\small Factor}} \tabularnewline
		\cline{2-4}
		\multirow{9}*{\rotatebox{90}{\small{Lines cut by the branch-cut}}} & & 
		${\rm PT}^\Lambda(\mathbb{I}_n)$ & $  {\rm Pf}^\prime A\, \times
  T_{a_1 b_1}\cdots T_{a_{p}  b_{p} }     $ \tabularnewline[1ex]
		\cline{2-4}
		&  &  &  \tabularnewline[0ex]
		& \bfseries 0 & $\L^0$ & $\L^0$ \tabularnewline[1ex]
		& \bfseries 1 & - & $\L^2$ \tabularnewline[1ex]
		& \bfseries 2 & $\L^2$ & $\L^2$ \tabularnewline[1ex]
		& \bfseries 3 & - & $\L^4$ \tabularnewline[1ex]
		& \bfseries 4 & $ \L^4$ & $\L^4$ \tabularnewline[1ex]
		\cline{2-4}
	\end{tabular}
	\caption{
	\label{intruleTable}
		$\L$-dependence when terms in the integrand terms are expanded around 
	$\L=0$. Empty entries mean that such terms do not occur.}
\end{table}

\begin{figure}[h]
\centering
\parbox[c]{7.9em}{\includegraphics[scale=0.25]{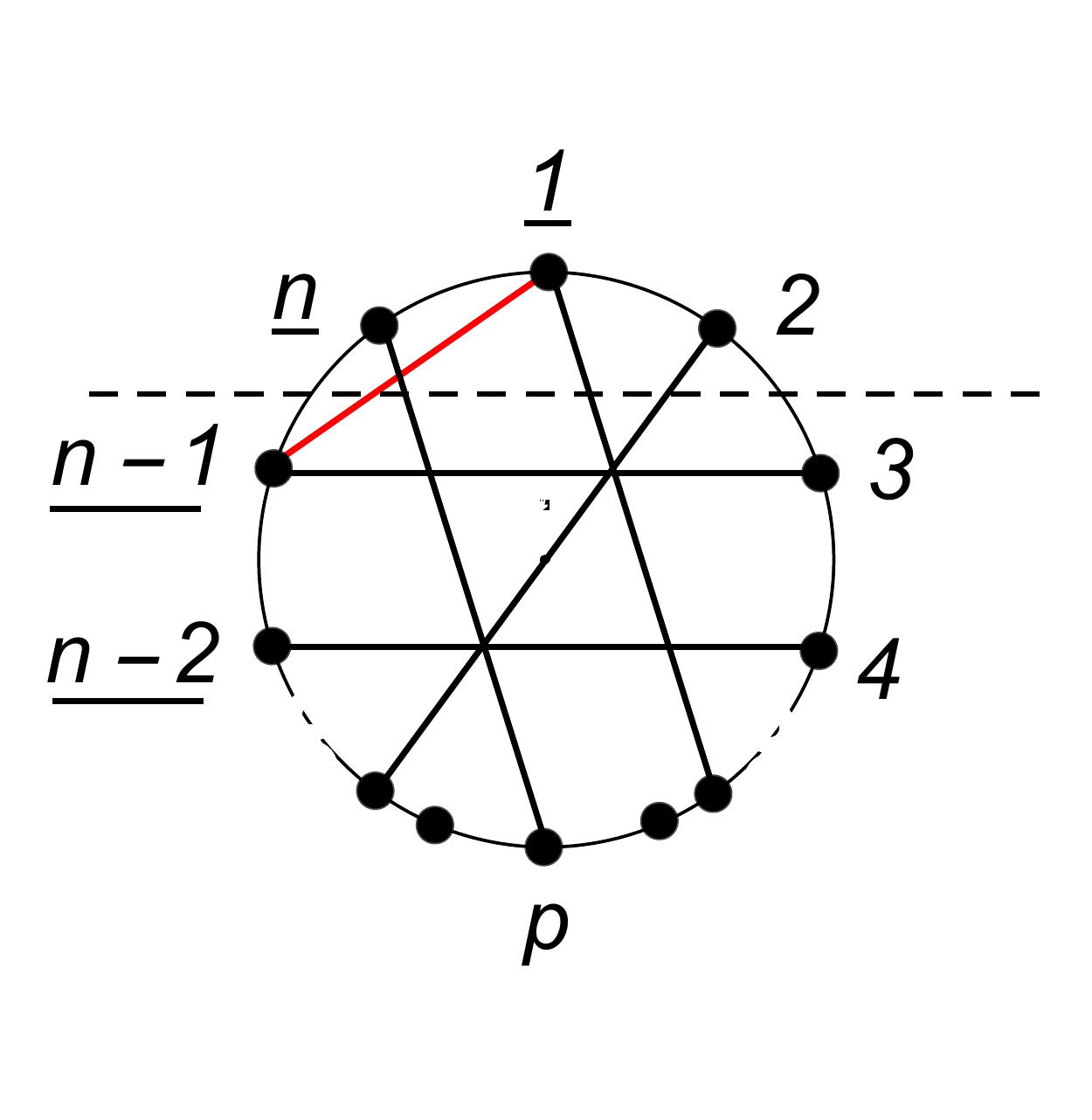}}= \,\,\, 0
\vspace{-0.4cm}
\caption{This factorization cut vanishes since the dashed black line cuts six lines (more than four), five black and one red (rule II).}\label{FigDcR2}
\end{figure}

\noindent Now, combining the expansions \eqref{eq:MeasureL0}, \eqref{eq:fpL0} and the Table \ref{intruleTable}, we state the following rule: 
\begin{itemize}
\item {\bf Rule II.} {\it  If the branch-cut (factorization) cuts more than four lines for a given configuration in the corresponding graph then, this contribution vanishes.}
\end{itemize}
In Fig. \ref{FigDcR2}, this rule is illustrated through a simple example where the dashed black line (factorization cut) is cutting more than four lines.
\vspace{-0.0cm}

Now that the integration rules have been properly established, we are ready to demonstrate their use through a couple of examples and present some general results. This is our focus in the next two sections.
 
\section{Four and Six Points} \label{examples4pt6}

In this section we will evaluate the worldsheet integral in \eqref{eq:totalphi4} at four and six points, showing that it generates the expected sum over Witten diagrams, and that there are no ambiguities in the integrand. This is a nontrivial feature given that we work with operatorial building blocks. The four and six point examples are simpler because they only involve ladder diagrams. More general diagrams contribute above six-points, as we discuss in the next section. To make the calculations more efficient, we use the integration rules discussed above. They can be extended to the single-cover approach, where the role of the branch-cut variable $\L$ is given by the infinitesimal parameter $\epsilon$  that controls the rate at which punctures approach each other in a factorization cut.

\subsection{Four points}\label{sec:fourpoint}

\begin{figure}[h]
\centering
(a)
\!\!\!\!\!\!
\parbox[c]{12.2em}{\includegraphics[scale=0.23]{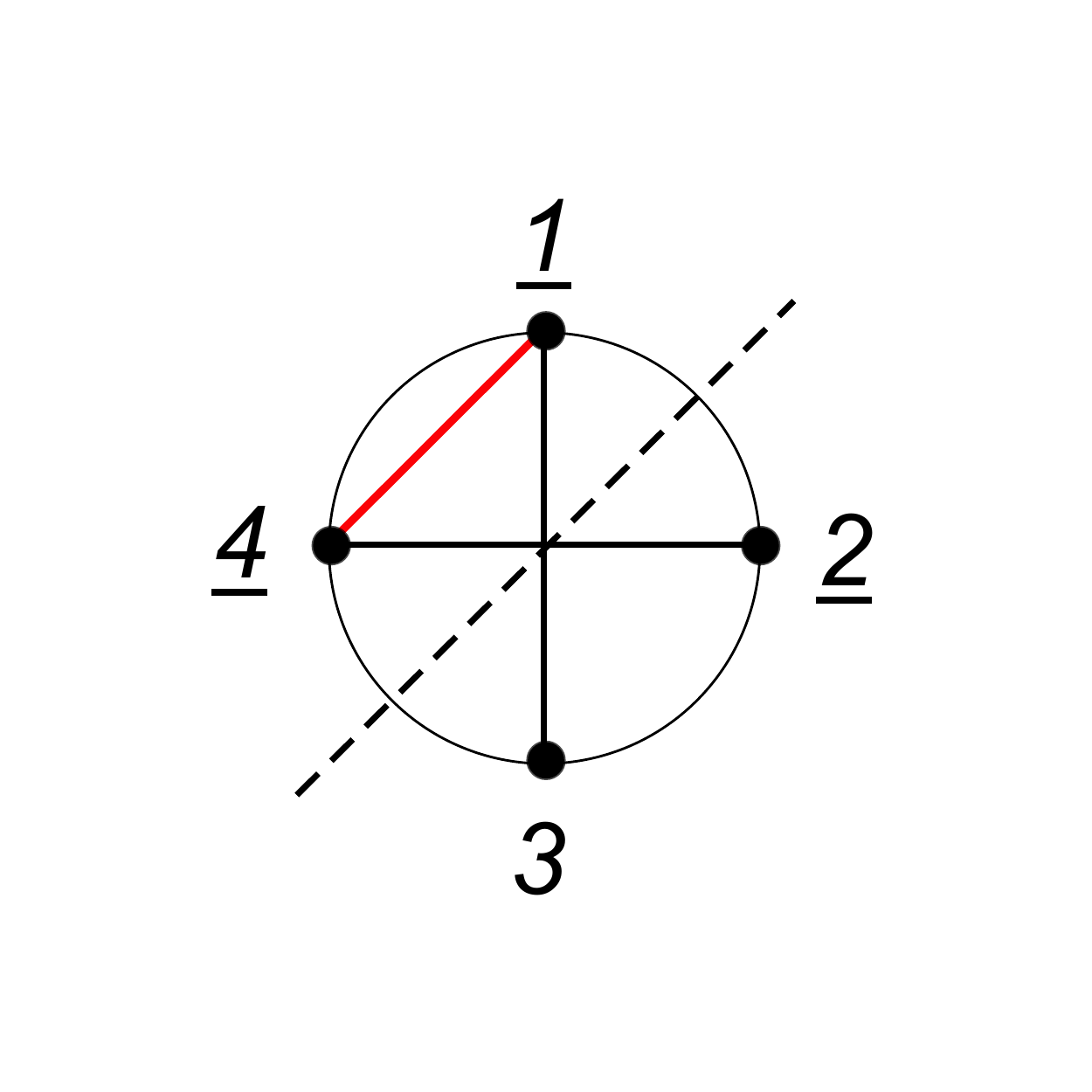}}
(b)
\!\!\!\!\!\!
\parbox[c]{8.5em}{\includegraphics[scale=0.23]{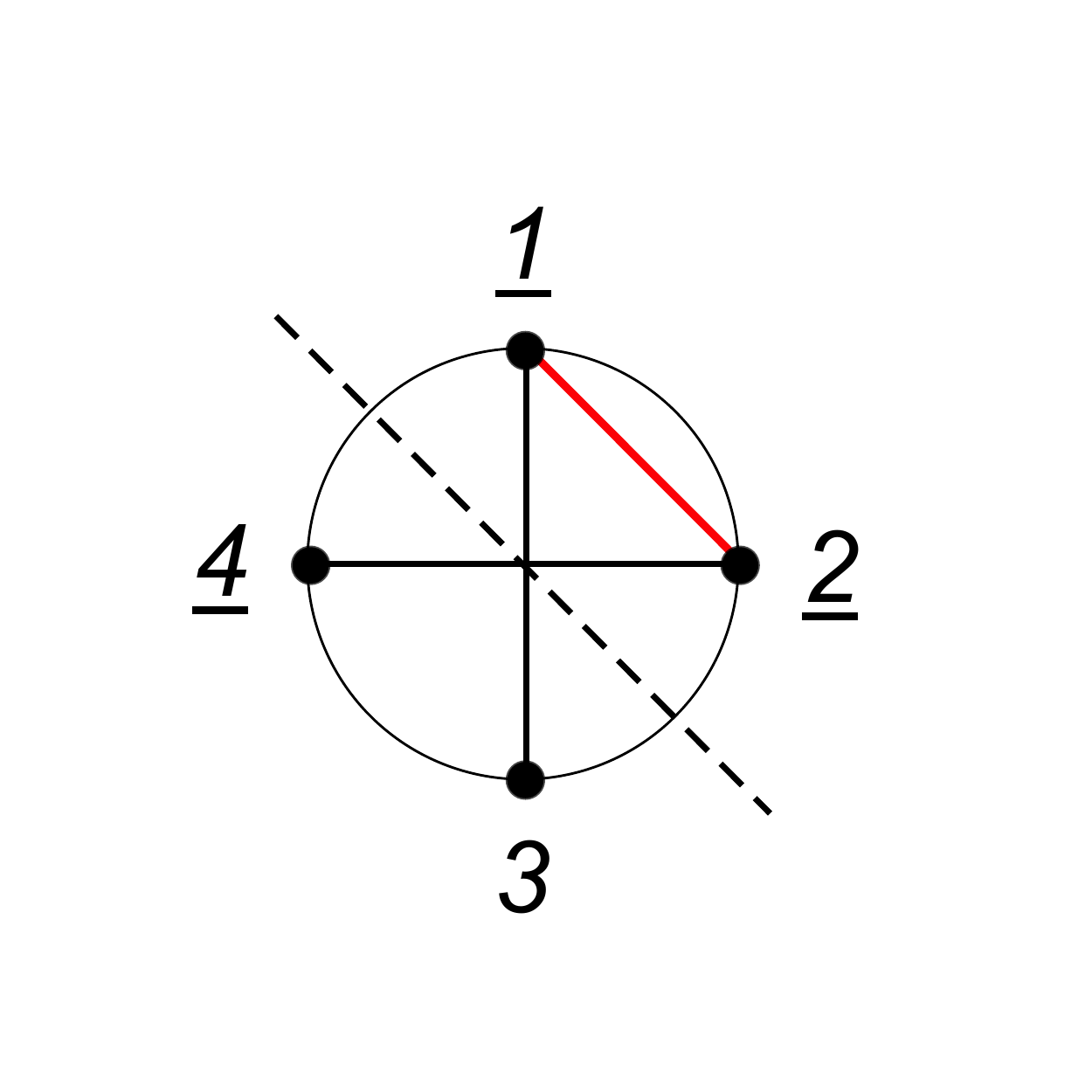}} 
\vspace{-0.45cm}
\caption{Factorization of ${\cal A}(\mathbb{I}_4)$ with (a) ${\rm Pf} A^{14}_{14}$, (b) ${\rm Pf} A^{12}_{12}$.}\label{Fig2} 
\end{figure}

\begin{figure}[h]
\centering
\text{$
\mathord{\begin{tikzpicture}[scale=0.45]
	\node at (-3,0) [circle,,fill=black,inner sep=0pt,minimum size=0mm,label=center:$ \text{${\scriptstyle \eta=0}$} $]  {};
	\node at (-2,0) [circle,,fill=black,inner sep=0pt,minimum size=0mm,label=above:$ $]  {};
	\node at (1,0) [circle,,fill=black,inner sep=0pt,minimum size=1.5mm,label=above:$1$]  {};
	\node at (2,0) [circle,,fill=black,inner sep=0pt,minimum size=0mm,label=above:$ $]{};
	\node at (3,0) [circle,,fill=black,inner sep=0pt,minimum size=1.5mm,label=above:$2$]{};
	\node at (4,0) [circle,,fill=black,inner sep=0pt,minimum size=0mm,label=above:$ $]{};
	\node at (5,0) [circle,,fill=black,inner sep=0pt,minimum size=1.5mm,label=above:$3$]{};
		\node at (6,0) [circle,,fill=black,inner sep=0pt,minimum size=0mm,label=above:$ $]{};
			\node at (7,0) [circle,,fill=black,inner sep=0pt,minimum size=1.5mm,label=above:$4$]{};
				\node at (4,-2) [circle,,fill=black,inner sep=0pt,minimum size=1.0mm,label=above:$ $]{};
	\draw[very thick,black] (-1,0) -- (9,0);
		\draw[ black] (1,0) -- (4,-2);
			\draw[ black] (4,-2) -- (3,0);
			   \draw[ black] (5,0) -- (4,-2);
			       \draw[black] (4,-2) -- (7,0);
	\end{tikzpicture}}
	$}
	\caption{Four-point contact Witten diagram}\label{4ptcontactfigure}
\end{figure}
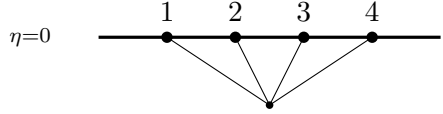

Let us first consider the ordered correlator
\begin{equation}\label{eq:11}
{\cal A}(\mathbb{I}_4) \, {\cal C}^{\Delta}_4 =  \int_{\gamma}  \dif\sigma_3 ( \sigma_{41} \sigma_{12} \sigma_{24} )^2 (S_3)^{-1}  
{\rm PT}(\mathbb{I}_4) \times  \frac{(-1)}{\sigma_{14} }\,   {\rm Pf} A^{14}_{14}\,  \frac{ (-1) }{ (\sigma_{13} \sigma_{24})} \,\, {\cal C}^{\Delta}_4.
\end{equation}
The corresponding graph of ${\cal A}(\mathbb{I}_4) $
is drawn in Fig. \ref{Fig2}(a). 
\vspace{-0.0cm}
We see  there is only one factorization contribution, $\sigma_3\rightarrow \sigma_2$, which is given by the  dashed black line. To perform the computation we choose the following parametrization:  $\s_a=\epsilon \, x_a + \sigma_2$, $a=2,3$, with $x_2=0$, $x_3=\text{constant}$ and $\s_2\equiv \s_L$. The measure and integrand of \eqref{eq:11} can be expanded as follows
\begin{align}
&\dif\s_3 = x_{32} \,  \dif \epsilon , \quad S_3=\frac{1}{\epsilon}  \left[ \hat S_3  + {\cal O}(\epsilon) \right], \quad \hat S_3 =\frac{\a_{32}}{x_{32}} , \nonumber \\
& ( \sigma_{41} \sigma_{12} \sigma_{24} )^2  {\rm PT}(\mathbb{I}_4) \times  \frac{(-1)}{\sigma_{14} }\,   {\rm Pf} A^{14}_{14}\,  \frac{ 1 }{ (\sigma_{13} \sigma_{24})}  = \frac{1}{\epsilon^2} \frac{1}{x_{23}^2} \, \a_{23}+ {\cal O}(\epsilon^{-1}),
\end{align}
such that ${\cal A}(\mathbb{I}_4) $ becomes
\begin{equation}\label{}
{\cal A}(\mathbb{I}_4) = - \int_{\gamma_{\hat S_3 + {\cal O}(\epsilon) }}  \frac{\dif\epsilon}{\epsilon} \left[ \hat S_3 + {\cal O}(\epsilon) \right]^{-1} \, \left(  \a_{32} + {\cal O}(\epsilon) \right).
\end{equation}
By the GRT, the contour $\gamma_{\hat S_3 + {\cal O}(\epsilon) }$ can be deformed into $\gamma_{\epsilon} = \{  |\epsilon| = \delta \}$. One then finds that 
\begin{equation}\label{eq:89}
{\cal A}(\mathbb{I}_4) =  \int_{\gamma_{\epsilon}}  \frac{\dif\epsilon}{\epsilon} \left[ \hat S_3 + {\cal O}(\epsilon) \right]^{-1} \, \left(  \a_{32} + {\cal O}(\epsilon) \right) = [ \a_{32}]^{-1} \a_{32} = \mathbb{I},
\end{equation}
and
\begin{equation}\label{eq:four-point}
{\cal A}(\mathbb{I}_4)  \,\, {\cal C}_{4}^\Delta = {\cal C}_{4}^\Delta
\end{equation}
which is the four-point contact diagram illustrated in Figure \ref{4ptcontactfigure}.

Notice that switching the order of the Pfaffian and the scattering equation in \eqref{eq:11}, leads to the same expression. Moreover, had we  made a different choice of the reduced Pfaffian, {\it e.g.} $\frac{(-1)}{\s_{12}}\, {\rm Pf} A^{12}_{12}$, the only non-zero factorization contribution would come from the dashed black line in Fig. \ref{Fig2}(b). Using the parametrization $\s_a=\epsilon \, x_a + \sigma_2$, $a=3,4$, with $x_4=0$, $x_3=\text{constant}$ and $\s_4\equiv \s_L$, we could then expand around $\epsilon=0$ and apply the GRT as we did in \eqref{eq:89} to obtain same result as \eqref{eq:four-point}. Hence, there are no ambiguities arising from the choice of Pfaffian or the ordering of terms in the integrand.

\subsection{Six points}\label{sec:sixpoint}

\begin{figure}[h]
\centering
\hspace{-0.7cm}
\parbox[c]{5.1em}{\includegraphics[scale=0.2]{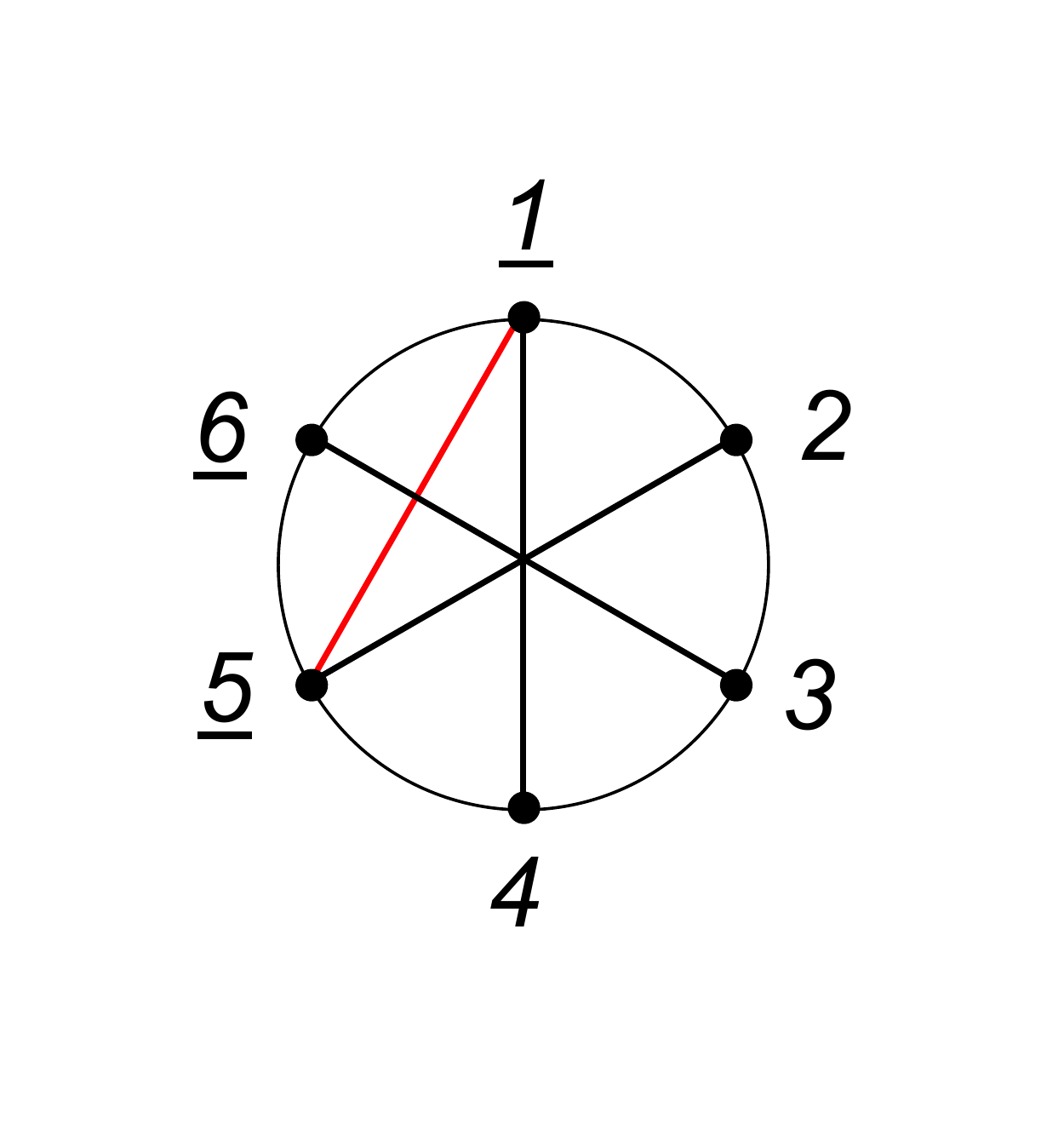}}\,\, \,
\parbox[c]{5.1em}{\includegraphics[scale=0.2]{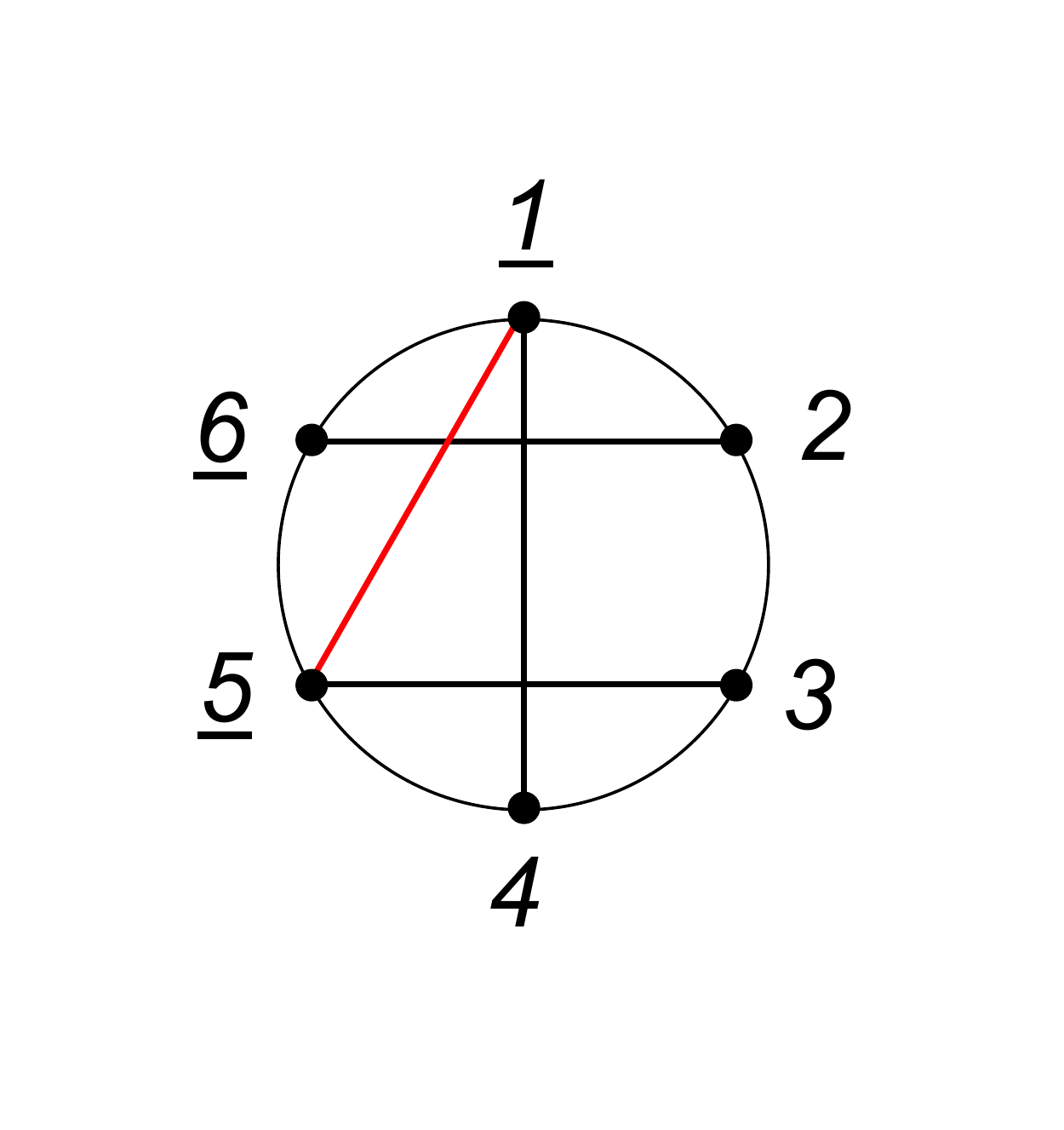}} \,  \,\, 
\parbox[c]{5.1em}{\includegraphics[scale=0.2]{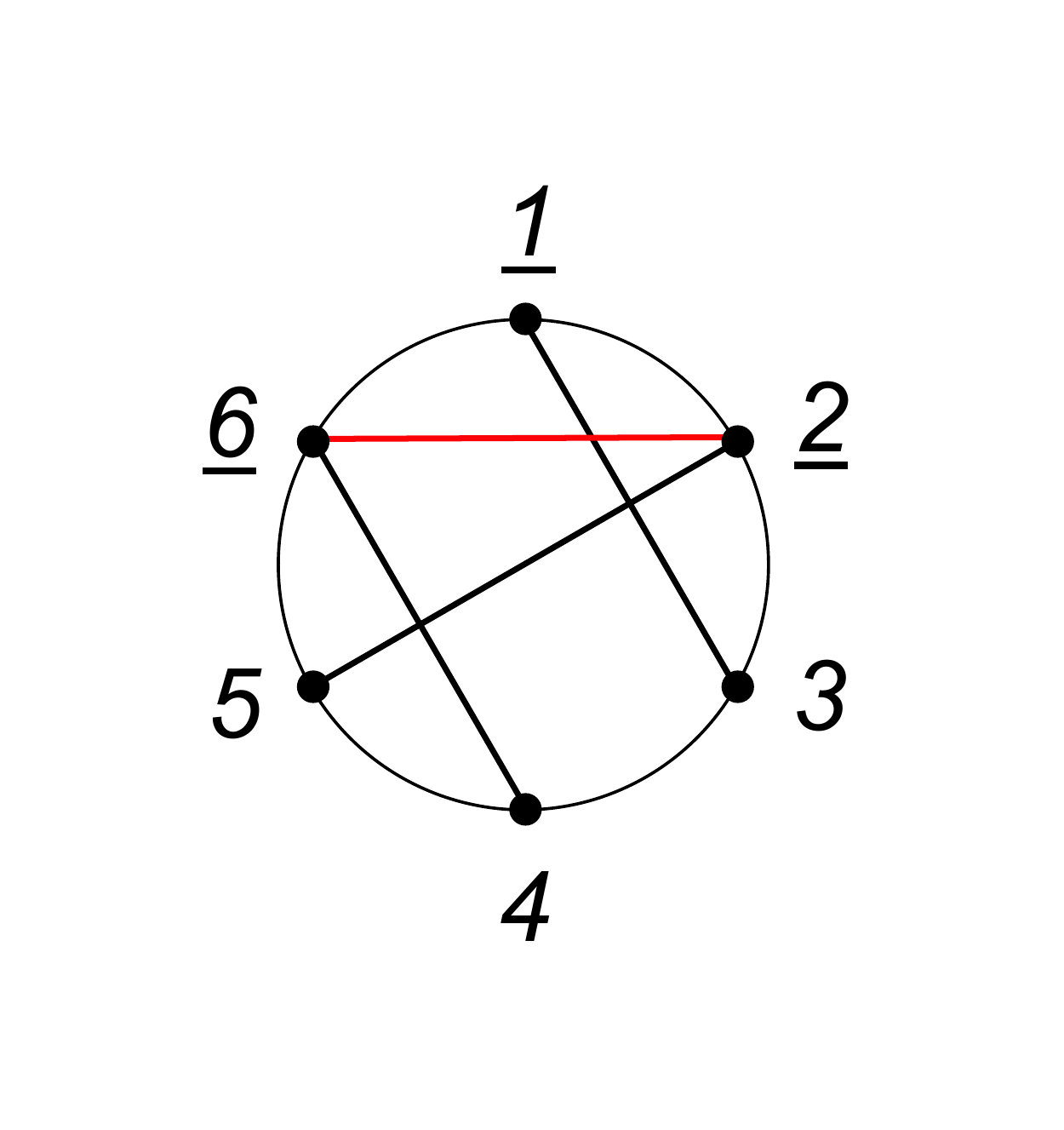}} \, \,\, 
\parbox[c]{5.1em}{\includegraphics[scale=0.2]{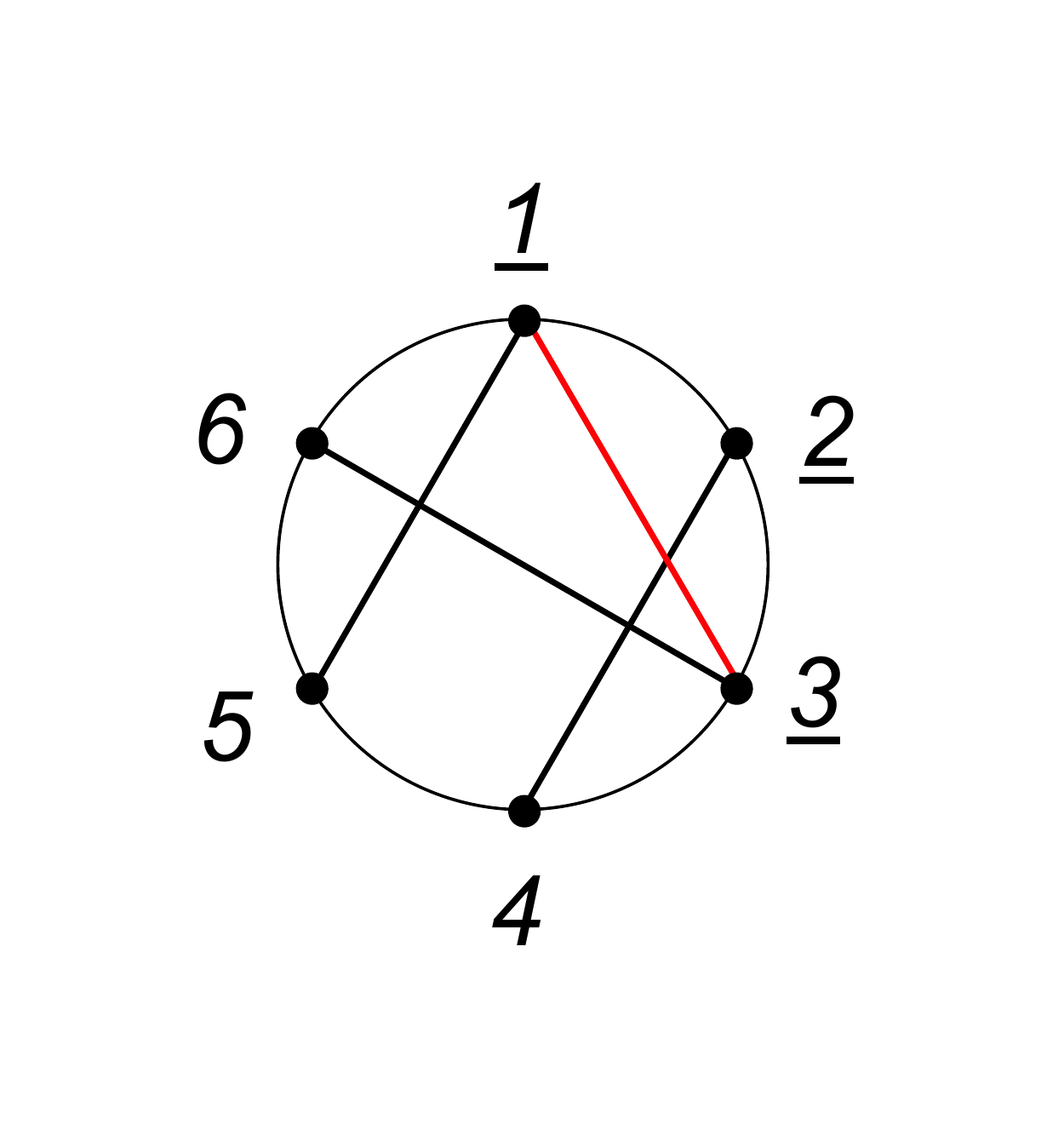}}   
\vspace{-0.45cm}
\caption{Diagrammatic representation of ${\cal A}(\mathbb{I}_6)$.}\label{Fig4} 
\end{figure}

From equation \eqref{eq:7}, we see that ${\cal A}(\mathbb{I}_6)$ 
is encoded in the four diagrams given in Fig. \ref{Fig4}.

In the first diagram, we have fixed the vertices $\{5,6,1\} $ and removed the rows and columns $\{ 1,5\}$ from the $A$-matrix in the reduced Pfaffian:
\begin{equation}\label{}
{\cal A}(\mathbb{I}_6: 14,25,36)  =  \int_{\gamma}  \dif\sigma_2\dif\sigma_3\dif\sigma_4 ( \sigma_{56} \sigma_{61} \sigma_{15} )^2 (S_2S_3S_4)^{-1}  
{\rm PT}(\mathbb{I}_6) \, \frac{{\rm Pf} A^{15}_{15}}{\sigma_{15} }\,  \frac{ (-1) }{ (\sigma_{14} \sigma_{25} \sigma_{36})},
\end{equation}
where the second argument in ${\cal A}$ denotes the perfect matching, and
$\gamma=\gamma_{S_2}\cap \gamma_{S_3} \cap \gamma_{S_4}$.
The integration rules imply that this diagram has only a vanishing factorization cut $\sigma_3\rightarrow\sigma_6$, as we will now show. Considering the parametrization $\s_a=\epsilon \, x_a + \sigma_2$, $a=3,6$, with $x_6=0$, $x_3=\text{constant}$ and $\s_6\equiv \s_L$, we obtain
\begin{align}
&\dif\s_3 = x_{36} \,  \dif \epsilon , \quad S_3=\frac{1}{\epsilon\, x_{36}}  \left[ \tilde S_3  + {\cal O}(\epsilon) \right], \quad \tilde S_3 =\a_{36} , \nonumber \\
& ( \sigma_{56} \sigma_{61} \sigma_{15} )^2 (S_4 S_5)^{-1}    \frac{ {\rm PT}(\mathbb{I}_6)   {\rm Pf} A^{15}_{15}}{\sigma_{15} \, (\sigma_{14} \sigma_{25} \sigma_{36})}  = \frac{( \sigma_{56} \sigma_{61} \sigma_{15} ) \s_{42} }{\epsilon^2\, x_{36}^2} \nonumber\\
&
\qquad\qquad\qquad \qquad\qquad\qquad\qquad\qquad\quad
\times
[(\tilde S_{4} + {\cal O}(\epsilon)) (\tilde S_5 + {\cal O}(\epsilon)) ]^{-1} \a_{24}\a_{36},
\end{align}
where
\begin{equation}
\tilde S_2 = \a_{21}  \s_{26} \s_{24} \s_{25}  + \s_{21}  (\a_{23}+\a_{26}) \s_{24} \s_{25} + \s_{21}  \s_{26} \a_{24} \s_{25} + \s_{21}  \s_{26} \s_{24} \a_{25}, \quad \tilde S_4= \tilde S_2\Big|_{2\leftrightarrow 4}  .\quad
\end{equation}
Using the GRT, we can deform the contour $\gamma=\gamma_{ \tilde S_2  + {\cal O}(\epsilon) }\cap \gamma_{\tilde S_3  + {\cal O}(\epsilon)} \cap \gamma_{\tilde S_4  + {\cal O}(\epsilon)}$  into $\tilde \gamma= \gamma_{\epsilon}\cap \gamma_{\tilde S_2  + {\cal O}(\epsilon)} \cap \gamma_{\tilde S_3  + {\cal O}(\epsilon)}$. After integrating $\epsilon$ around $\gamma_\epsilon$,  the new contour $\tilde \gamma$ becomes, $\tilde \gamma|_{\epsilon=0} =  \gamma_{\tilde S_2 } \cap \gamma_{\tilde S_3 }$.  Since $\tilde S_3$ is independent of $\sigma_a$, this contour is empty and ${\cal A}(\mathbb{I}_6: 14,25,36)  $ vanishes.

\begin{figure}[h]
\centering
\parbox[c]{8.2em}{\includegraphics[scale=0.23]{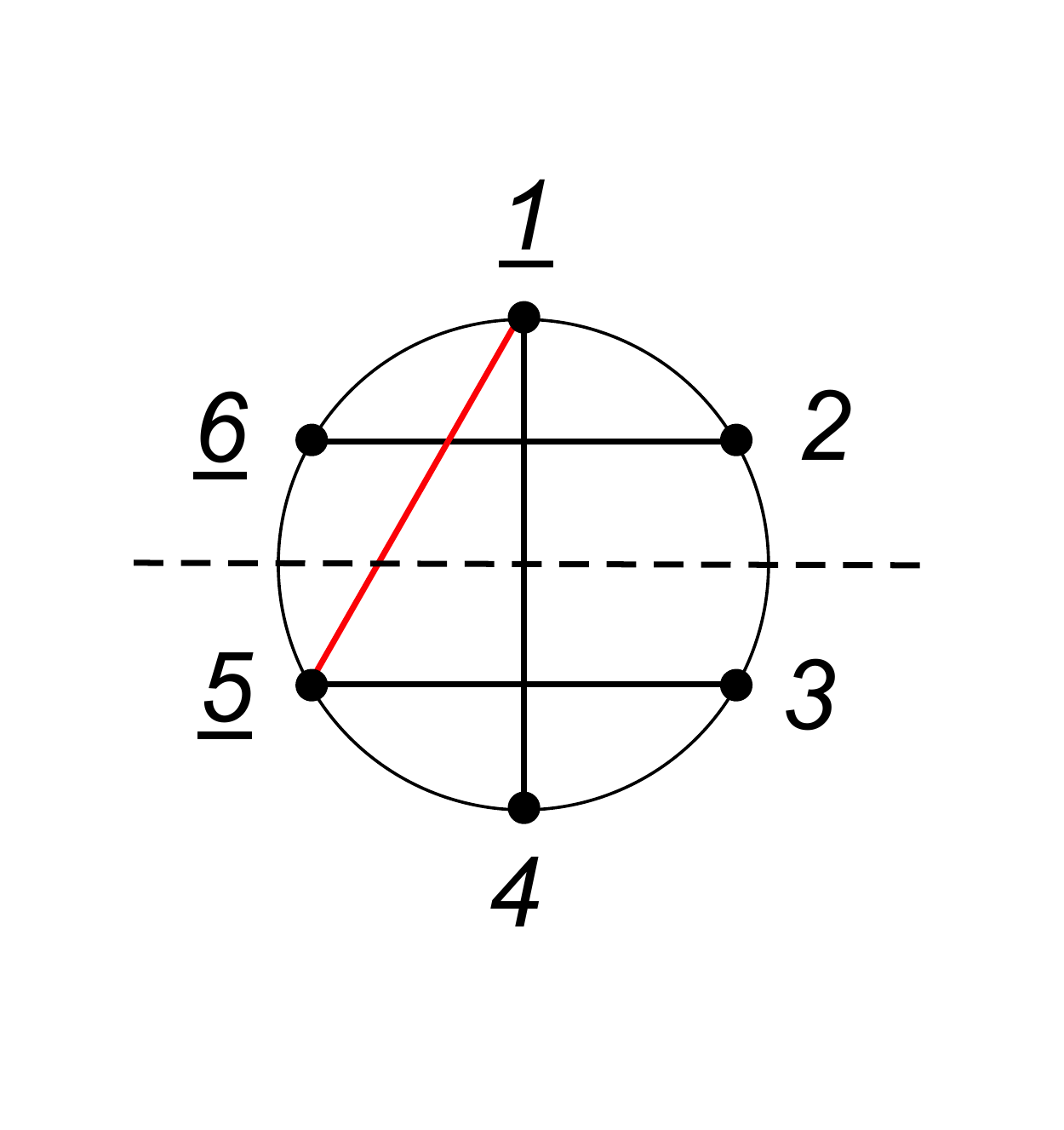}}
$\Rightarrow$
\qquad
\parbox[c]{10.2em}{\includegraphics[scale=0.13]{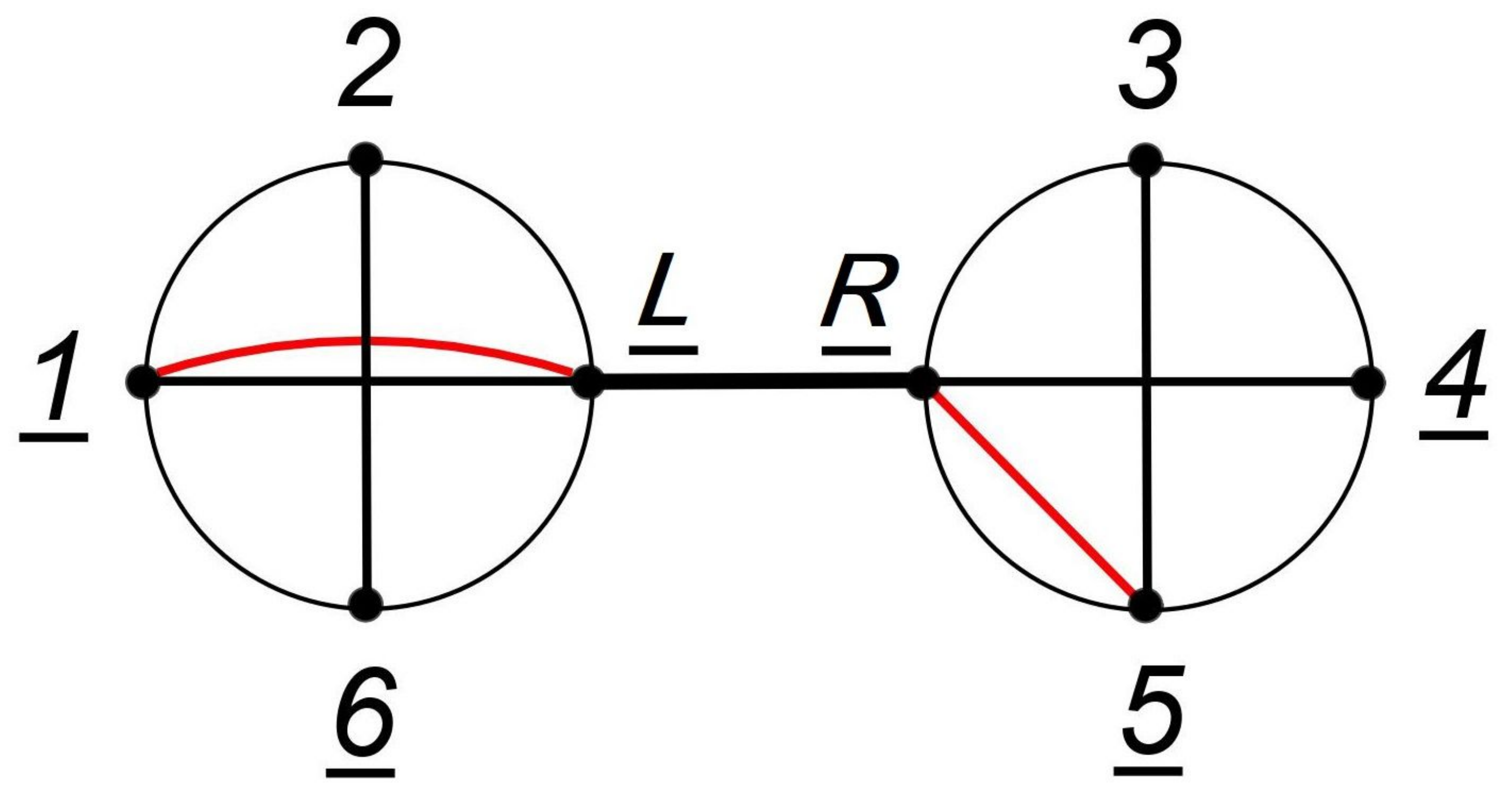}} 
\vspace{-0.45cm}
\caption{${\cal A}(\mathbb{I}_6: 14,26,35)$ diagram and its factorization contribution.}\label{Fig5} 
\vspace{-0.2cm}
\end{figure}

The last three diagrams in Figure \ref{Fig4} are identical up to cyclic permutations, so we will focus on the second one ${\cal A}(\mathbb{I}_6:14,26,35) $. Fixing the punctures $(\s_5,\s_6,\s_1) $ and removing rows and columns $\{ 1,5\}$ from the $A$-matrix leads to
\begin{equation}\label{eq:94}
{\cal A}(\mathbb{I}_6: 14,26,35)  =  \int_{\gamma}  \dif\sigma_2\dif\sigma_3\dif\sigma_4 ( \sigma_{56} \sigma_{61} \sigma_{15} )^2 (S_2S_3S_4)^{-1}  
{\rm PT}(\mathbb{I}_6) \, \frac{{\rm Pf} A^{15}_{15}}{\sigma_{15} }\,  \frac{ 1 }{ (\sigma_{14} \sigma_{26} \sigma_{35})} .
\end{equation}
By the integration rules, the second diagram in Fig. \ref{Fig4} has two factorization cuts: 
 $\sigma_2\rightarrow \sigma_6$ and $\sigma_3\rightarrow \sigma_4 \rightarrow \sigma_5$. If we apply the GRT on the support of $\gamma_{S_4}$ as before, the first factorization $\sigma_2\rightarrow \sigma_6$ vanishes. Thus,
the only contribution comes from the second one (see Fig. \ref{Fig5}).
\vspace{-0.0cm}
\noindent

In order to compute the contribution in Fig. \ref{Fig5}, we consider the parametrization, $\sigma_a=\epsilon x_a+\sigma_5$,  
with $a=3,4,5$, $x_4=\text{constant}$, $x_5=0$, $\sigma_5=\sigma_L$,
and expand around $\epsilon=0$, obtaining
\begin{align}\label{eq:95}
&
\dif\sigma_2\wedge\dif\sigma_3\wedge \dif\sigma_4=\dif\sigma_2\wedge (\epsilon \, x_{45}\,\dif x_3\wedge \dif\epsilon)
\nonumber \\
&
(\sigma_{15} \sigma_{56} \sigma_{61})^2 \, {\rm PT}(\mathbb{I}_6)\, 
\frac{    {\rm Pf}A^{15}_{15}  }{  \sigma_{15} } \,  \frac{1}{(\sigma_{14} \sigma_{26} \sigma_{35} )}
= \frac{(\sigma_{1L} \sigma_{L 6} \sigma_{61})^2 }{ ( \sigma_{12}\sigma_{2 L }\sigma_{L  6}\sigma_{61} ) }
 \frac{\a_{26}}{\sigma_{1 L }\sigma_{26}}     \frac{1 }{(\sigma_{1 L }\sigma_{26})}  \qquad\nonumber\\
&
\qquad\qquad \qquad\qquad\qquad\qquad
\times
\frac{x^2_{5R} x_{R4} }{ (x_{R3} x_{34} x_{45})} \frac{1}{(x_{35} x_{R4})} \frac{1}{x_{R5}}\frac{\a_{34}}{x_{34}} \, \frac{1}{\epsilon^4}
+ {\cal O}(\epsilon^{-3}), 
\end{align}
and
\begin{align}\label{}
& 
S_2 = \left[\hat S_2 + {\cal O}(\epsilon) \right] , \qquad  
S_3 = \frac{1}{\epsilon}   \left[ \hat S_3  + {\cal O}(\epsilon) \right] , \qquad
S_4 = \frac{1}{\epsilon}   \left[ \hat S_4 + {\cal O}(\epsilon) \right] , \nonumber\\
&
\hat S_2=\frac{\a_{21}}{\sigma_{21}}  +\frac{\a_{26}}{\sigma_{26}} + \frac{\a_{2L}}{\sigma_{2L}} ,
  \qquad \hat S_3=\frac{\a_{34}}{x_{34}}  +\frac{\a_{35}}{x_{35}} + \frac{\a_{3R}}{x_{3R}} , 
 \qquad \hat S_4=\frac{\a_{43}}{x_{43}}  +\frac{\a_{45}}{x_{45}} + \frac{\a_{4R}}{x_{4R}} , 
\end{align}
where $x_{R}=\infty$, $\a_{2L}= \a_{23}+\a_{24}+\a_{25}$ and $\a_{a R}= \a_{a6}+\a_{a1}+\a_{a2}$, $a=3,4$. From \eqref{eq:95}, it is straightforward to check there are two real poles, $\sigma_{26}=0$ and $\epsilon=0$. The contour $\gamma$ can then be deformed into $\hat\gamma=\hat\gamma_1+\hat\gamma_2$, with $\hat\gamma_1=\gamma_{26}\cap\gamma_{ S_2}\cap\gamma_{ S_3}$, $\gamma_{26} = \{ |\s_2-\s_6 |=\delta \}$, and $\hat\gamma_2= \gamma_{\epsilon}\cap\gamma_{ S_2}\cap\gamma_{ S_3}$. As explained above, the integration over $\gamma_{26}$ vanishes so the only contribution comes from the second contour $\hat \gamma_2$. 

After integrating over $\gamma_\epsilon$, the full integral is split into two parts: one with $\{\sigma_1,\sigma_2,\sigma_L,\sigma_6\}$ and the other with $\{x_3,x_4,x_5,x_{R}\}$. Moreover, using the identity
\begin{equation}
{\rm PT}(4,5,R )\,\hat S_4 + {\rm PT}(3,5,R )\, \hat S_3 =  {\rm PT}(5,R )\, \left[ ( {\cal D}_3+ {\cal D}_4 + {\cal D}_5)^2 + m^2 \right],
\end{equation}
we find that on the support of $\gamma_{S_3}$, $S_4$ reduces to
\begin{equation}
\hat S_4\Big|_{\gamma_{\hat S_3}} = \frac{{\rm PT}(5,R) }{{\rm PT}(4,5,R)} \, \left[ ( {\cal D}_3+ {\cal D}_4 + {\cal D}_5)^2 + m^2 \right].
\end{equation}
Putting everything together, ${\cal A}(\mathbb{I}_6:14,26,35)$ can then be written as
\begin{align}
&
{\cal A}(\mathbb{I}_6:14,26,35) 
= \int_{\gamma_{\hat S_2}} \Big[\dif \s_{2}   (\hat S_2)^{-1}  \Big] 
(\s_{61} \s_{1L}   \s_{L6}  )^2 \, {\rm PT}(6,1,2,L) 
 \left[ ( {\cal D}_3+ {\cal D}_4 + {\cal D}_5)^2 + m^2
\right]^{-1}
\nonumber
\\
&
\times
\int_{\gamma_{\hat S_3}}  \Big[ \dif x_{3}   (\hat S_3)^{-1}   \Big]
(x_{R4}  x_{45}  x_{5R})^2  \, {\rm PT}(R,3,4,5)
 \left[
\frac{1}{\s_{1L}} \frac{\a_{26}}{\s_{26}} \, \frac{1}{\s_{26} \s_{1L}} \right]\left[
\frac{(-1)}{x_{R5}} \frac{\a_{34}}{x_{34}} \, \frac{1}{x_{35} x_{R4}} \right].
 \label{}
\end{align}
Since $\a_{3 R}= \a_{31}+\a_{32}+\a_{36} = -(\a_{34}+\a_{35})$ by the Ward identities, $\a_{26}$ commutes with $\hat S_3$ and $ [ ( {\cal D}_3+ {\cal D}_4 + {\cal D}_5)^2 + m^2]$, and we obtain
\begin{align}\label{eq:14}
{\cal A}(\mathbb{I}_6:14,26,35) = {\cal A}(6,1,2,L:1L,26)  
[ ( {\cal D}_{3}+{\cal D}_{4}+{\cal D}_{5})^2+m^2 ]^{-1}
{\cal A}(R,3,4,5:R4,35) ,\qquad 
\end{align}
where
\begin{align}\label{eq:15}
{\cal A}(6,1,2,L:1L,26) = 
\int_{\gamma_{\hat S_2}} \!\!\! \dif \s_2 (\s_{1L} \s_{L 6} \s_{61})^2 ({\hat S}_2)^{-1}  {\rm PT}(6,1,2,L) \,
\frac{ {\rm Pf}A^{1L}_{1L}}{\s_{1L}} \frac{1}{(\s_{1L} \s_{26})}  ,
\end{align}
\begin{align}\label{eq:16}
{\cal A}(R,3,4,5:R4,35)  = 
\int_{\gamma_{ \hat S_3}} \!\!\!\dif x_3 (x_{45} x_{5R} x_{R4})^2 (\hat{S}_3)^{-1} \, {\rm PT}(R,3,4,5) \,
 \frac{(-1) {\rm Pf}A^{R5}_{R5}}{  x_{R5}} \frac{1}{(x_{R4}x_{35})}  ,
\end{align}
and
\begin{align}\label{}
{\rm Pf}A^{1L}_{1L} = \frac{\a_{26}}{\s_{26} } \, ,\qquad 
{\rm Pf}A^{R5}_{R5} = \frac{\a_{34}}{x_{34} }.
\end{align}

\begin{figure}[h]
\centering
\text{$
\mathord{\begin{tikzpicture}[scale=0.4]
\node at (-3,0) [circle,,fill=black,inner sep=0pt,minimum size=0mm,label=center:$ \text{${\scriptstyle \eta=0}$} $]  {};
	\node at (0,0) [circle,,fill=black,inner sep=0pt,minimum size=0mm,label=above:$ $]  {};
	\node at (1,0) [circle,,fill=black,inner sep=0pt,minimum size=1.5mm,label=above:$6$]  {};
	\node at (2,0) [circle,,fill=black,inner sep=0pt,minimum size=0mm,label=above:$ $]{};
	\node at (3,0) [circle,,fill=black,inner sep=0pt,minimum size=1.5mm,label=above:$1$]{};
	\node at (4,0) [circle,,fill=black,inner sep=0pt,minimum size=0mm,label=above:$ $]{};
	\node at (5,0) [circle,,fill=black,inner sep=0pt,minimum size=1.5mm,label=above:$2$]{};
		\node at (6,0) [circle,,fill=black,inner sep=0pt,minimum size=0mm,label=above:$ $]{};
			\node at (7,0) [circle,,fill=black,inner sep=0pt,minimum size=1.5mm,label=above:$3$]{};
				\node at (8,0) [circle,,fill=black,inner sep=0pt,minimum size=0mm,label=above:$ $]{};
				   \node at (9,0) [circle,,fill=black,inner sep=0pt,minimum size=1.5mm,label=above:$4$]{};
				\node at (3,-2) [circle,,fill=black,inner sep=0pt,minimum size=1.0mm,label=above:$ $]{};
				 \node at (11,0) [circle,,fill=black,inner sep=0pt,minimum size=1.5mm,label=above:$5$]{};
				 \node at (9,-2) [circle,,fill=black,inner sep=0pt,minimum size=1.0mm,label=above:$ $]{};
	\draw[very thick,black] (-1,0) -- (13,0);
		\draw[black] (1,0) -- (3,-2);
			\draw[black] (3,-2) -- (3,0);
			   \draw[black] (5,0) -- (3,-2);
			       \draw[black] (9,-2) -- (7,0);
			        \draw[black] (9,-2) -- (9,0);
			        \draw[black] (9,-2) -- (11,0);
			        	\draw[black] (3,-2) -- (9,-2);
	\end{tikzpicture}}
	$}
	\caption{Six-point $\phi^4$ Witten diagram}\label{6ptfig}
\end{figure}
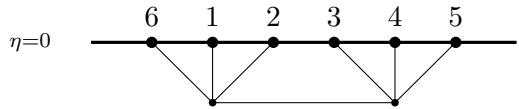

The expression in \eqref{eq:14} is the analytic representation of the factorization given in Fig. \ref{Fig5}.
In section \ref{sec:fourpoint}, we showed that  ${\cal A}(6,1,2,L:1L,26) = {\cal A}(R,3,4,5:R4,35) = \mathbb{I}$. It is then straightforward to see that 
\begin{equation}\label{eq:17}
{\cal A}(\mathbb{I}_6:14,26,35) \, \,  {\cal C}_6^\Delta \!= \!  [ ( {\cal D}_{3}+{\cal D}_{4}+{\cal D}_{5})^2+m^2 ]^{-1} \, \,  {\cal C}_6^\Delta,
\end{equation}
which is the Witten diagram for two four-point vertices connected by a bulk-to-bulk propagator, illustated in Figure \ref{6ptfig}.

Since all terms in \eqref{eq:14} commute, and the CSE commute with the Pfaffian in each four-point integrand, this implies that shuffling terms in the Pfaffian with the CSE leaves the final result unchanged. Finally, note that the other diagrams in Fig. \ref{Fig4} can be computed just by relabelling ${\cal A}(\mathbb{I}_6:14,26,35)$. After summing over permutations, we see that \eqref{eq:totalphi4} reproduces all ten Witten diagrams contributing to the six-point correlator with the correct coefficients. In Appendix \ref{sixpoint-16}, we re-compute the six-point correlator using a less optimal choice of Pfaffian than the one used in \eqref{eq:94}. The calculation turns out to be a lot more laborious but leads to the same result, demonstrating the gauge invariance of the method used in this section and the power of the integration rules. 

\section{Eight Points and Beyond}\label{sec:eightpoint}

In this section we will use the proposed formula \eqref{eq:totalphi4} to compute the tree-level eight-point correlator. This correlator involves perfect matchings with non-ladder topologies and therefore exhibits more generic structure than the lower-point correlators considered in the previous section. We then sketch how to extend our eight-point calculations to $n$-points.

\begin{figure}[h]
\centering
\parbox[c]{5.1em}{\includegraphics[scale=0.2]{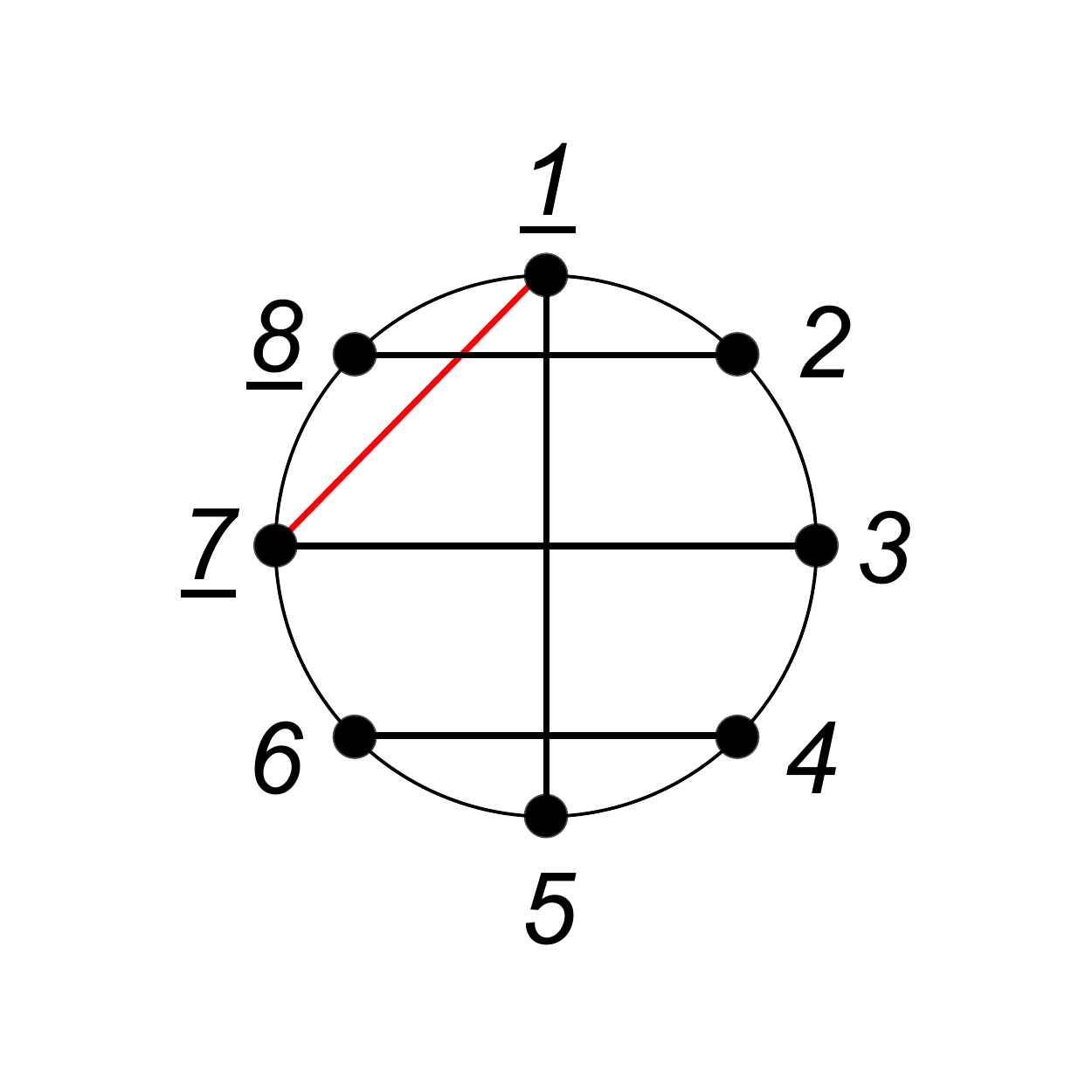}}\,\, \,
\parbox[c]{5.1em}{\includegraphics[scale=0.2]{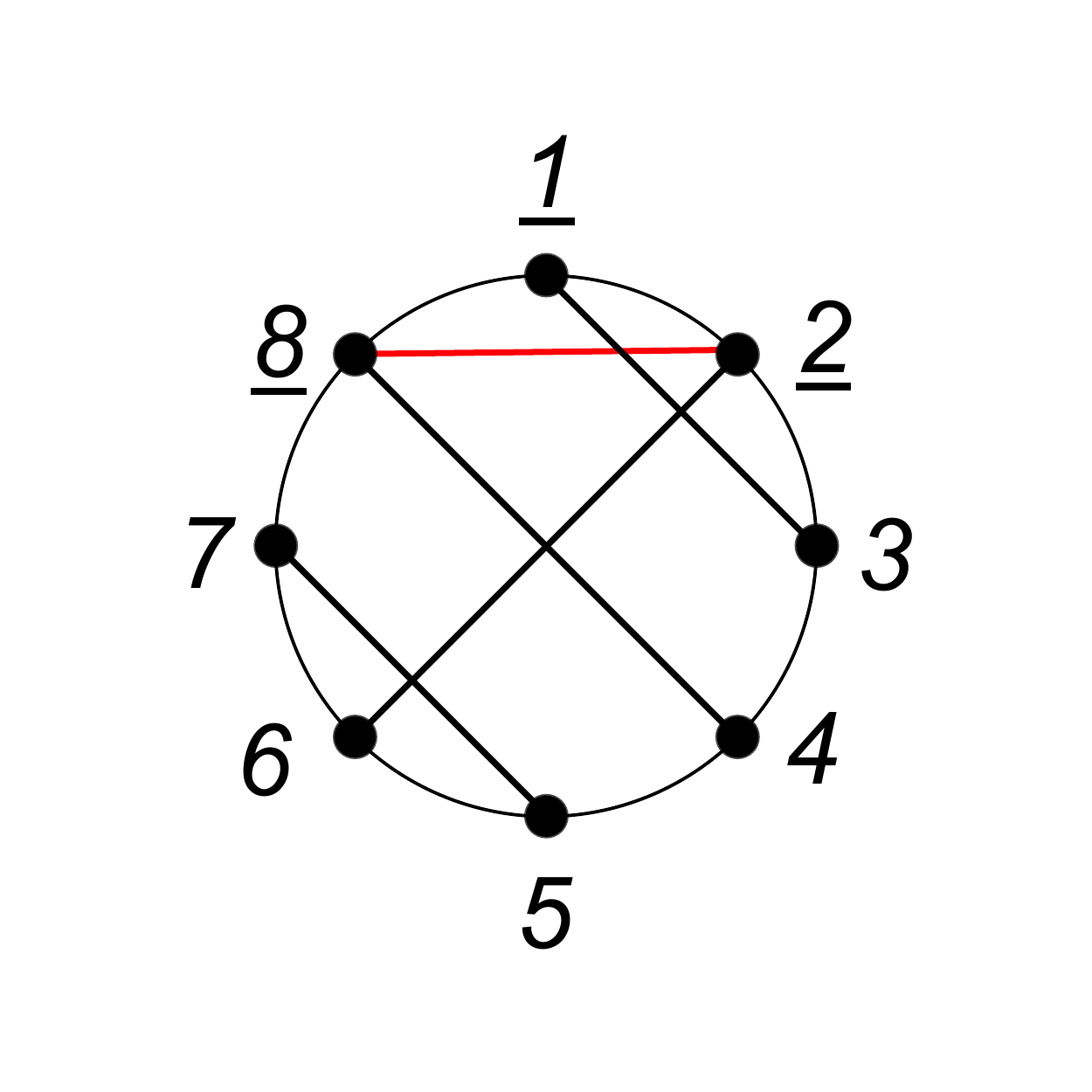}} \,  \,\, 
\parbox[c]{5.1em}{\includegraphics[scale=0.2]{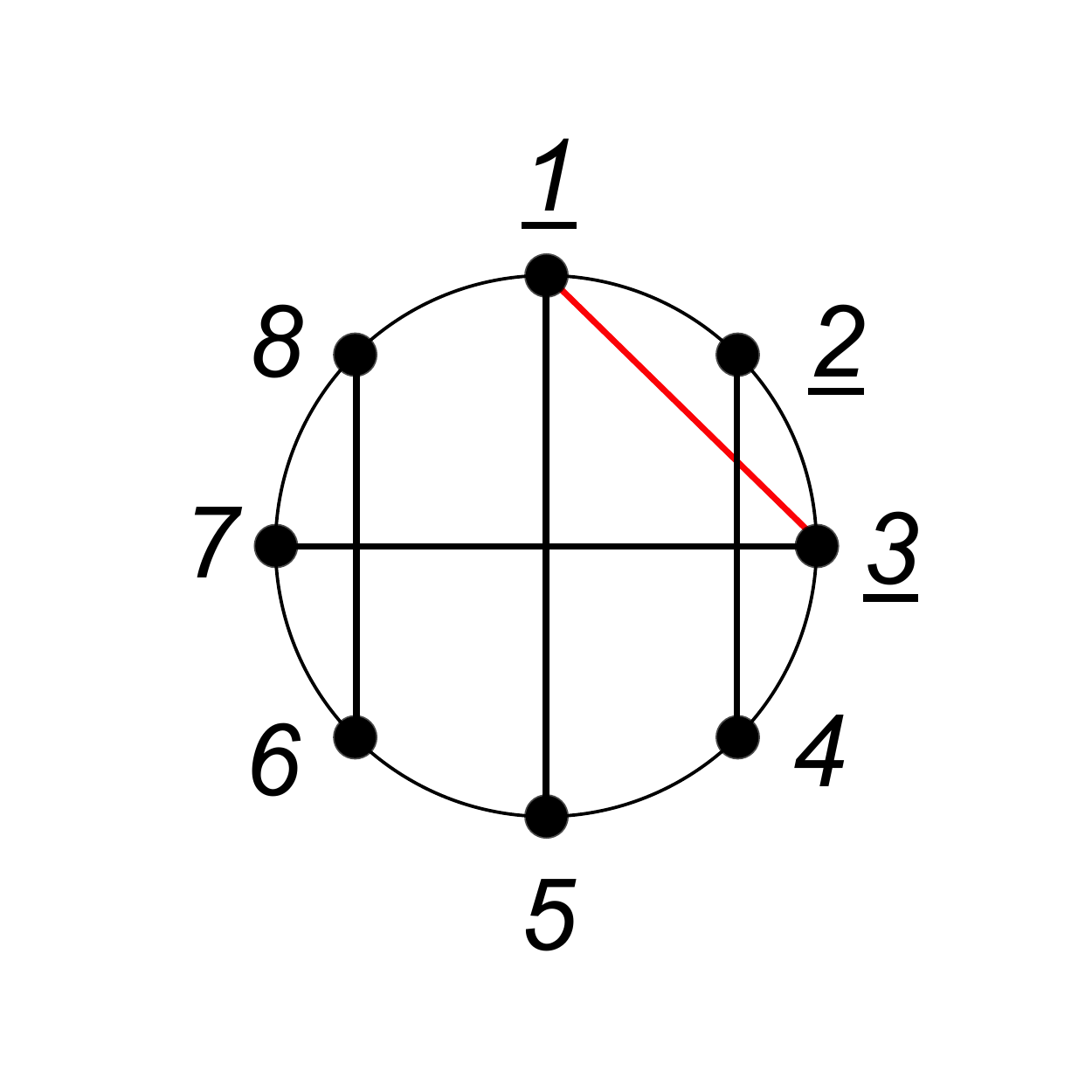}} \, \,\, 
\parbox[c]{5.1em}{\includegraphics[scale=0.2]{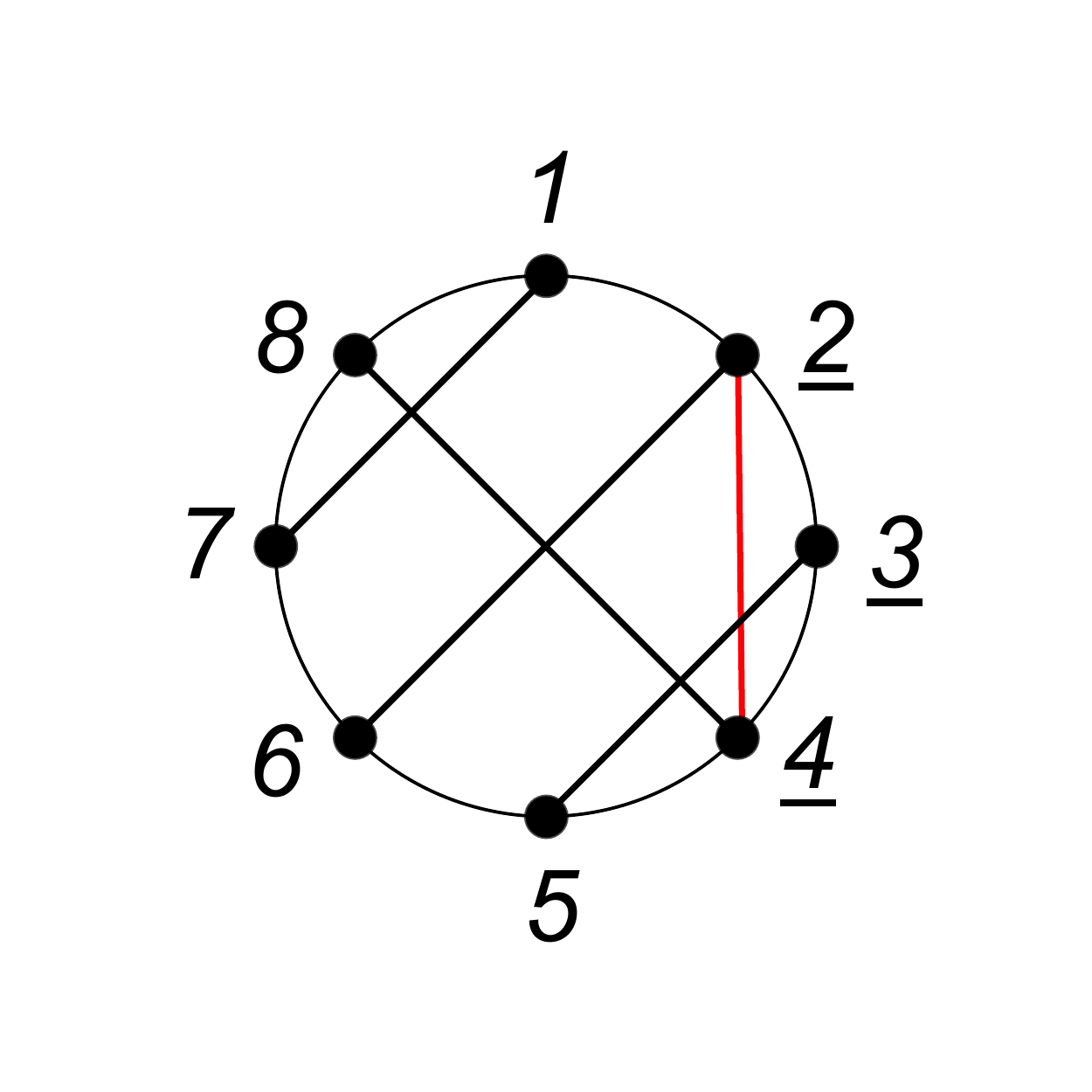}}   \\
\parbox[c]{4.7em}{\includegraphics[scale=0.2]{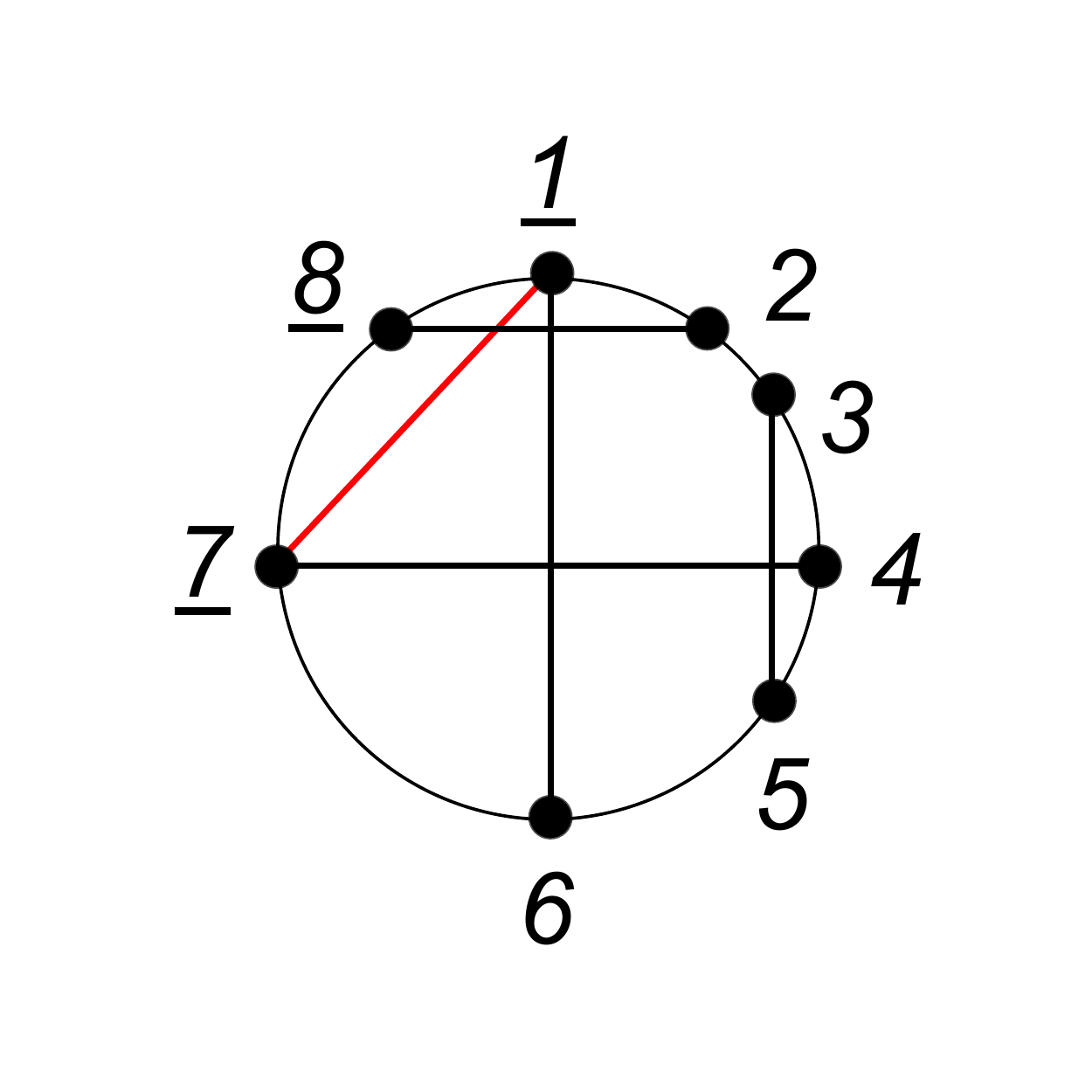}}  
\parbox[c]{4.7em}{\includegraphics[scale=0.2]{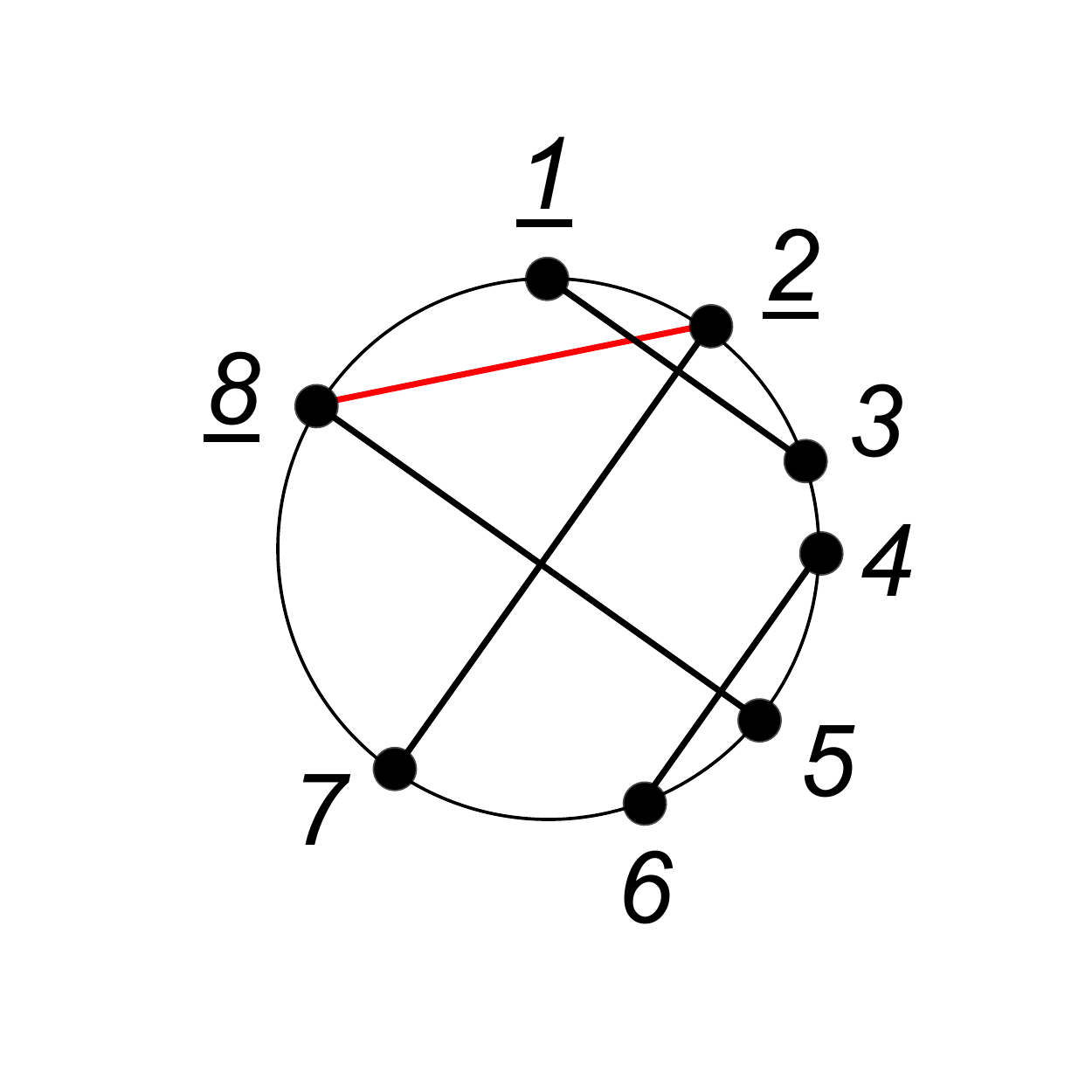}}  
\parbox[c]{4.7em}{\includegraphics[scale=0.2]{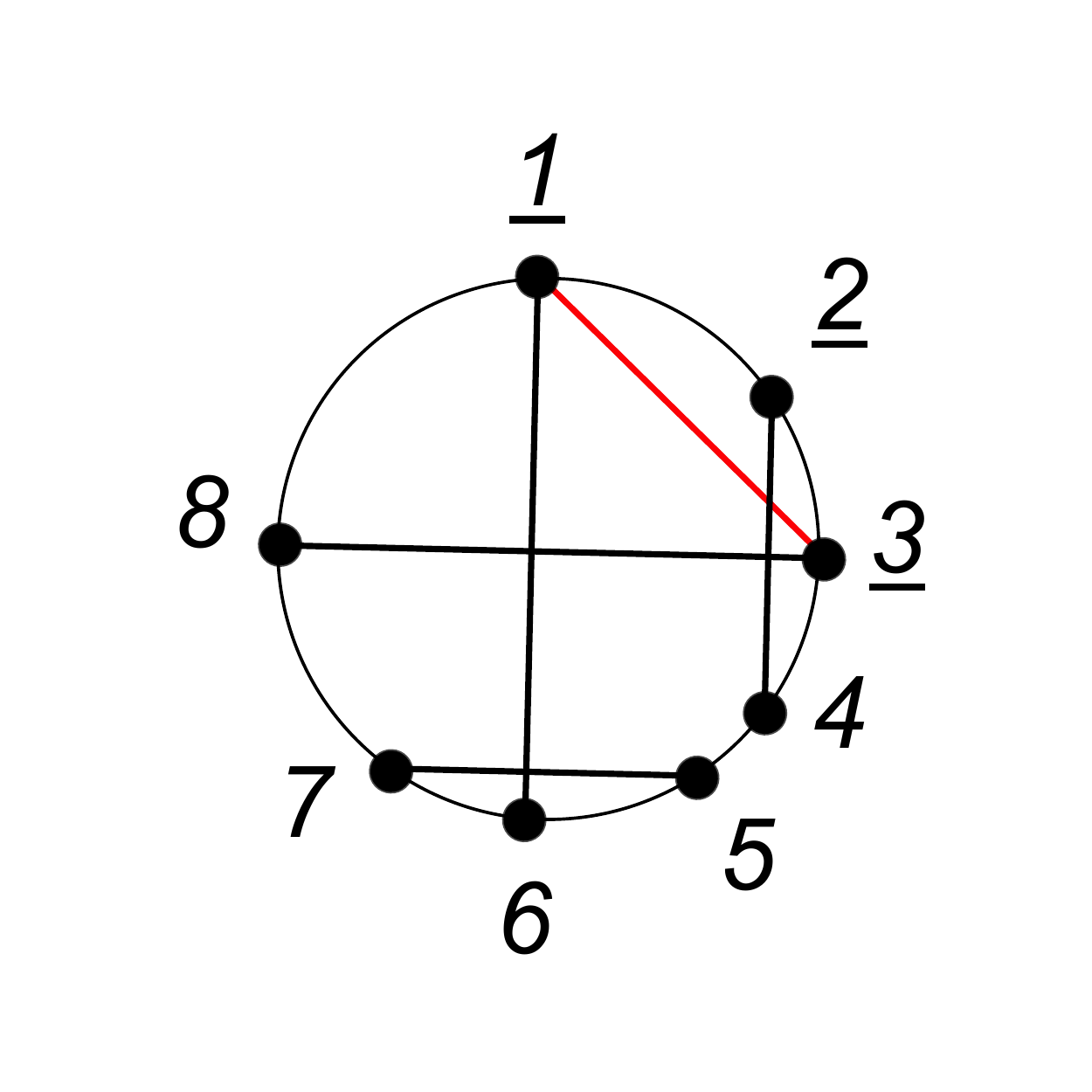}}  
\parbox[c]{4.7em}{\includegraphics[scale=0.2]{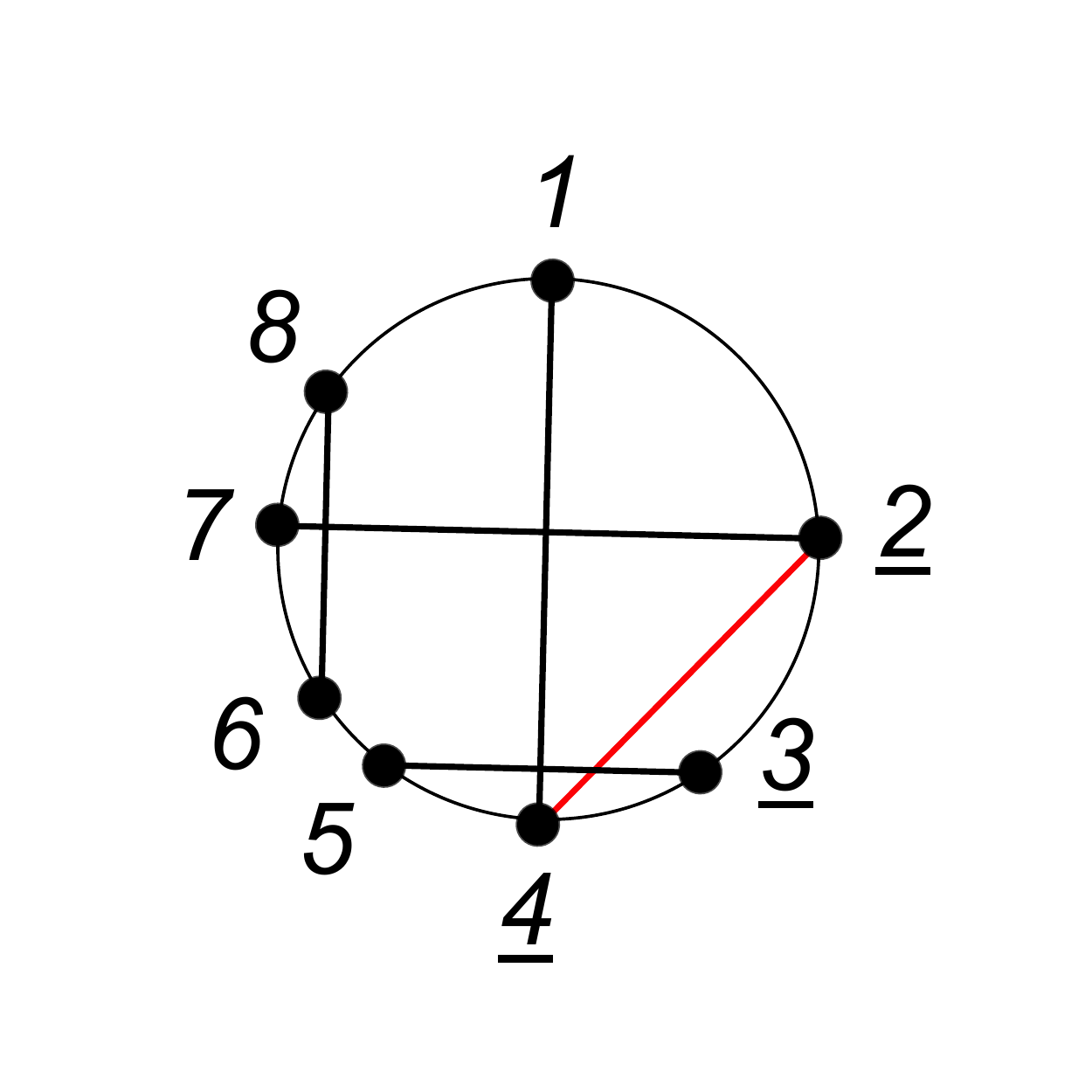}}  
\parbox[c]{4.7em}{\includegraphics[scale=0.2]{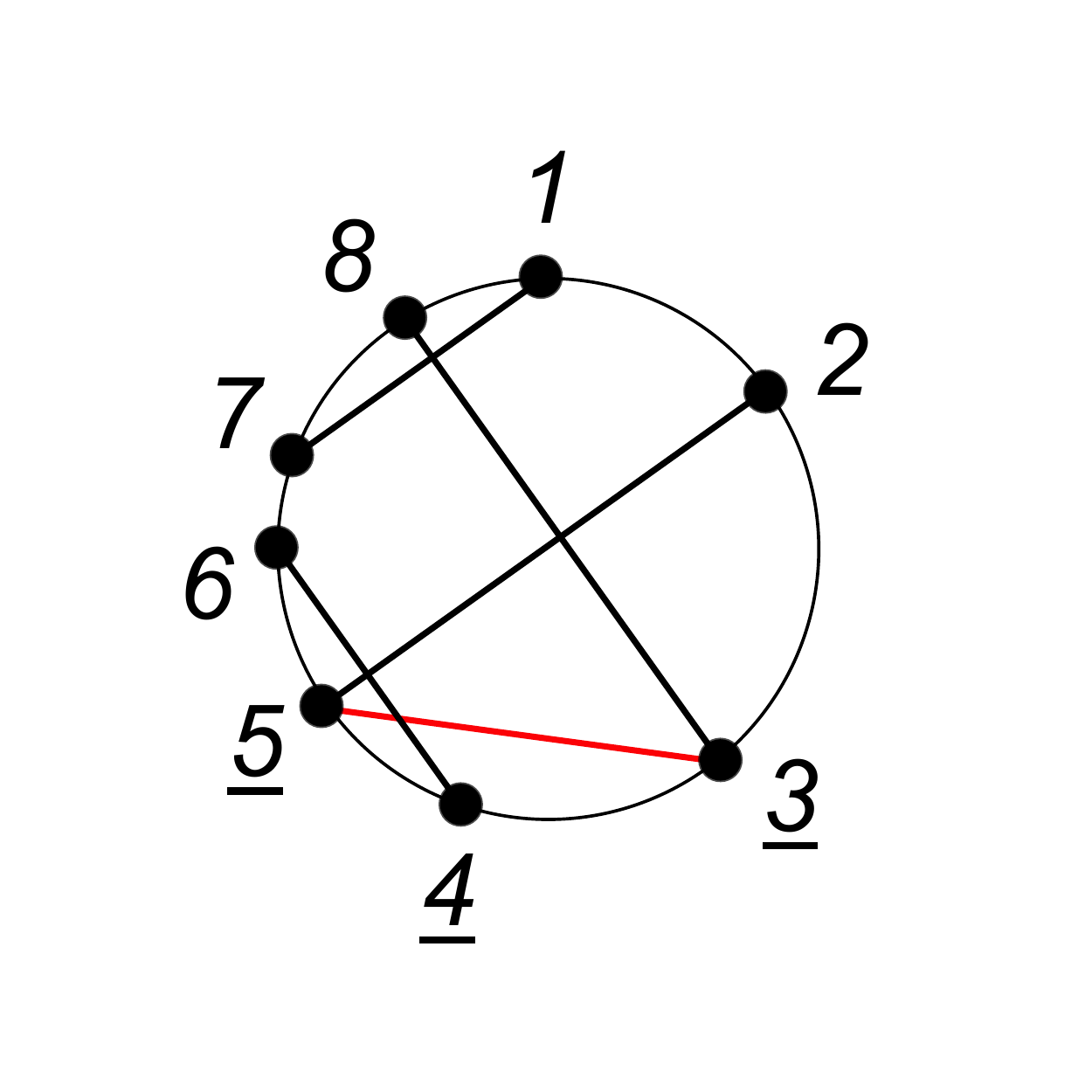}}  
\parbox[c]{4.7em}{\includegraphics[scale=0.2]{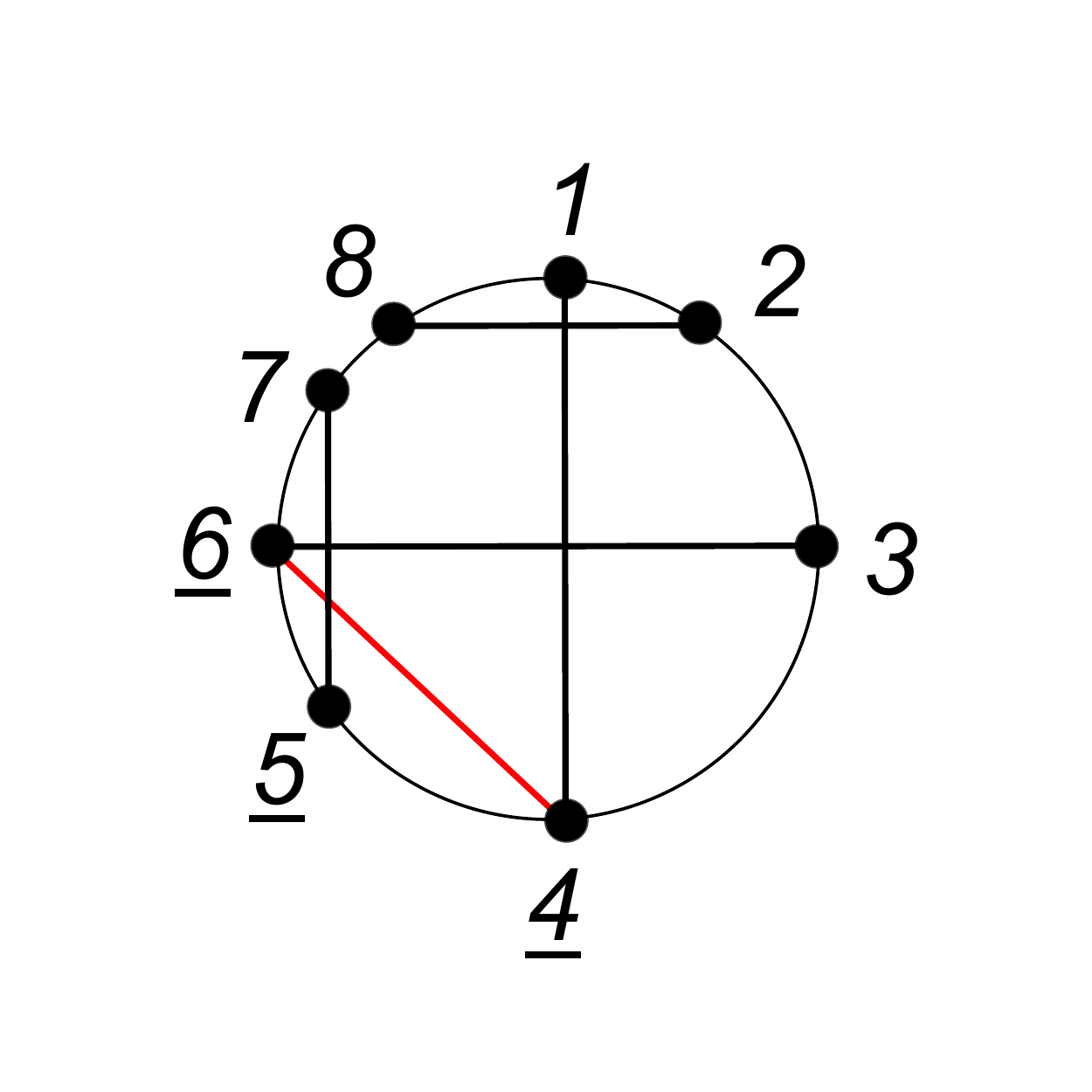}}  
\parbox[c]{4.7em}{\includegraphics[scale=0.2]{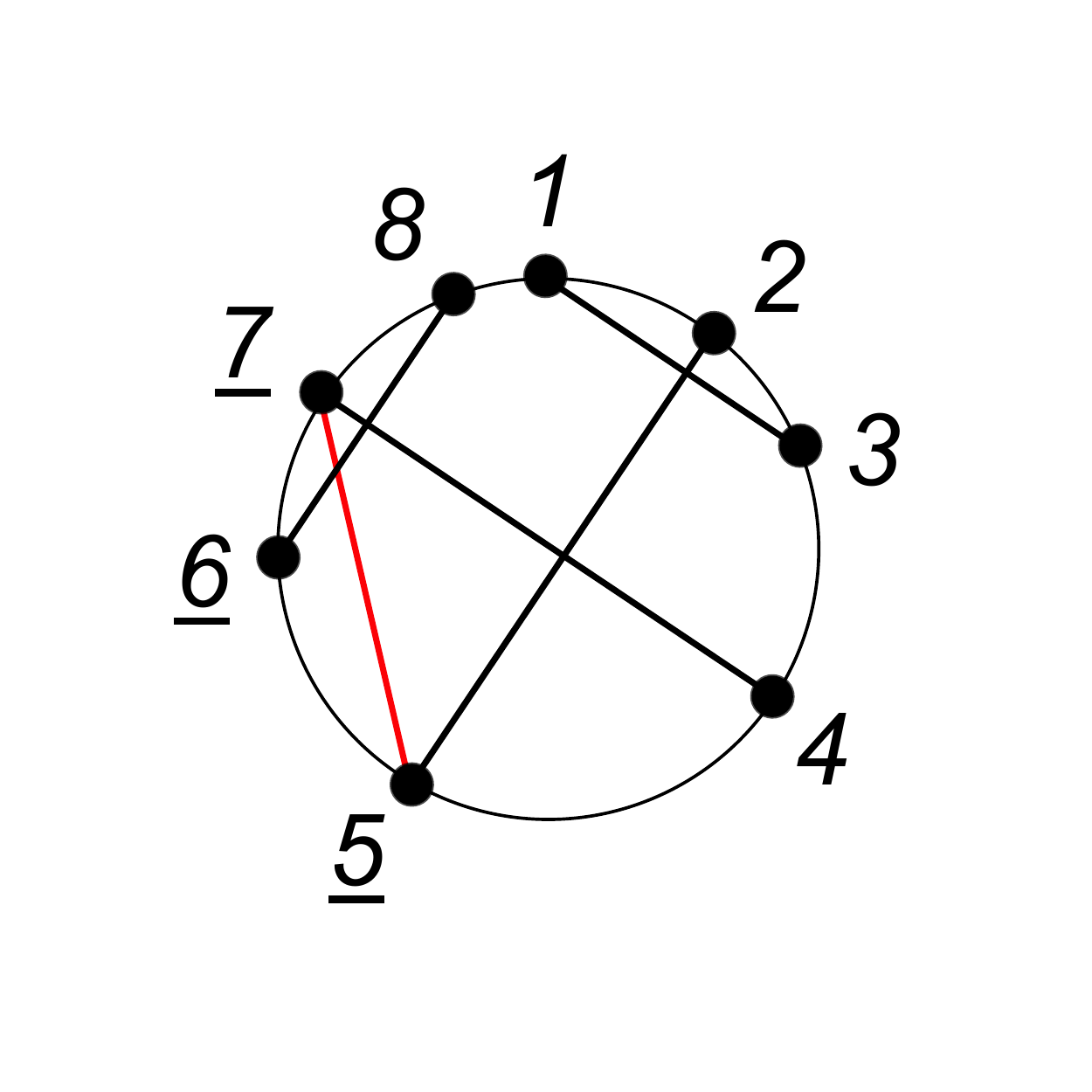}}  
\parbox[c]{4.7em}{\includegraphics[scale=0.2]{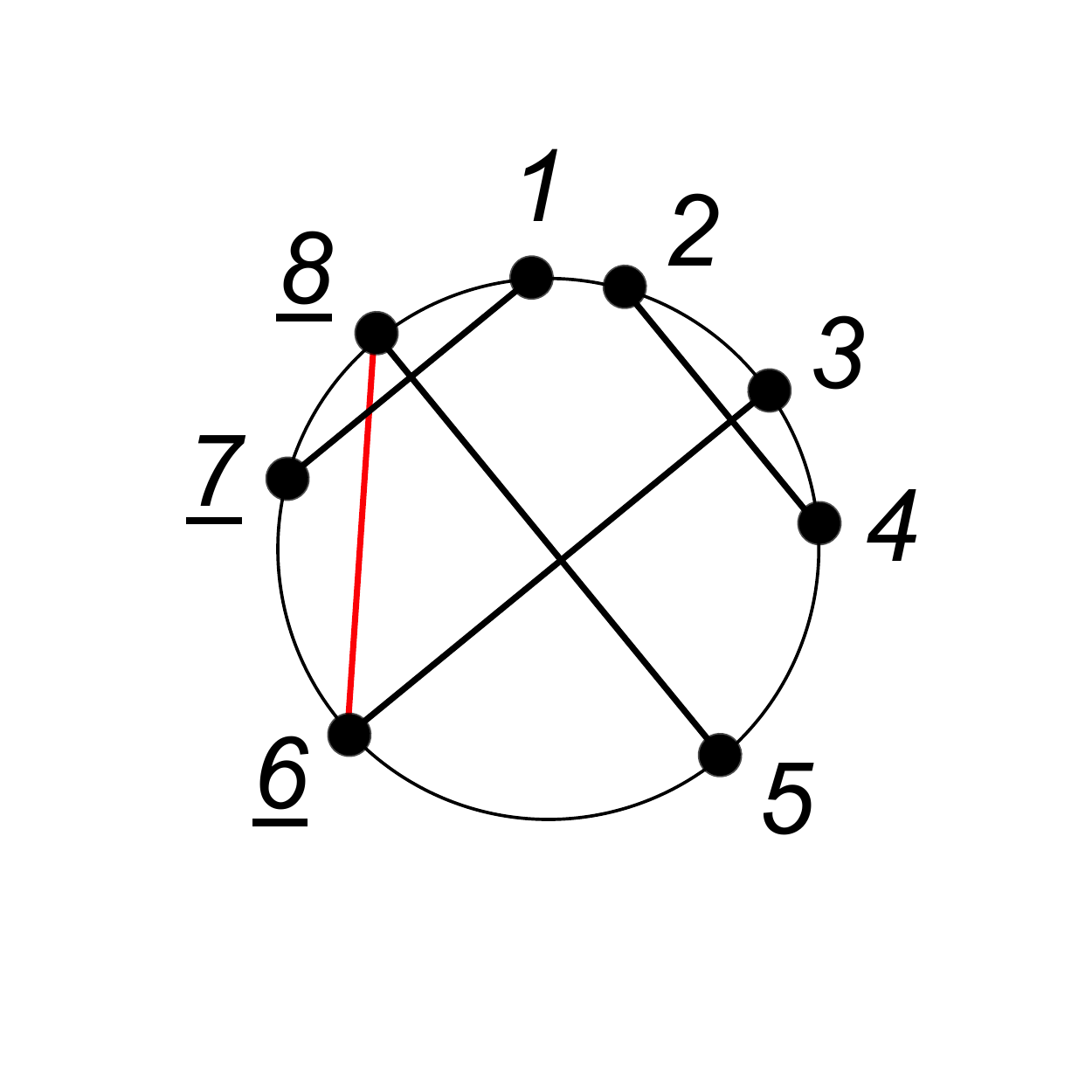}}  
\caption{All non-zero diagrammatic contributions for ${\cal A}(\mathbb{I}_8)$.}\label{Fig6} 
\end{figure}

From the integration rules of subsection \ref{IntegrationRules}, we find there are twelve non-zero contributions at eight-points, given in Fig. \ref{Fig6}.

Observe that there are only two different topologies: ladder and non-ladder diagrams, respectively first and second rows in Fig. \ref{Fig6}. Furthermore, several graphs are simply related by relabelings, so it is enough to sample one in each row.

\subsection{Ladder Contributions}\label{Ladder-8}

\begin{figure}[h]
\centering
\vspace{-0.2cm}
\hspace{-2.7cm}
\parbox[c]{7.1em}{\includegraphics[scale=0.25]{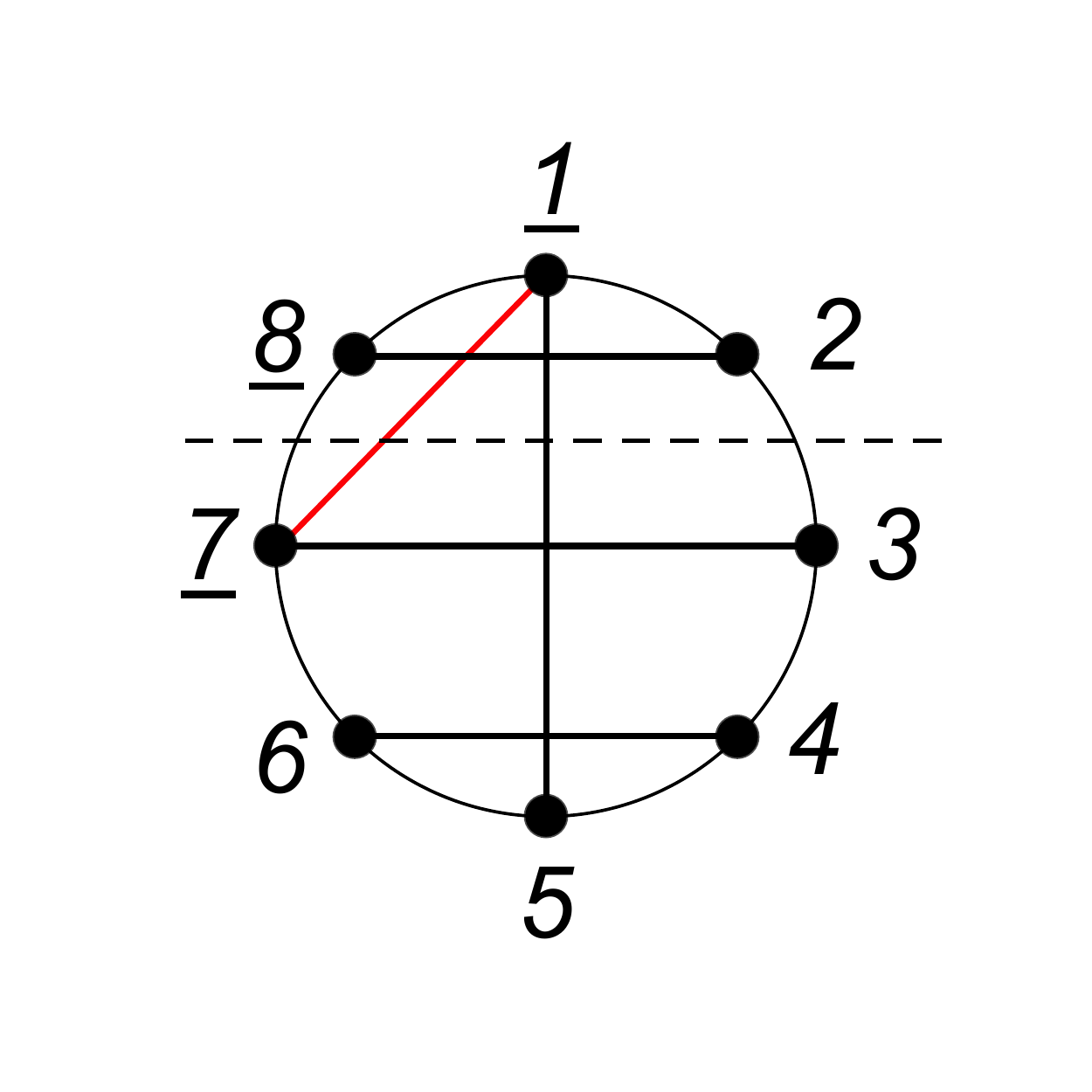}}\,\qquad
$\Rightarrow$
\quad
\parbox[c]{5.1em}{\includegraphics[scale=0.34]{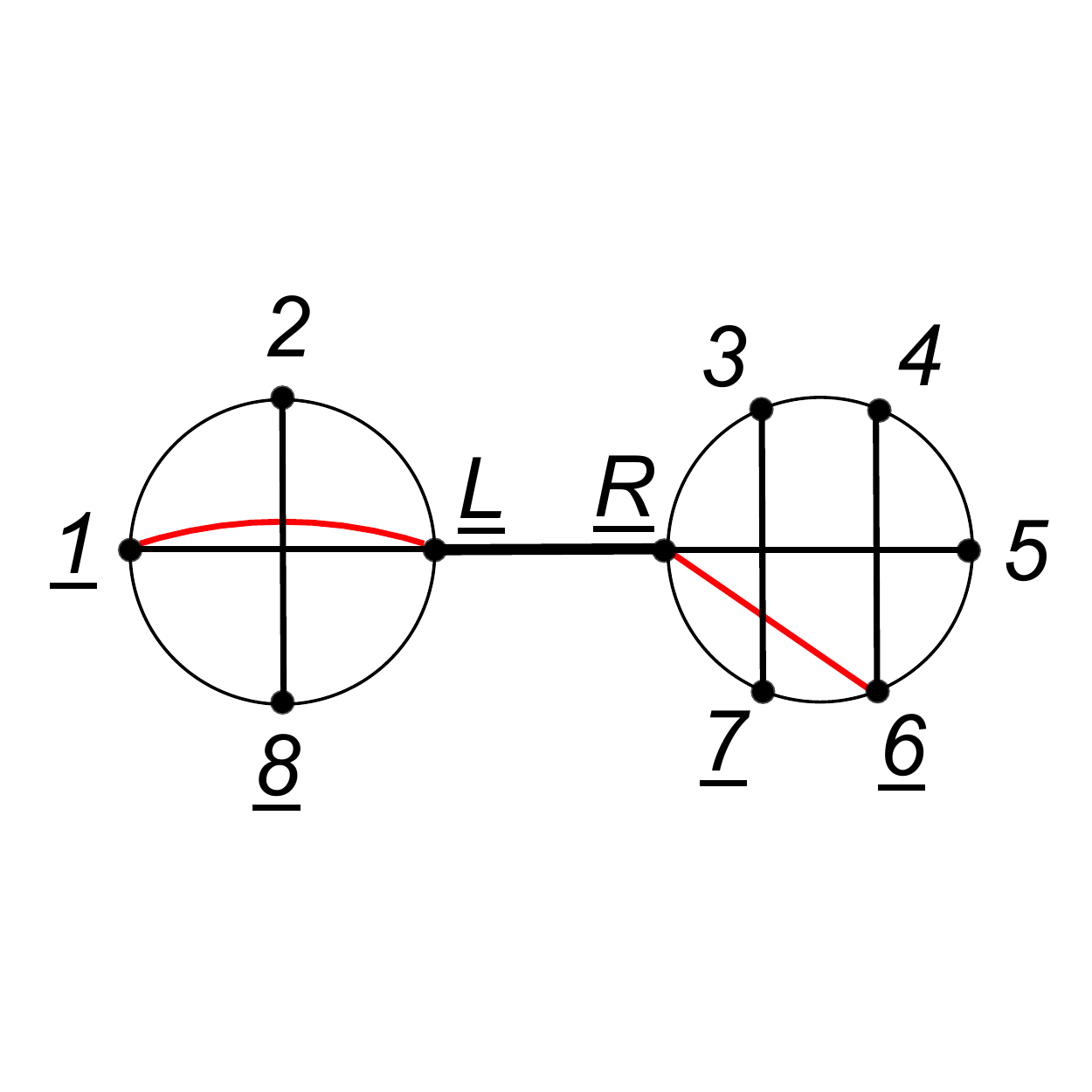}}\,\, \, 
\vspace{-0.7cm}
\caption{Ladder diagram, ${\cal A}(\mathbb{I}_8:15,28,37,46)$, and its factorization contribution.}\label{Fig7} 
\end{figure}

Let us consider the first diagram in Fig. \ref{Fig6}, namely  ${\cal A}(\mathbb{I}_8, 15,28,37,46)$, with fixed punctures $ (z_7,z_8,z_1)$:
\begin{equation}\label{}
{\cal A}(\mathbb{I}_8: 15,28,37,46)  =  \int_{\gamma}  \prod_{a=2}^6 \dif\sigma_a ( \sigma_{78} \sigma_{81} \sigma_{17} )^2 \prod_{a=2}^6  ( S_a )^{-1}  
{\rm PT}(\mathbb{I}_8) \, \frac{{\rm Pf} A^{17}_{17}}{\sigma_{17} }\,  \frac{ (-1) }{ (\sigma_{15} \sigma_{28} \sigma_{37}  \s_{46}  )} ,
\end{equation}
where $\gamma=\cap_{a=2}^{6}\gamma_{S_{a}}$. The integration rules imply that ${\cal A}(\mathbb{I}_8: 15,28,37,46)$ has only  two factorization contributions: 
$\s_2 \rightarrow \s_8$ and  $\s_3 \rightarrow \s_4 \rightarrow \s_5 \rightarrow \s_6 \rightarrow \s_7 $. We saw in the previous section that the first one vanishes after choosing the contour $\gamma_{S_6}$ to perform the GRT. Thus, we focus on the second one, Fig. \ref{Fig7}. 

Considering  the parametrization 
\begin{equation}\label{parametrization8}
\s_a=\epsilon x_a + \s_L, \quad a=3,4,5,6,7, \quad \text{ with}\,\, x_6=\text{constant},\,\, x_7=0 \,\,\text{and} \,\, \s_{7} = \s_{L},
\end{equation}
 the measure, the  integrand, and the scattering equations become
\begin{align}\label{eq:107}
&
\dif\sigma_2\wedge\dif\sigma_3\wedge \dif\sigma_4\wedge \dif\sigma_5\wedge \dif\sigma_6=\dif\sigma_2\wedge (\epsilon^3 \, x_{67}\,\dif x_3\wedge \dif x_4\wedge \dif x_5\wedge \dif\epsilon)
\nonumber \\
&
(\sigma_{17} \sigma_{78} \sigma_{81})^2 \, {\rm PT}(\mathbb{I}_8)\, 
\frac{    {\rm Pf}A^{17}_{17}  }{  \sigma_{17} } \,  \frac{1}{(\sigma_{15} \sigma_{28} \sigma_{37} \s_{46} )}
= \frac{(\sigma_{1L} \sigma_{L 8} \sigma_{81})^2 }{ ( \sigma_{12}\sigma_{2 L }\sigma_{L  8}\sigma_{81} ) }
 \frac{\a_{28}}{\sigma_{1 L }\sigma_{28}} \frac{1}{ (\s_{1L} \s_{28} )}\qquad\nonumber\\
&
\qquad\qquad \qquad\qquad\qquad\qquad
\times
\frac{x^2_{R7} x_{R6} }{ (x_{R3} x_{34} x_{45}  x_{56} x_{67} )}  \frac{ {\rm Pf} A^{R7}_{R7}  }{x_{R7}} \frac{1}{(x_{37} x_{46} x_{R5})} \, \frac{1}{\epsilon^8}
+ {\cal O}(\epsilon^{-3}), 
\end{align}
where $x_R=\infty$,
\begin{equation}\label{eq:108}
 {\rm Pf}A^{1L}_{1L} =
\frac{\a_{28}}{\s_{28}}, \qquad
 {\rm Pf}A^{R7}_{R7} = \frac{\a_{34} \a_{56} }{x_{34} x_{56}} -\frac{\a_{35} \a_{46} }{x_{35} x_{46}} +\frac{\a_{36} \a_{45} }{x_{36} x_{45}},
\end{equation}
and
\begin{align}\label{}
& 
S_2 = \left[\hat S_2 + {\cal O}(\epsilon) \right] , \qquad  
S_a = \frac{1}{\epsilon}   \left[ \hat S_3  + {\cal O}(\epsilon) \right] , \quad
a=3,4,5,6 , \nonumber\\
&
\hat S_2=\frac{\a_{21}}{\sigma_{21}}  +\frac{\a_{28}}{\sigma_{28}} + \frac{\a_{2L}}{\sigma_{2L}} ,
  \quad \hat S_3=\frac{\a_{34}}{x_{34}}  +\frac{\a_{35}}{x_{35}} +\frac{\a_{36}}{x_{36}} +\frac{\a_{37}}{x_{37}} + \frac{\a_{3R}}{x_{3R}} , 
 \quad \hat S_a=\hat S_3\Big|_{a\leftrightarrow 3} , \,\,\, a=4,5,6.
\end{align}
Here we have used the notation, $\a_{2L}= \a_{23}+\a_{24}+\a_{25} +\a_{26}+\a_{27}$ and $\a_{a R}= \a_{a8}+\a_{a1}+\a_{a2}$, with $a=3,4,5,6$. 

From the above expansions, we identify two poles: $\sigma_{28}=0$ and $\epsilon=0$. We then deform $\gamma$ into $\hat\gamma=\hat\gamma_1+\hat\gamma_2$, with $\hat\gamma_1=\gamma_{28}\cap\gamma_{ S_2}\cap\gamma_{ S_3} \cap\gamma_{ S_4}\cap\gamma_{ S_5}  $, $\gamma_{28} = \{ |\s_2-\s_8 |=\delta \}$, and $\hat\gamma_2= \gamma_{\epsilon}\cap\gamma_{ S_2}\cap\gamma_{ S_3} \cap\gamma_{ S_4}\cap\gamma_{ S_5}  $. The integral over $\hat\gamma_1$ then vanishes as we previously saw so the only contribution comes from $\hat \gamma_2$. After integrating $\epsilon$ around $\epsilon=0$, the full integral breaks into two parts: one with $\{\sigma_1,\sigma_2,\sigma_L,\sigma_8\}$ and the other one with $\{x_3,x_4,x_5,x_6,x_7,x_{R}\}$. Using the identity
\begin{align}
&{\rm PT}(6,7,R )\,\hat S_6 + {\rm PT}(5,7,R )\,\hat S_5 +{\rm PT}(4,7,R )\,\hat S_4 + {\rm PT}(3,7,R )\,\hat S_3 \nonumber \\
&
=  {\rm PT}(7,R )\, \left[ ( {\cal D}_3+ {\cal D}_4 + {\cal D}_5+ {\cal D}_6+ {\cal D}_7)^2 + m^2 \right],
\end{align}
it is clear that
\begin{equation}
\hat S_6\Big|_{\gamma_{\hat S_3} \cap \gamma_{\hat S_4} \cap \gamma_{\hat S_5} } = \frac{{\rm PT}(7,R) }{{\rm PT}(6,7,R)} \, \left[ ( {\cal D}_3+ {\cal D}_4 + {\cal D}_5+ {\cal D}_6+ {\cal D}_7)^2 + m^2 \right].
\end{equation}
Then ${\cal A}(\mathbb{I}_8: 15,28,37,46) $ becomes
\begin{align}
&
{\cal A}(\mathbb{I}_8: 15,28,37,46)
= \int_{\gamma_{\hat S_2}} \Big[\dif \s_{2}   (\hat S_2)^{-1}  \Big] 
(\s_{81} \s_{1L}   \s_{L8}  )^2 \, {\rm PT}(8,1,2,L) 
 \left[ ( {\cal D}_8+ {\cal D}_1 + {\cal D}_2)^2 + m^2
\right]^{-1}
\nonumber
\\
&
\int_{\tilde \gamma   }  \prod_{a=3}^5  \Big[ \dif x_{a}   (\hat S_a)^{-1}   \Big]
(x_{R6}  x_{67}  x_{7R})^2  \, {\rm PT}(R,3,4,5,6,7)
 \left[
\frac{1}{\s_{1L}} \frac{\a_{28}}{\s_{28}}  \frac{1}{\s_{28} \s_{1L}} \right] \!\! \left[
\frac{(-1) \, {\rm Pf} A_{R7}^{R7}   }{x_{R7}} \frac{1}{x_{37} x_{46} x_{R5}} \right],
 \label{}
\end{align}
where we have used the CWI and defined $\tilde \gamma=\gamma_{\hat S_3} \cap\gamma_{\hat  S_4}\cap\gamma_{\hat  S_5} $.

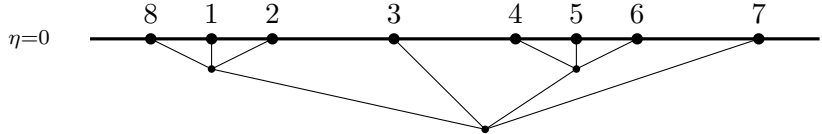
\begin{figure}[h]
\centering
\text{$
\mathord{\begin{tikzpicture}[scale=0.4]
\node at (-3,0) [circle,,fill=black,inner sep=0pt,minimum size=0mm,label=center:$ \text{${\scriptstyle \eta=0}$} $]  {};
	\node at (0,0) [circle,,fill=black,inner sep=0pt,minimum size=0mm,label=above:$ $]  {};
	\node at (1,0) [circle,,fill=black,inner sep=0pt,minimum size=1.5mm,label=above:$8$]  {};
	\node at (2,0) [circle,,fill=black,inner sep=0pt,minimum size=0mm,label=above:$ $]{};
	\node at (3,0) [circle,,fill=black,inner sep=0pt,minimum size=1.5mm,label=above:$1$]{};
	\node at (4,0) [circle,,fill=black,inner sep=0pt,minimum size=0mm,label=above:$ $]{};
	\node at (5,0) [circle,,fill=black,inner sep=0pt,minimum size=1.5mm,label=above:$2$]{};
		\node at (6,0) [circle,,fill=black,inner sep=0pt,minimum size=0mm,label=above:$ $]{};
			\node at (7,0) [circle,,fill=black,inner sep=0pt,minimum size=0mm,label=above:$ $]{};
				\node at (8,0) [circle,,fill=black,inner sep=0pt,minimum size=0mm,label=above:$ $]{};
				   \node at (9,0) [circle,,fill=black,inner sep=1.5pt,minimum size=0mm,label=above:$3$]{};
				\node at (10,0) [circle,,fill=black,inner sep=0pt,minimum size=0mm,label=above:$ $]{};
				 \node at (3,-1) [circle,,fill=black,inner sep=0pt,minimum size=1.0mm,label=above:$ $]{};
				 \node at (15,-1) [circle,,fill=black,inner sep=0pt,minimum size=1.0mm,label=above:$ $]{};
				  \node at (13,0) [circle,,fill=black,inner sep=1.5pt,minimum size=0mm,label=above:$4$]{};
				   \node at (14,0) [circle,,fill=black,inner sep=0pt,minimum size=0mm,label=above:$ $]{};
				    \node at (15,0) [circle,,fill=black,inner sep=0pt,minimum size=1.5mm,label=above:$5$]{};
				     \node at (17,0) [circle,,fill=black,inner sep=0pt,minimum size=1.5mm,label=above:$6$]{};
				        \node at (21,0) [circle,,fill=black,inner sep=0pt,minimum size=1.5mm,label=above:$7$]{};
				       \node at (12,-3) [circle,,fill=black,inner sep=0pt,minimum size=1.0mm,label=above:$ $]{};
	\draw[very thick,black] (-1,0) -- (23,0);
		\draw[black] (1,0) -- (3,-1);
			\draw[black] (3,-1) -- (3,0);
			   \draw[black] (5,0) -- (3,-1);
			        \draw[black] (15,-1) -- (13,0);
			              \draw[black] (15,-1) -- (15,0);
			               \draw[black] (15,-1) -- (17,0);
			                 \draw[black] (15,-1) -- (12,-3);
			               \draw[black] (12,-3) -- (21,0);
			               \draw[black] (12,-3) -- (9,0);
			               \draw[black] (12,-3) -- (3,-1);
	\end{tikzpicture}}
$}
\caption{Eight-point Witten digram corresponding to the ladder graph, ${\cal A}(\mathbb{I}_8:15,28,37,46)$ }\label{abc1}
\end{figure}

On the support of $\gamma_{\hat S_3} \cap \gamma_{\hat S_4} \cap \gamma_{\hat S_5}$, one has the following identity,
\begin{equation}
\frac{(-1)}{x_{R7}} \, {\rm Pf}A^{R7}_{R7} = \frac{1}{x_{R6}} \, {\rm Pf}A^{R6}_{R6},
\end{equation}
which is illustrated  by an example in Appendix \ref{sixpoint-16}. We then obtain
\begin{align}\label{eq:113}
{\cal A}(\mathbb{I}_8, 15,28,37,46) =& {\cal A}( 8,1,2,L: 28,1L) \left[ ( {\cal D}_3+ {\cal D}_4 + {\cal D}_5+ {\cal D}_6+ {\cal D}_7)^2 + m^2 \right]^{-1} \nonumber \\
& {\cal A}( R,3,4,5,6,7: R5,37,46) ,
\end{align}
where
\begin{align}\label{}
{\cal A}(8,1,2,L:1L,28) = 
\int_{\gamma_{S_2}} \! \dif \sigma_2 (\s_{1L} \s_{L 8} \s_{81})^2 (\hat{S}_2)^{-1}  {\rm PT}(8,1,2,L) \,
\frac{ {\rm Pf}A^{1L}_{1L}}{\s_{1L}} \frac{1}{(\s_{1L}\s_{28})}  ,
\end{align}
and
\begin{align}\label{}
{\cal A}(R,3,4,5,6,7 \! : \! R5,37,46)  = \!
\int_{\tilde \gamma}  \prod_{a=3}^5 \dif x_a\,  \hat{S}_a^{-1}  (x_{67} x_{7R} x_{R6})^2 \, {\rm PT}(R,3,4,5,6,7) 
  \frac{ {\rm Pf}A^{R6}_{R6}  }{  x_{R6}}    \frac{1}{(x_{R5}x_{37} x_{46} )}.
\end{align}
The factorization in \eqref{eq:113} is represented in Figure \ref{Fig7}. Since
$$
{\cal A}(8,1,2,L:1L,28)  = \mathbb{I}_4, \qquad {\cal A}(R,3,4,5,6,7:R5,37,46)    = \left[ (  {\cal D}_4 + {\cal D}_5+ {\cal D}_6)^2 + m^2 \right]^{-1},
$$
we can finally show that
\begin{align}
{\cal A}(\mathbb{I}_8, 15,28,37,46) \,  {\cal C}_8^{\Delta}= \! \left[ ( {\cal D}_3+ {\cal D}_4 + {\cal D}_5+ {\cal D}_6+ {\cal D}_7)^2 + m^2 \right]^{-1}  \!
\left[ (  {\cal D}_4 + {\cal D}_5+ {\cal D}_6)^2 + m^2 \right]^{-1} \!  {\cal C}_8^{\Delta} .
\end{align}
Using equation \eqref{eq:prop-insertion} and the CWI, it is straightforward to see that the above expression reproduces the Witten diagram in Figure \ref{abc1}.
The other three diagrams in the top line of Fig. \ref{Fig6} are obtained by relabeling.

Since there are no ambiguities in ${\cal A}( 8,1,2,L: 28,1L)$ and $ {\cal A}( R,3,4,5,6,7: R5,37,46) $, and all terms in \eqref{eq:113} commute, we conclude that the integrand of ${\cal A}(\mathbb{I}_8, 15,28,37,46) $ is also free of ambiguities.

\subsection{Non-Ladder Contributions}\label{Non-Ladder-8}

\begin{figure}[h]
\centering
\hspace{-2.7cm}
\parbox[c]{7.1em}{\includegraphics[scale=0.25]{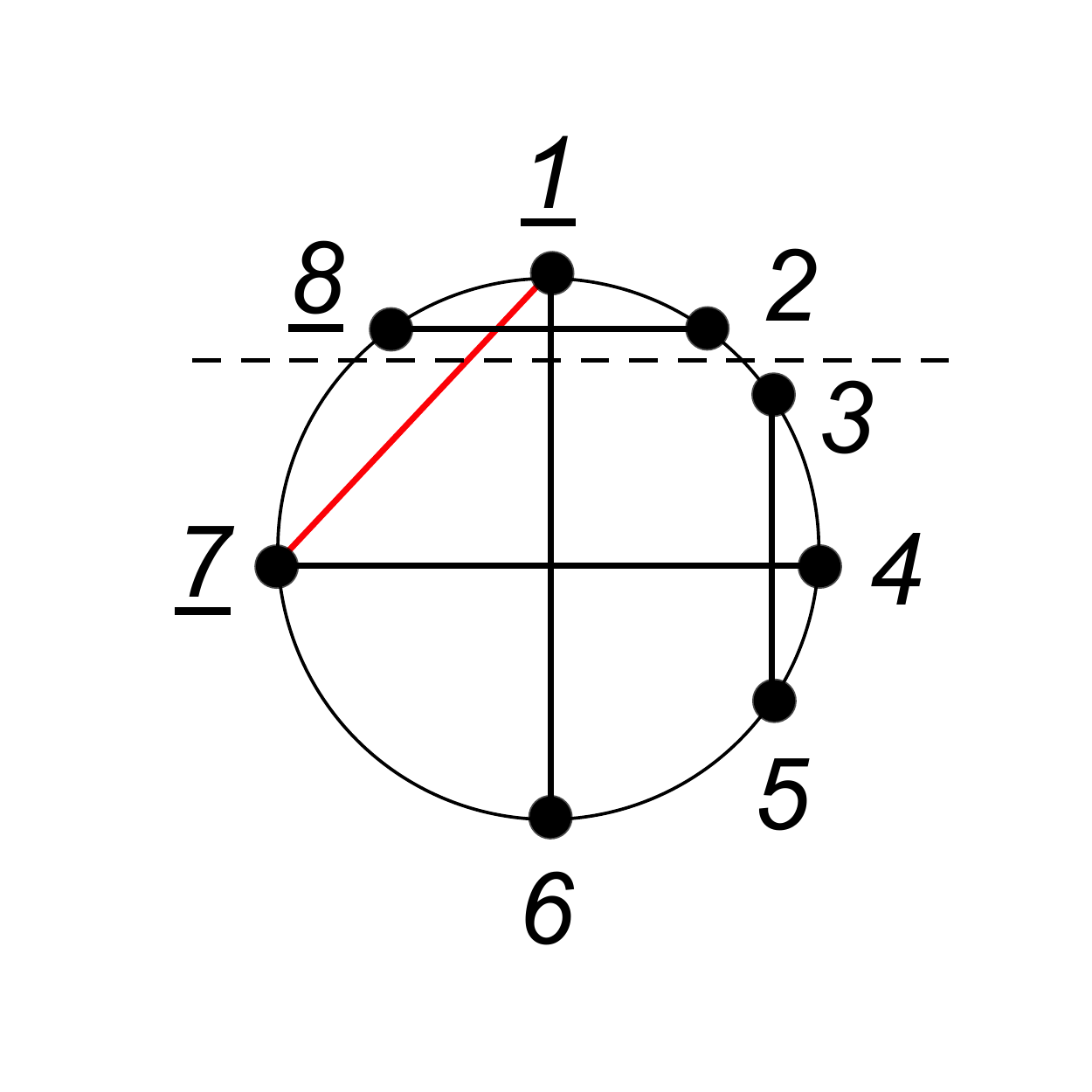}}\,\qquad
$\Rightarrow$
\quad
\parbox[c]{5.1em}{\includegraphics[scale=0.34]{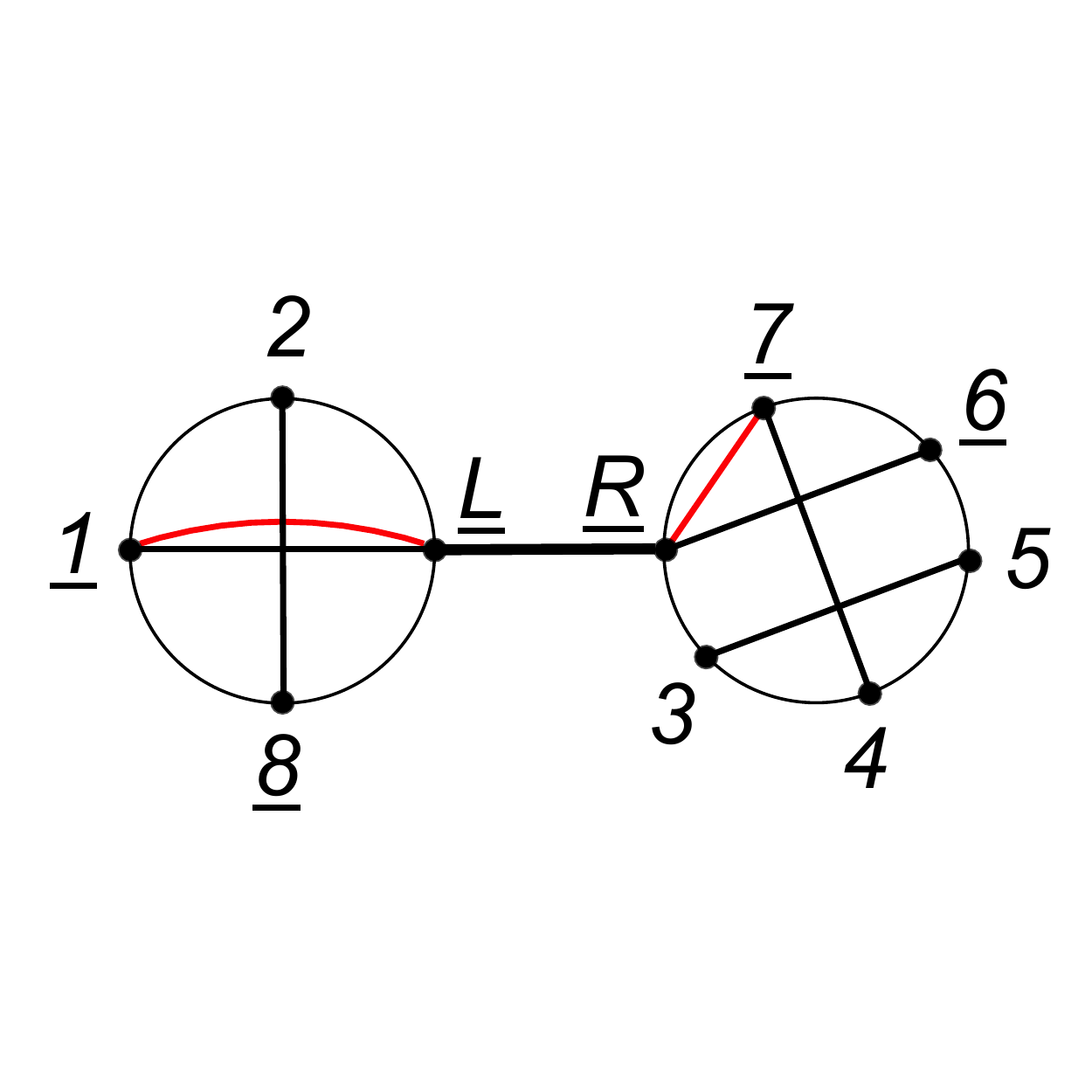}}\,\, \, 
\vspace{-0.7cm}
\caption{Non-ladder diagram, ${\cal A}(\mathbb{I}_8:16,28,35,47)$, and its factorization contribution.}\label{Fig8} 
\end{figure}

Now we are going to compute the contributions of the non-ladder diagrams given by the graphs on the bottom row of Fig. \ref{Fig6}. We focus on the first diagram, {\it i.e.} ${\cal A}(\mathbb{I}_8, 16,28,35,47)$, since the others are obtained by relabelling. It can be expressed as
\begin{equation}\label{}
{\cal A}(\mathbb{I}_8, 16,28,35,47) =  \int_{\gamma}  \prod_{a=2}^6 \dif\sigma_a ( \sigma_{78} \sigma_{81} \sigma_{17} )^2 \prod_{a=2}^6  ( S_a )^{-1}  
{\rm PT}(\mathbb{I}_8) \, \frac{{\rm Pf} A^{17}_{17}}{\sigma_{17} }\,  \frac{ (-1) }{ (\sigma_{16} \sigma_{28} \sigma_{35}  \s_{47}  )} ,
\end{equation}
with $\gamma = \bigcap_{a=2}^6 \gamma_{S_a}$.
Again, we identify two factorization contributions via the integration rules, given by $\s_{2}\rightarrow \s_8$ and $\s_{3}\rightarrow \s_4 \rightarrow \s_5 \rightarrow \s_6 \rightarrow \s_7$.  Using the contour $\gamma_{S_6}$ for the GRT, the first factorization cut vanishes trivially. For the  second contribution we use the parametrization given in \eqref{parametrization8}, with
$\s_a=\epsilon x_a + \s_L, \, a=3,4,5,6,7$, $x_6=\text{constant}$, $ x_7=0$, and $\s_{7} = \s_{L}$. Next, we obtain a similar expansion to \eqref{eq:107}:
\begin{align}\label{eq:119}
&{\cal A}(\mathbb{I}_8:16,28,35,47)\nonumber\\
&= {\cal A}( 8,1,2,L:1L,28) \left[ ( {\cal D}_3+ {\cal D}_4 + {\cal D}_5+ {\cal D}_6+ {\cal D}_7)^2 + m^2 \right]^{-1}  {\cal A}( R,3,4,5,6,7: R6,35,47) ,
\end{align}
with 
\begin{align}\label{eq:121}
{\cal A}(8,1,2,L:1L,28) = 
\int_{\gamma_{S_2}} \! \dif \sigma_2 (\s_{1L} \s_{L 8} \s_{81})^2 (\hat{S}_2)^{-1}  {\rm PT}(8,1,2,L) \,
\frac{ {\rm Pf}A^{1L}_{1L}}{\s_{1L}} \frac{1}{(\s_{1L}\s_{28})}  ,
\end{align}
and
\begin{align}\label{eq:122}
{\cal A}(R,3,4,5,6,7:R6,35,47)  = & 
\int_{\tilde \gamma}  \prod_{a=3}^5 \dif x_a\,  (\hat{S}_a)^{-1}\,  (x_{67} x_{7R} x_{R6})^2 \, {\rm PT}(R,3,4,5,6,7) \,
  \nonumber \\
 &  \frac{(-1) }{  x_{R7}} \,   {\rm Pf}A^{R7}_{R7}  \,  \frac{1}{(x_{R6}x_{35} x_{47} )},
\end{align}
where ${\rm Pf}A_{1L}^{1L}$ and ${\rm Pf}A_{R7}^{R7}$ are given in equation \eqref{eq:108}. This factorization is depicted in Fig \ref{Fig8}.
\vspace{-0.8cm}

In order to compute ${\cal A}(R,3,4,5,6,7:R6,35,47)$  in \eqref{eq:122} we will use the ${\rm SL}(2,\mathbb{C})$ symmetry to choose a more convenient puncture fixing. First we rewrite the integrand in way that makes this symmetry manifest. This can be accomplished using the following deformation of the CSE:
\begin{align}\label{eq:123}
&
 \hat S_3=\frac{\a_{34}}{x_{34}}  +\frac{\a_{35}}{x_{35}} +\frac{\a_{36}}{x_{36}} +\frac{\a_{37}}{x_{37}} + \frac{\a_{3R}}{x_{3R}} , 
 \quad \hat S_a=\hat S_3\Big|_{a\leftrightarrow 3} , \,\,\, a=4,5, \nonumber \\
 &
  \hat S_6=\frac{\a_{63}}{x_{63}}  +\frac{\a_{64}}{x_{64}} +\frac{\a_{65} }{x_{65}} +\frac{\a_{67}+\Delta_{67}}{x_{67}} + \frac{\a_{6R} +\Delta_{6R}}{x_{6R}} , \nonumber\\
  &
  \hat S_7=\frac{\a_{73}}{x_{73}}  +\frac{\a_{74}}{x_{74}} +\frac{\a_{75} }{x_{75}} +\frac{\a_{76}+\Delta_{76}}{x_{76}} + \frac{\a_{7R} +\Delta_{7R}}{x_{7R}} , \nonumber\\
  &
  \hat S_R=\frac{\a_{R3}}{x_{R3}}  +\frac{\a_{R4}}{x_{R4}} +\frac{\a_{R5} }{x_{R5}} +\frac{\alpha_{R6}+\Delta_{R6}}{x_{R6}} + \frac{\a_{R7} +\Delta_{R7}}{x_{R7}} ,
\end{align}
with $\Delta_{ab}=\Delta_{ba}$ and
\begin{equation}
\Delta_{67} = -m^2  - {\cal D}_R\cdot {\cal D}_R , \quad 
\Delta_{6R} = {\cal D}_R\cdot {\cal D}_R  +m^2 , \quad
\Delta_{7R} = {\cal D}_R\cdot {\cal D}_R  + m^2, 
\end{equation}
where ${\cal D}_R= {\cal D}_8+{\cal D}_1+{\cal D}_2$. Since $\a_{ab}=\a_{ba}$, then \eqref{eq:123} is the only possible deformation.
It is simple to show that these scattering equations  are ${\rm SL}(2,\mathbb{C})$ covariant. Note that the $\Delta_{ab}$ parameters were inspired by ones in \cite{Naculich:2014naa,Naculich:2015zha,Bjerrum-Bohr:2019nws}.

The deformations in \eqref{eq:123} correspond to the following deformed $A$-matrix of ${\cal A}(R,3,4,5,6,7:R6,35,47) $:
\begin{equation}\label{eq:125}
A=
\left[
\begin{matrix}
0& \frac{\a_{34}}{x_{34}}  & \frac{\a_{35}}{x_{35}} & \frac{\a_{36}}{x_{36}} &\frac{\a_{37}}{x_{37}} & \frac{\a_{3R}}{x_{3R}} \\
\frac{\a_{43}}{x_{43}}  & 0 & \frac{\a_{45}}{x_{45}} & \frac{\a_{46}}{x_{46}} &\frac{\a_{47}}{x_{47}} & \frac{\a_{4R}}{x_{4R}}\\
\frac{\a_{53}}{x_{53}}  &  \frac{\a_{54}}{x_{54}} & 0 & \frac{\a_{56}}{x_{56}} &\frac{\a_{57}}{x_{57}} & \frac{\a_{5R}}{x_{5R}} \\
\frac{\a_{63}}{x_{63}}  & \frac{\a_{64}}{x_{64}} & \frac{\a_{65} }{x_{65}} & 0 & \frac{\a_{67}+\Delta_{67}}{x_{67}} & \frac{\a_{6R} +\Delta_{6R}}{x_{6R}}  \\
 \frac{\a_{73}}{x_{73}}  & \frac{\a_{74}}{x_{74}}  & \frac{\a_{75} }{x_{75}}  & \frac{\a_{76}+\Delta_{76}}{x_{76}} & 0 & \frac{\a_{7R} +\Delta_{7R}}{x_{7R}} \\
 \frac{\a_{R3}}{x_{R3}}  & \frac{\a_{R4}}{x_{R4}} & \frac{\a_{R5} }{x_{R5}} & \frac{\alpha_{R6}+\Delta_{R6}}{x_{R6}}  & \frac{\a_{R7} +\Delta_{R7}}{x_{R7}} & 0 
\end{matrix}
\right].
\end{equation}
On the support of the deformed CSE we then have the identity
\begin{equation}
\frac{(-1)}{x_{R7}} \, {\rm Pf}A^{R7}_{R7} = \frac{1}{x_{57}} \, {\rm Pf}A^{57}_{57},
\end{equation}
with
\begin{equation}\label{eq:128}
 {\rm Pf}A^{57}_{57} = \frac{\a_{34} (\a_{6R} + \Delta_{6R}) }{x_{34} x_{6R}} -\frac{\a_{36} \a_{4R} }{x_{36} x_{4R}} +\frac{\a_{3R} \a_{46} }{x_{3R} x_{46}}.
\end{equation}

\begin{figure}
\centering
\hspace{-1.4cm}
\parbox[c]{11.1em}{\includegraphics[scale=0.35]{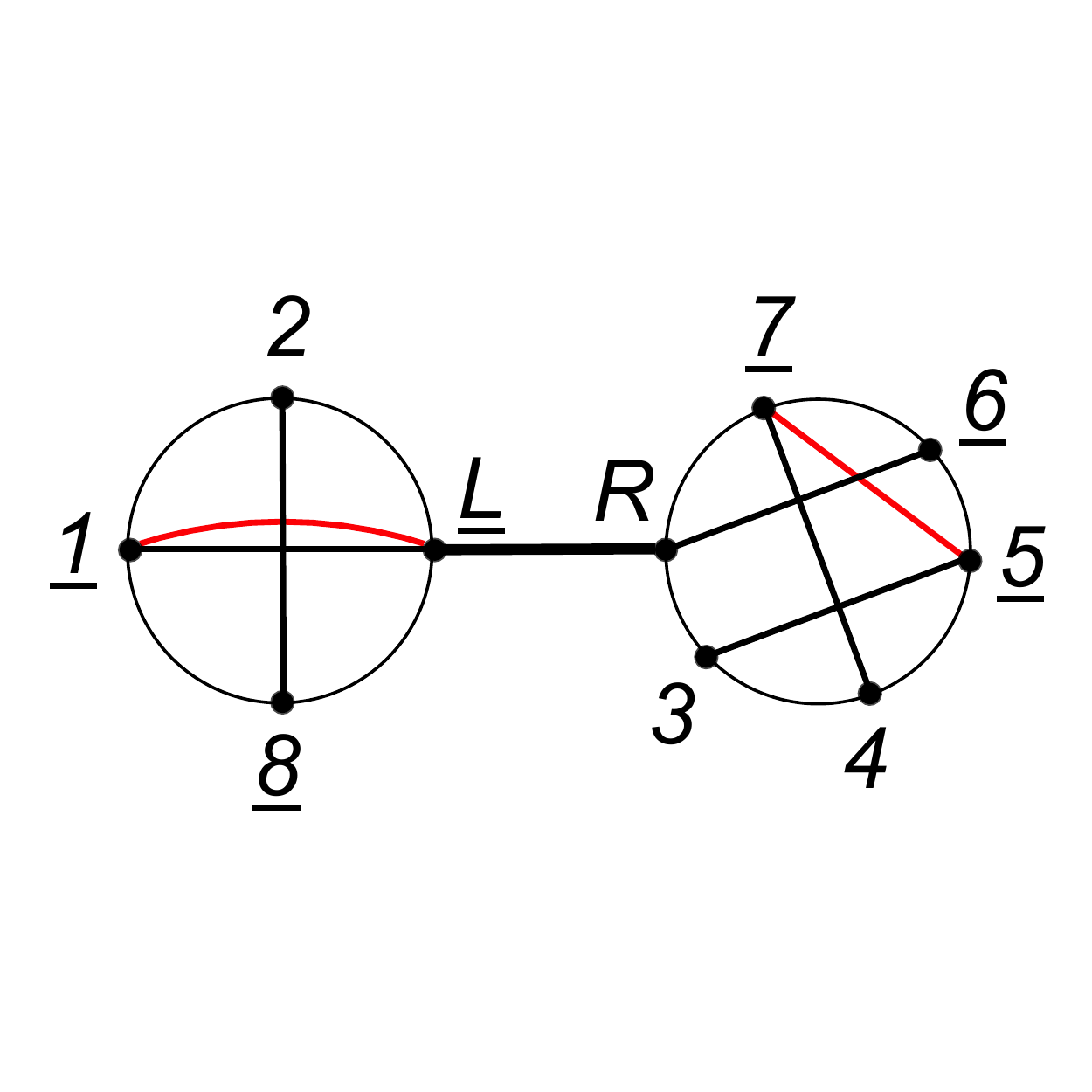}}\,  \qquad
\quad
\parbox[c]{5.1em}{\includegraphics[scale=0.35]{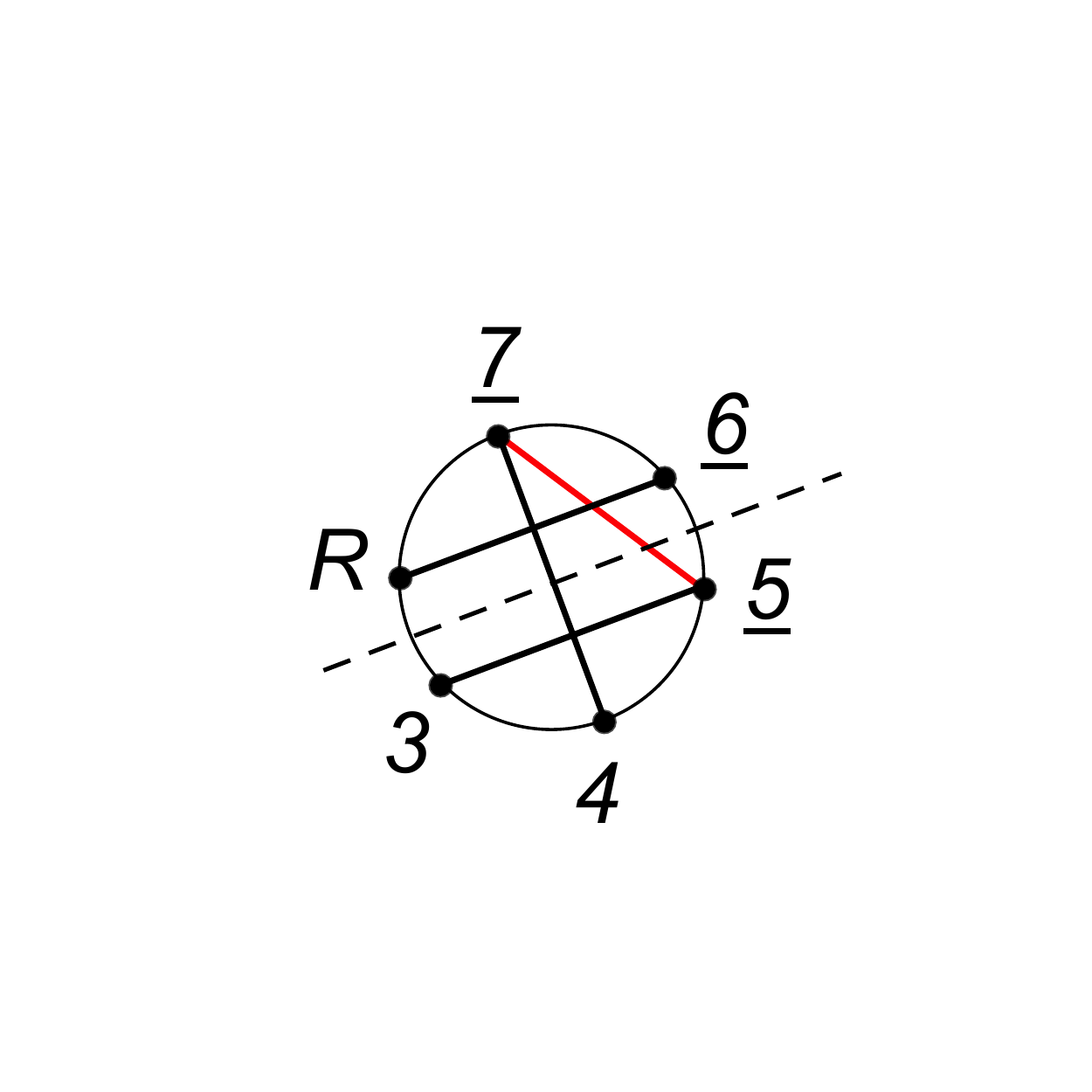}}\,\, \, 
\vspace{-0.7cm}
\caption{A new gauge fixing  and its factorization contribution.}\label{Fig9} 
\end{figure}

Now that we have figured out how to rewrite the CHY integrand of ${\cal A}(R,3,4,5,6,7:R6,35,47) $ in a manifestly ${\rm SL}(2,\mathbb{C})$ invariant way, we can choose the previously used gauge of section \ref{sec:sixpoint}:
\begin{align}\label{}
{\cal A}(R,3,4,5,6,7:R6,35,47)  = & 
\int_{\tilde \gamma}  \prod_{a=R }^4 \dif x_a\,  (\hat{S}_a)^{-1}\,  (x_{56} x_{67} x_{75})^2 \, {\rm PT}(R,3,4,5,6,7) \,
  \nonumber \\
 &  \frac{1 }{  x_{57}} \,   {\rm Pf}A^{57}_{57}  \,  \frac{1}{(x_{R6}x_{35} x_{47} )},
\end{align}
where $\tilde \gamma = \gamma_{\hat S_{R} } \cap \gamma_{\hat S_{3}} \cap \gamma_{\hat S_{4} } $.
We depict the new gauge in Fig. \ref {Fig9}.
This six-point calculation was already performed in section \ref{sec:sixpoint} so it is straightforward to see that ${\cal A}(R,3,4,5,6,7:R6,35,47)  $ becomes
\begin{align}\label{eq:624}
& {\cal A}(R,3,4,5,6,7:R6,35,47) \nonumber \\
&= {\cal A}( R,7,6,L: 7L,R6) \left[ ( {\cal D}_3+ {\cal D}_4 + {\cal D}_5)^2 + m^2 \right]^{-1} 
 {\cal A}( \tilde R,3,4,5: \tilde{R}4,35) .
\end{align}
Here we have
\begin{align}\label{eq:131}
{\cal A}( R,7,6,L: 7L,R6) = 
\int_{\gamma_{\tilde S_R}} \!\!\! \dif x_R (x_{67} x_{7L} x_{L6})^2 ({\tilde S}_R)^{-1}  {\rm PT}(R,7,6,L) \,
\frac{ {\rm Pf}A^{7L}_{7L}}{x_{7L}} \frac{1}{(x_{7L} x_{R6})}  ,
\end{align}
\begin{align}\label{eq:132}
 {\cal A}( \tilde R,3,4,5: \tilde{R}4,35)  = 
\int_{\gamma_{ \tilde S_3}} \!\!\!\dif y_3 (y_{45} y_{5\tilde R} y_{\tilde R4})^2 (\tilde{S}_3)^{-1} \, {\rm PT}(\tilde R,3,4,5) \,
 \frac{(-1) {\rm Pf}A^{\tilde R5}_{\tilde R5}}{  y_{\tilde R5}} \frac{1}{(y_{\tilde R4}y_{35})}  ,
\end{align}
with
\begin{eqnarray}
 {\rm Pf}A^{7L}_{7L}=\frac{\a_{6R}+\Delta_{6R}}{x_{6R}}, \qquad  {\rm Pf}A^{\tilde R5}_{\tilde R5}=\frac{\a_{34}}{y_{34}},
\end{eqnarray}
 and
\begin{eqnarray}
\tilde S_R=\frac{\a_{R6}+\Delta_{R6}}{x_{R6}}  +\frac{\a_{R7}+\Delta_{R7}}{x_{R7}} + \frac{\a_{RL}}{x_{RL}} , \qquad
\tilde S_3=\frac{\a_{34}}{y_{34}}  +\frac{\a_{35}}{y_{35}} + \frac{\a_{3 \tilde R}}{y_{3 \tilde R}} ,
\end{eqnarray}
where
$\a_{RL}=\a_{R3}+\a_{R4}+\a_{R5}$, $\a_{3\tilde R}=\a_{3R}+\a_{37}+\a_{36}$.

\begin{figure}
\centering
\text{$
\mathord{\begin{tikzpicture}[scale=0.4]
\node at (-3,0) [circle,,fill=black,inner sep=0pt,minimum size=0mm,label=center:$ \text{${\scriptstyle \eta=0}$} $]  {};
	\node at (0,0) [circle,,fill=black,inner sep=0pt,minimum size=0mm,label=above:$ $]  {};
	\node at (1,0) [circle,,fill=black,inner sep=0pt,minimum size=1.5mm,label=above:$8$]  {};
	\node at (2,0) [circle,,fill=black,inner sep=0pt,minimum size=0mm,label=above:$ $]{};
	\node at (3,0) [circle,,fill=black,inner sep=0pt,minimum size=1.5mm,label=above:$1$]{};
	\node at (4,0) [circle,,fill=black,inner sep=0pt,minimum size=0mm,label=above:$ $]{};
	\node at (5,0) [circle,,fill=black,inner sep=0pt,minimum size=1.5mm,label=above:$2$]{};
		\node at (6,0) [circle,,fill=black,inner sep=0pt,minimum size=0mm,label=above:$ $]{};
			\node at (7,0) [circle,,fill=black,inner sep=0pt,minimum size=1.5mm,label=above:$3$]{};
				\node at (8,0) [circle,,fill=black,inner sep=0pt,minimum size=0mm,label=above:$ $]{};
				   \node at (9,0) [circle,,fill=black,inner sep=0pt,minimum size=1.5mm,label=above:$4$]{};
				\node at (10,0) [circle,,fill=black,inner sep=0pt,minimum size=0mm,label=above:$ $]{};
				 \node at (11,0) [circle,,fill=black,inner sep=0pt,minimum size=1.5mm,label=above:$5$]{};
				 \node at (12,0) [circle,,fill=black,inner sep=0pt,minimum size=0mm,label=above:$ $]{};
				  \node at (3,-1) [circle,,fill=black,inner sep=0pt,minimum size=1.0mm,label=above:$ $]{};
				   \node at (9,-1) [circle,,fill=black,inner sep=0pt,minimum size=1.0mm,label=above:$ $]{};
				    \node at (15,0) [circle,,fill=black,inner sep=0pt,minimum size=1.5mm,label=above:$6$]{};
				        \node at (21,0) [circle,,fill=black,inner sep=0pt,minimum size=1.5mm,label=above:$7$]{};
				       \node at (12,-3) [circle,,fill=black,inner sep=0pt,minimum size=1.0mm,label=above:$ $]{};
	\draw[very thick,black]  (-1,0) -- (23,0);
		\draw[black] (1,0) -- (3,-1);
			\draw[black] (3,-1) -- (3,0);
			   \draw[black] (5,0) -- (3,-1);
			       \draw[black] (9,-1) -- (7,0);
			        \draw[black] (9,-1) -- (9,0);
			        \draw[black] (9,-1) -- (11,0);
			              \draw[black] (12,-3) -- (15,0);
			               \draw[black] (12,-3) -- (21,0);
			               \draw[black] (12,-3) -- (9,-1);
			               \draw[black] (12,-3) -- (3,-1);
	\end{tikzpicture}}
	$}
\caption{Eight point Witten diagram associated to the non-ladder graph ${\cal A}(\mathbb{I}_8:16,28,35,47)$.}\label{abc2}	
\end{figure}
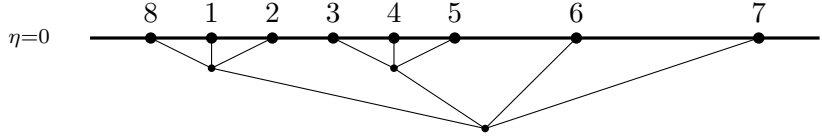

Since the four point subdiagrams given in \eqref{eq:121}, \eqref{eq:131} and \eqref{eq:132} are just the identity operator, the non-ladder diagram ${\cal A}(\mathbb{I}_8, 16,28,35,47) $ reduces to
\begin{align}\label{}
{\cal A}(\mathbb{I}_8: \! 16,28,35,47) {\cal C}_8^\Delta \!= \!\! \left[ ( {\cal D}_3+ {\cal D}_4 + {\cal D}_5+ {\cal D}_6+ {\cal D}_7)^2 + m^2 \right]^{-1} \! \left[ ( {\cal D}_3+ {\cal D}_4 + {\cal D}_5)^2 + m^2 \right]^{-1}  {\cal C}_8^\Delta,
\end{align}
which corresponds to the Witten diagram in Figure \ref{abc2}. The other non-ladder diagrams in Fig. \ref{Fig6} are obtained from ${\cal A}(\mathbb{I}_8: \! 16,28,35,47)$ by relabelling. Given that all terms in \eqref{eq:119} and \eqref{eq:624} commute, we conclude that the integrand of the 8-point correlator is free of ambiguities. 

\subsection{General Diagrams}

\begin{figure}[h]
\centering
\parbox[c]{7.7em}{\includegraphics[scale=0.25]{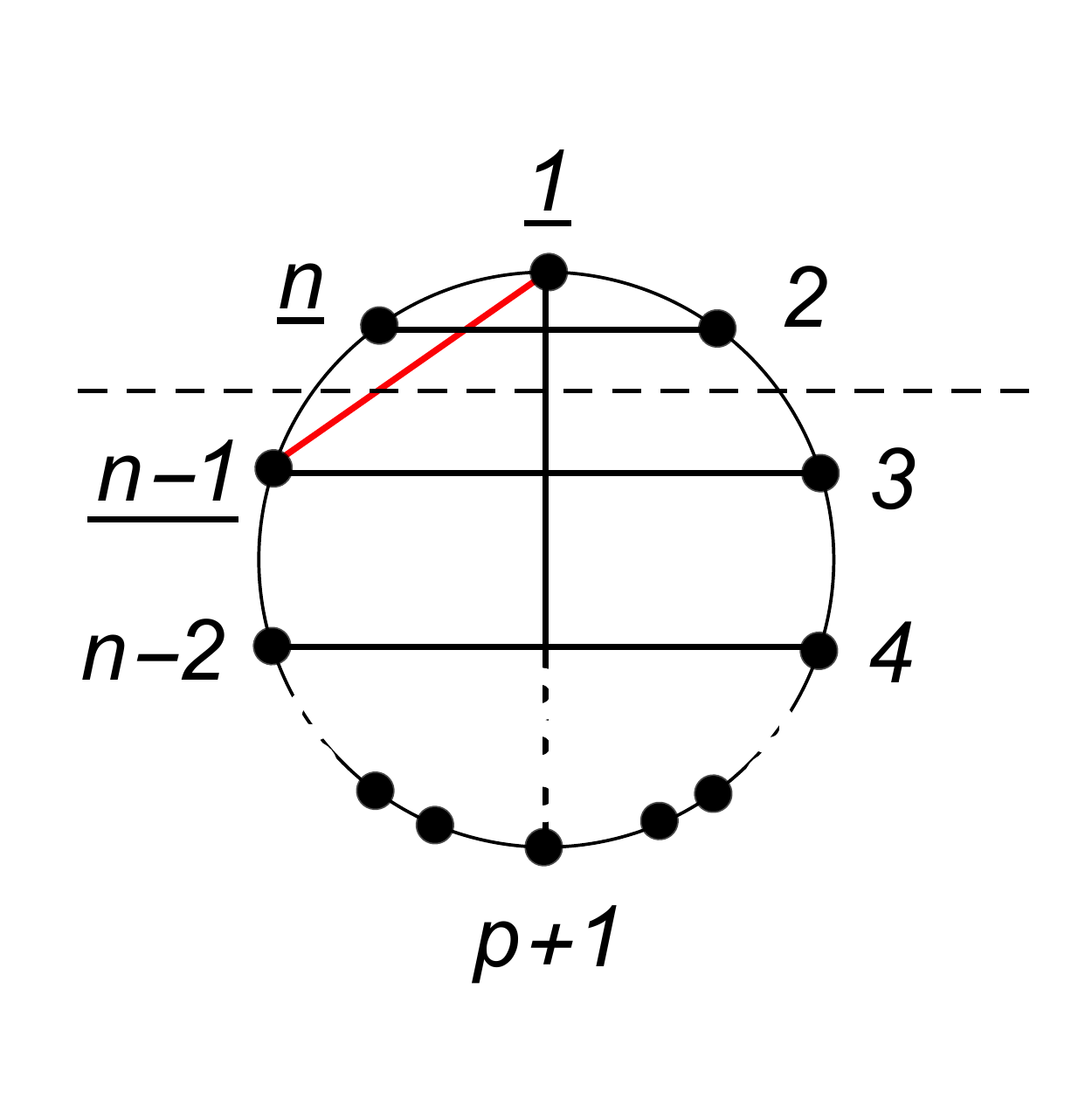}}
\vspace{-0.2cm}
\caption{$n$-point ladder diagram, ${\cal A}(\mathbb{I}_n \!: \! 1  (p+1),2n,... , p (p+2))$, and its
factorization contribution.}\label{Fig10} 
\vspace{-0.2cm}
\end{figure}

Let us comment on the evaluation of our worldsheet formula at $n$-points. 
First we point out that the six and eight-point results in sections \ref{sec:sixpoint} and \ref{Ladder-8} can be straightforwardly extended to ladder diagrams with any number of points. In particular, let us consider the ladder diagram in Fig. \ref{Fig10}, where we fix the positions of legs $\left\{n-1,n,1\right\} $ and remove rows and columns $\left\{ 1,n-1\right\} $ from the $A$-matrix in the reduced Pfaffian. Like in previous cases, only one factorization contributes, notably $\sigma_3\rightarrow\sigma_4 \rightarrow \cdots \rightarrow\sigma_{n-1}$.
\vspace{-0.5cm}

\begin{figure}[h]
\centering
\text{$
\mathord{\begin{tikzpicture}[scale=0.35]
\node at (-3,0) [circle,,fill=black,inner sep=0pt,minimum size=0mm,label=center:$ \text{${\scriptstyle \eta=0}$} $]  {};
	\node at (0,0) [circle,,fill=black,inner sep=0pt,minimum size=0mm,label=above:$ $]  {};
	\node at (1,0) [circle,,fill=black,inner sep=0pt,minimum size=1.5mm,label=above:$n$]  {};
	\node at (2,0) [circle,,fill=black,inner sep=0pt,minimum size=0mm,label=above:$ $]{};
	\node at (3,0) [circle,,fill=black,inner sep=0pt,minimum size=1.5mm,label=above:$1$]{};
	\node at (4,0) [circle,,fill=black,inner sep=0pt,minimum size=0mm,label=above:$ $]{};
	\node at (5,0) [circle,,fill=black,inner sep=0pt,minimum size=1.5mm,label=above:$2$]{};
		\node at (6,0) [circle,,fill=black,inner sep=0pt,minimum size=0mm,label=above:$ $]{};
			\node at (7,0) [circle,,fill=black,inner sep=0pt,minimum size=1.5mm,label=above:$3$]{};
				\node at (8,0) [circle,,fill=black,inner sep=0pt,minimum size=0mm,label=above:$ $]{};
				   \node at (9,0) [circle,,fill=black,inner sep=1.5pt,minimum size=0mm,label=above:$4$]{};
				\node at (15,-3.2) [circle,,fill=black,inner sep=0pt,minimum size=1.0mm,label=above:$ $]{};
				 \node at (15,-1) [circle,,fill=black,inner sep=0pt,minimum size=1.0mm,label=above:$ $]{};
				 \node at (3,-1) [circle,,fill=black,inner sep=0pt,minimum size=1.0mm,label=above:$ $]{};
				  \node at (13,0) [circle,,fill=black,inner sep=1.5pt,minimum size=0mm,label=above:$\!\!\!\!\! p$]{};
				   \node at (14,0) [circle,,fill=black,inner sep=0pt,minimum size=0mm,label=above:$ $]{};
				    \node at (15,0) [circle,,fill=black,inner sep=0pt,minimum size=1.5mm,label=above:$\!\!\! p+1$]{};
				     \node at (17,0) [circle,,fill=black,inner sep=0pt,minimum size=1.5mm,label=above:$\,\,\, p+2$]{};
				    \node at (21,0) [circle,,fill=black,inner sep=0pt,minimum size=1.5mm,label=above:$\!\!\! n-2$]{};
\node at (23,0) [circle,,fill=black,inner sep=0pt,minimum size=1.5mm,label=above:$\,\,\,\,\,\,\,\,  n-1$]{};
      \node at (15,-4.5) [circle,,fill=black,inner sep=0pt,minimum size=1.0mm,label=above:$ $]{};
	\draw[very thick,black] (-1,0) -- (9,0);
	\draw[very thick,black,dashed] (9,0) -- (13,0);
	\draw[very thick,black] (13,0) -- (17,0);
	\draw[very thick,black,dashed] (17,0) -- (21,0);
	\draw[very thick,black] (21,0) -- (25,0);
		\draw[black] (1,0) -- (3,-1);
			\draw[black] (3,-1) -- (3,0);
			   \draw[black] (5,0) -- (3,-1);
			        \draw[black] (15,-1) -- (13,0);
			              \draw[black] (15,-1) -- (15,0);
			               \draw[black] (15,-1) -- (17,0);
			                 \draw[dashed] (15,-1) -- (15,-3);
			                   \draw[black,] (15,-3) -- (15,-4.5);
			               \draw[black] (15,-3.2) -- (21,0);
			                  \draw[black] (15,-4.5) -- (7,0);
			               \draw[black] (15,-3.2) -- (9,0);
			               \draw[black] (15,-4.5) -- (3,-1);
			                     \draw[black] (15,-4.5) -- (23,0);
	\end{tikzpicture}}
	$}
\caption{Witten diagram corresponding to ${\cal A}(\mathbb{I}_n \!: \! 1  (p+1),2n,... , p (p+2))\, {\cal C}_n^\Delta $.}\label{abc3}	
\end{figure}
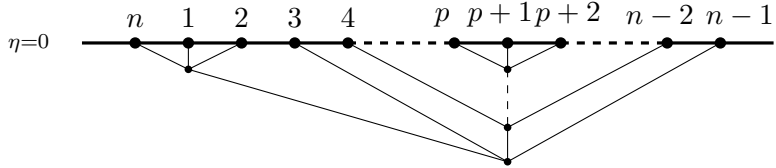

Using the parametrization,  $\sigma_a=\epsilon x_a+\sigma_{n-1}$,  
with $a=3,4,\ldots,n-1$, $x_{n-2}=\textrm{constant}$, $x_{n-1}=0$, $\sigma_{n-1}\equiv\sigma_L$, and expanding around $\epsilon=0$, one obtains a generalization of  \eqref{eq:14}:  
\begin{align}\label{eq:135}
&
{\cal A}(\mathbb{I}_n \!: \! 1  (p+1),2n,... , p (p+2)) \, {\cal C}^\Delta_n
=\nonumber  \\
&
 {\cal A}(n,1,2,L:1L,2 n) 
[( {\cal D}_{n}+{\cal D}_{1}+{\cal D}_{2})^2+m^2 ]^{-1} 
{\cal A}(R,3,..., n-1:R(p+1),..., p (p+2)) \, {\cal C}^\Delta_n ,
\end{align}
where  ${\cal A}(n,1,2,L:1L,2 n)$ is similar to \eqref{eq:15} and
\begin{multline}\label{eq:19}
{\cal A}(R,3,..., n-1:R(p+1),..., p (p+2)) = \\
\int_{\hat \gamma}  \prod_{a=3}^{n-3}  \dif x_a  \hat{S}^{-1}_a \frac{ {\rm Pf}A^{R(n-2)}_{R(n-2)}}{ x_{R(n-2)}}
\frac{
[x_{(n-2)(n-1)} x_{(n-1)R} x_{R(n-2)} ]^2   {\rm PT}(R,..., n-1)}{ (x_{R(p+1)}x_{3(n-1)}x_{4(n-2)} \cdots x_{p(p+2)}   )}, 
\end{multline}
with $x_R=\infty$. Here we have used the identity (see Appendix \ref{sixpoint-16})
\begin{equation}\label{pfaffianidentity}
\frac{ (-1) {\rm Pf}A^{R(n-1)}_{R(n-1)}}{ x_{R(n-1)}}  =  \frac{ {\rm Pf}A^{R(n-2)}_{R(n-2)}}{ x_{R(n-2)}}.
\end{equation} 
The integrand in \eqref{eq:19} reproduces the same ladder diagram as in Fig. \ref{Fig10} but with $(n-2)$ points, so equation \eqref{eq:135} provides a recursion relation. Finally, from \eqref{eq:135} and the identity \eqref{eq:prop-insertion}, it is not hard to see that ${\cal A}(\mathbb{I}_n \!: \! 1  (p+1),2n,... , p (p+2)) \, {\cal C}^\Delta_n$ reproduces the Witten diagram in Figure \ref{abc3}.
Since all terms in \eqref{eq:135} commute, this provides an inductive proof that we are free to shuffle terms in the Pfaffian with CSE in the integrands of ladder diagrams.

\begin{figure}[h]
\centering
\parbox[c]{9.2em}{\includegraphics[scale=0.15]{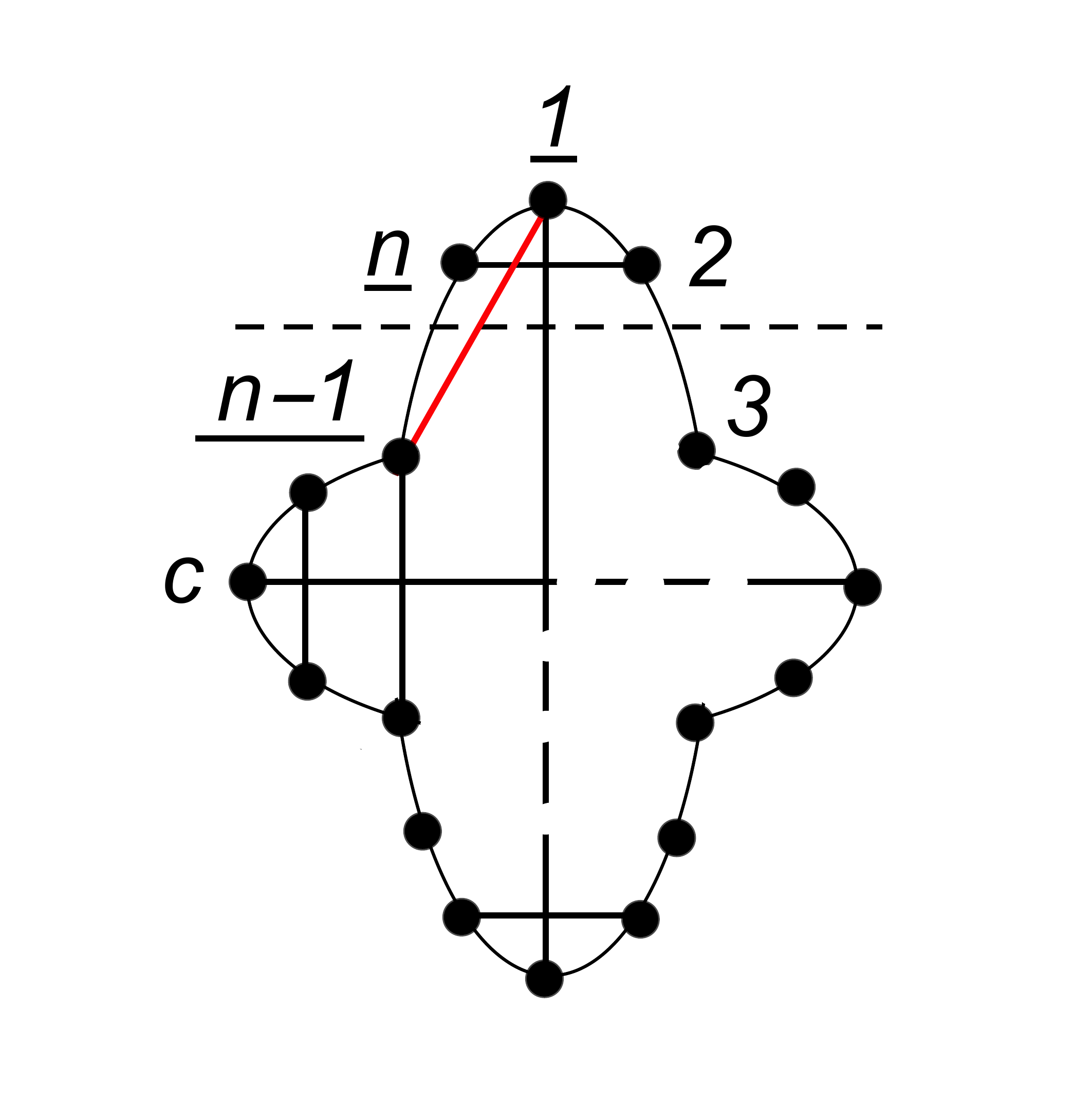}} 
\vspace{-0.2cm}
\caption{Factorization contribution of a non-ladder diagram.}\label{Fig11} 
\vspace{-0.2cm}
\end{figure}

We do not have yet a recursion for non-ladder diagrams, though the method we developed in section \ref{Non-Ladder-8} can be systematically applied. Recall that in section \ref{Non-Ladder-8}, we decomposed the eight-point non-ladder digram in Fig. \ref{Fig8} into a product of two ladder diagrams, one with four external legs and the other with six legs. We then deformed the CSE and Pfaffian of the 6-point diagram to make its ${\rm SL}(2,\mathbb{C})$ symmetry manifest 
and chose a different gauge fixing to obtain a graph (integrand) with the structure Fig. \ref{Fig10}. The generalization of this procedure to higher-point non-ladder diagrams is straightforward. Consider the non-ladder diagram in Fig. \ref{Fig11}.
After applying the integration rules and GRT, we obtain just one factorization contribution,
where the subgraphs graphs are given by a four-point diagram and a non-ladder diagram with $(n-2)$ points as shown in Fig. \ref{Fig11}. We then deform the CSE and Pfaffian of the $(n-2)$-point diagram and choose a new gauge fixing such that the procedcure is repated. For example, the new gauge fixing would correspond to fixing legs $(n-1)$ and $c$ and removing the corresponding rows and columns for the Pfaffian.  
This way we can break up a non-ladder diagram into a product of ladder diagrams connected by bulk-to-bulk propagators. Since the deformation parameters commute  with the CSE and elements of the Pfaffian, they do not introduce any ambiguity. This implies that after factorizing the non-ladder diagram all terms commute, providing an inductive proof that the worldsheet integrand is free of ambiguities. 


\section{One-loop Scattering Equations} \label{1loop}

In this section, we will generalize the tree-level construction in section \ref{wsformulasc} to 1-loop. Our proposal for the 1-loop $n$-point correlator is to take a tree-level $(n+2)$-point correlator, deform it, and paste together two of the
legs. 
This is analogous to the Feynman tree theorem for scattering amplitudes \cite{Feynman:1963ax,Caron-Huot:2010fvq} and similar formulae have appeared in the context of ambitwistor strings \cite{Geyer:2015bja,Cachazo:2015aol,Farrow:2020voh} and loop-level recursion \cite{Arkani-Hamed:2010zjl,Baadsgaard:2015twa}. 
More precisely, the 1-loop $n$-point cosmological correlator for $\phi^4$ theory in dS is given by
\begin{equation}
\Psi_{n}^{1-{\rm loop}}=-\frac{1}{\pi}\int {\rm d}^{d} \ell\int_{-\infty}^{\infty}\frac{\omega^{2} {\rm d}\omega}{\omega^{2}+\nu^{2}}\lim_{\vec{k}_{\pm}\rightarrow\pm\vec{\ell}}\tilde{\Psi}_{n+2}\left((\nu_{+},\vec{k}_{+}),(\nu_{1},\vec{k}_{1}),...,(\nu_{n},\vec{k}_{n}),(\nu_{-},\vec{k}_{-})\right),
\label{1loopmain}
\end{equation}
where $\tilde{\Psi}$ is a deformed tree-level correlator which we will describe below, $\vec{\ell}$ is the loop momentum projected onto the boundary, $\nu_{\pm}=\pm i\omega$, and $\nu_{1}=....=\nu_{n}=\nu=\Delta-d/2$. Legs $1$ to $n$ are external with fixed mass $m^{2}=\Delta\left(d-\Delta\right)$,
while legs $\pm$ are internal and their mass is integrated over:
\begin{equation}
\mathcal{D}_{i}^{2}=-m^{2},\,\,\,i=1,...,n
\end{equation}
\begin{equation}
\mathcal{D}_{\pm}^{2}=-M^{2}.
\end{equation}
The mass of the internal legs is given by $M^{2}=\Delta_{\pm}(d-\Delta_{\pm})=(\frac{d}{2}-\nu_{\pm})(\frac{d}{2}+\nu_{\pm})=\frac{d^{2}}{4}+\omega^{2}$. As we will explicitly see  in the next subsection, the integral over $\omega$ essentially connects the bulk-to-boundary propagators of the $\pm$ legs into a bulk-to-bulk propagator. Note that the tree-level correlator encodes a sum over Witten diagrams with the internal legs appearing in different places, so after pasting them together this gives a sum over 1-loop Witten diagrams. As we will explain in the next subsection, we expect our formula to work for scalar fields with mass $0\leq m \leq d/2$ in dS, but after Wick rotation it should work for any $m^2 \geq-d^2/4$ in AdS.

In more detail, the deformed tree-level correlator in \eqref{1loopmain} is essentially the same as the one in \eqref{eq:totalphi4} except for a few modifications:
\begin{equation}
\tilde{\Psi}_{n+2}=\frac{\delta^{d}(\vec{k}_{T})}{2(3!)^{p-1}}\!\!\sum_{\rho\in{\rm S}_{n+1}}\!\!\!{\rm sgn}_{\rho}\,\,\left.{\cal \tilde{A}}(\rho)\right|_{\rm{1PI}}\,\tilde{\mathcal{C}}_{n+2},
\label{deformedtree}
\end{equation}
where $\rho$ labels permutations of legs $\left\{ 1,...,n,-\right\}$ and $p=\frac{n}{2}+1$. The worldsheet integral $\tilde{\mathcal{A}}$ is similar to \eqref{eq:7}. However, we deform the differential operators and contact diagram as described below and discard perfect matchings in which the $\pm$ legs are attached to the same vertex. This is indicated by the subscript 1PI. The motivation for removing such contributions will become clear soon. Moreover, we fix the punctures $\left\{ \sigma_{\pm},\sigma_{1}\right\}$. This can be motivated by 1-loop formulae arising from ambitwistor string theory, where one starts with a genus-one worldsheet with a single fixed puncture ($\sigma_1$) which degenerates to a spherical worldsheet with two more fixed punctures ($\sigma_{\pm}$) \cite{Adamo:2013tsa,Geyer:2015bja}. Finally, note that \eqref{deformedtree} has an extra factor of $\frac{1}{2}$ compared to \eqref{eq:totalphi4}, which is simply  a symmetry factor of the 1-loop Witten diagrams that will arise after pasting the $\pm$ legs together. 

Let us describe the integrand in more detail. The deformed contact diagram is a product of bulk-to-boundary propagators
where two legs have a different scaling dimension than the others:
\begin{equation}
\tilde{\mathcal{C}}_{n+2}=\int\frac{{\rm d}\eta}{\eta^{d+1}}\mathcal{K}_{\nu_{+}}(\vec{k}_{+},\eta)\mathcal{K}_{\nu_{-}}(\vec{k}_{-},\eta)\prod_{i=1}^{n}\mathcal{K}_{\nu_{i}}(\vec{k}_{i},\eta).\label{contactd}
\end{equation}
The differential operators in the integrand and CSE are deformed as follows: 
\begin{eqnarray}
2\mathcal{D}_{1}\cdot\mathcal{D}_{+} &\rightarrow& \left(\mathcal{D}_{1}+\mathcal{\mathcal{D}}_{+}\right)^{2}+2m^{2} \nonumber \\
2\mathcal{D}_{1}\cdot\mathcal{D}_{-} &\rightarrow& -\left(\mathcal{D}_{1}-\mathcal{\mathcal{D}}_{-}\right)^{2}-2m^{2} \nonumber \\
2\mathcal{D}_{n}\cdot\mathcal{D}_{-} &\rightarrow& \left(\mathcal{D}_{n}+\mathcal{\mathcal{D}}_{-}\right)^{2}+2m^{2} \nonumber \\
2\mathcal{D}_{n}\cdot\mathcal{D}_{+} &\rightarrow& -\left(\mathcal{D}_{n}-\mathcal{\mathcal{D}}_{+}\right)^{2}-2m^{2}.
\label{deformation}
\end{eqnarray}
This was inspired by the deformed 1-loop SE proposed in \cite{Farrow:2020voh}. Prior to the deformation, the CSE for particles $\left\{ 1,n,+,-\right\} $ are given by
\begin{eqnarray}
S_{1}&=&\frac{2\mathcal{D}_{1}\cdot\mathcal{D}_{2}-m^{2}}{\sigma_{12}}+\sum_{j=3}^{n}\frac{2\mathcal{D}_{1}\cdot\mathcal{D}_{2}}{\sigma_{1j}}+\frac{2\mathcal{D}_{1}\cdot\mathcal{D}_{-}}{\sigma_{1-}}+\frac{2\mathcal{D}_{1}\cdot\mathcal{D}_{+}-m^{2}}{\sigma_{1+}} \nonumber \\
S_{n}&=&\frac{2\mathcal{D}_{n}\cdot\mathcal{D}_{+}}{\sigma_{n+}}+\sum_{j=1}^{n-2}\frac{2\mathcal{D}_{n}\cdot\mathcal{D}_{j}}{\sigma_{nj}}+\frac{2\mathcal{D}_{n}\cdot\mathcal{D}_{n-1}-m^{2}}{\sigma_{nn-1}}+\frac{2\mathcal{D}_{n}\cdot\mathcal{D}_{-}-m^{2}}{\sigma_{n-}}\nonumber \\
S_{+}&=&\frac{2\mathcal{D}_{+}\cdot\mathcal{D}_{1}-m^{2}}{\sigma_{+1}}+\sum_{j=2}^{n}\frac{2\mathcal{D}_{+}\cdot\mathcal{D}_{j}}{\sigma_{+j}}+\frac{2\mathcal{D}_{+}\cdot\mathcal{D}_{-}-2M^{2}+m^{2}}{\sigma_{+-}}\nonumber \\
S_{-}&=&\sum_{j=1}^{n-1}\frac{2\mathcal{D}_{-}\cdot\mathcal{D}_{j}}{\sigma_{-j}}+\frac{2\mathcal{D}_{-}\cdot\mathcal{D}_{n}-m^{2}}{\sigma_{-n}}+\frac{2\mathcal{D}_{-}\cdot\mathcal{D}_{+}-2M^{2}+m^{2}}{\sigma_{-+}}.
\end{eqnarray}
It is not difficult to check that they are ${\rm SL}(2,\mathbb{C})$ invariant, and the deformation in \eqref{deformation} preserves this symmetry. Since we are fixing legs $\left\{ 1,+,-\right\}$, the deformed CSE for these legs will not be needed in practice. 

In summary, a 1-loop correlator can essentially be obtained from a tree-level correlator with two additional legs. One important difference compared to the tree-level formula discussed earlier is that the spectral parameters of two bulk-to-boundary propagators in the contact term are different than the rest. Nevertheless, many previous formulas can be generalized. In particular, using \eqref{eq:confgenboundary} we find
\begin{equation}
(\mathcal{D}_{a}\cdot\mathcal{D}_{b})\mathcal{K}_{\nu_{a}}\mathcal{K}_{\nu_{b}}=\eta^{2}[\partial_{\eta}\mathcal{K}_{\nu_{a}}\partial_{\eta}\mathcal{K}_{\nu_{b}}+(\vec{k}_{a}\cdot\vec{k}_{b})\mathcal{K}_{\nu_{a}}\mathcal{K}_{\nu_{b}}],
\label{eq:DaDb2}
\end{equation}
where the spectral parameters for the two propagators can be different. Using this identity, we also find  
\begin{align}
\mathcal{D}_{1...p}^{2}U_{1,n}&= (\mathcal{D}_{1}+\ldots + \mathcal{D}_{p})^2 U_{1,n}, \nonumber \\
& = -\sum_{a=1}^{p}m_{a}^{2}U_{1,n}+2\sum_{1\leq a<b\leq p}(\mathcal{D}_{a}\cdot\mathcal{D}_{b})U_{1,n}, \label{eq:bulkvsboundary-props2}
\end{align}
where $\mathcal{D}_{1...p}^{2}$ is defined in \eqref{eq:D2def} with $k=k_{1...p}$ and the contact term consists of bulk-to-boundary propagators with different spectral parameters:
\begin{equation}
U_{1,n}(\eta)\equiv\prod_{a=1}^{n}\mathcal{K}_{\nu_{a}}(k_{a},\eta).
\end{equation}
Note that the differential operator on the left-hand-side of \eqref{eq:bulkvsboundary-props2} does not depend on the spectral parameters of the bulk-to-boundary propagators on which it acts.

In the next subsection, we will verify that \eqref{1loopmain} gives the correct 1-loop 4-point correlator. It is also not difficult to see that it solves the CWI for any number of legs. In the limit $\vec{k}_{\pm} \rightarrow \pm \vec{\ell}$, the momentum delta function becomes that of an $n$-point correlator. Since the conformal generators for each external leg can be commuted past the Pfaffian and CSE to act on the bulk-to-boundary propagator for that leg in the contact diagram, the CWI are equivalent to those of a tree-level $n$-point correlator.

\subsection{4-point Example} \label{1loopexample}

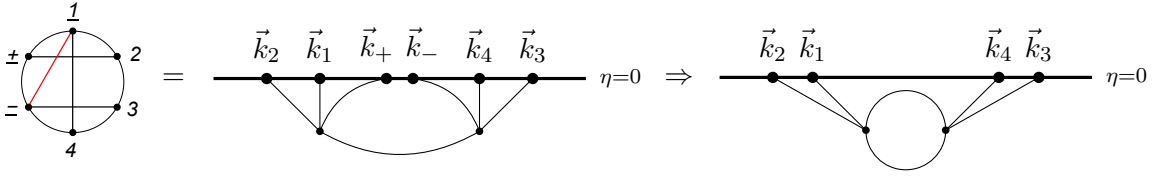
\begin{figure}
\centering
\text{$
\raisebox{0.59\height}{\parbox[c]{6.5em}{\includegraphics[scale=0.23]{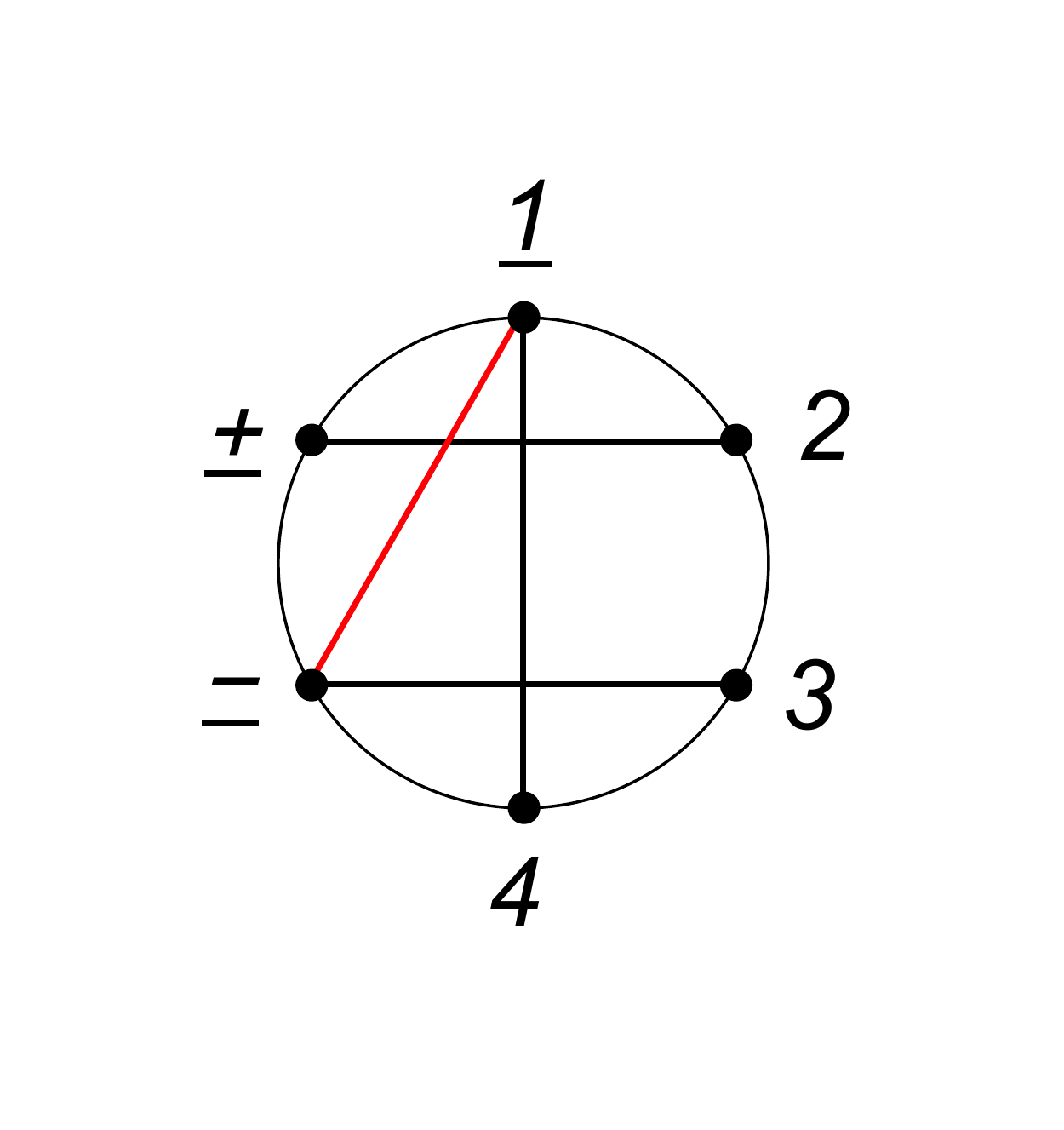}}}
\mathord{\begin{tikzpicture}[scale=0.35]
\node at (-2.5,0) [circle,,fill=black,inner sep=0pt,minimum size=0mm,label=center:$ \text{$=$} $]  {};
\node at (14.3,0) [circle,,fill=black,inner sep=0pt,minimum size=0mm,label=center:$ \text{${\scriptstyle \eta = 0}$} $]  {};
\node at (16.5,0) [circle,,fill=black,inner sep=0pt,minimum size=0mm,label=center:$ \text{$\Rightarrow $} $]  {};
	\node at (0,0) [circle,,fill=black,inner sep=0pt,minimum size=0mm,label=above:$ $]  {};
	\node at (1,0) [circle,,fill=black,inner sep=0pt,minimum size=1.5mm,label=above:$\vec{k}_2$]  {};
	\node at (2,0) [circle,,fill=black,inner sep=0pt,minimum size=0mm,label=above:$ $]{};
	\node at (3,0) [circle,,fill=black,inner sep=0pt,minimum size=1.5mm,label=above:$\vec{k}_1$]{};
	\node at (4,0) [circle,,fill=black,inner sep=0pt,minimum size=0mm,label=above:$ $]{};
	\node at (5.5,0) [circle,,fill=black,inner sep=0pt,minimum size=1.5mm,label=above:$\!\!\!\!\vec{k}_+$]{};
		\node at (6,0) [circle,,fill=black,inner sep=0pt,minimum size=0mm,label=above:$ $]{};
			\node at (6.5,0) [circle,,fill=black,inner sep=0pt,minimum size=1.5mm,label=above:$\,\,\,\,\vec{k}_-$]{};
				\node at (8,0) [circle,,fill=black,inner sep=0pt,minimum size=0mm,label=above:$ $]{};
				   \node at (9,0) [circle,,fill=black,inner sep=0pt,minimum size=1.5mm,label=above:$\vec{k}_4$]{};
				\node at (10,0) [circle,,fill=black,inner sep=0pt,minimum size=0mm,label=above:$ $]{};
				 \node at (11,0) [circle,,fill=black,inner sep=0pt,minimum size=1.5mm,label=above:$\vec{k}_3$]{};
				  \node at (3,-2) [circle,,fill=black,inner sep=0pt,minimum size=1.0mm,label=above:$ $]{};
				 \node at (9,-2) [circle,,fill=black,inner sep=0pt,minimum size=1.0mm,label=above:$ $]{};
	\draw[very thick,black] (-1,0) -- (13,0);
		\draw[black] (1,0) -- (3,-2);
			\draw[black] (3,-2) -- (3,0);
   \draw (5.5,0) to [bend right]   (3,-2);			   
		\draw (9,-2) to [bend right]   (6.5,0);	
			        \draw[black] (9,-2) -- (9,0);
			        \draw[black] (9,-2) -- (11,0);
			          \draw (9,-2) to [bend left]   (3,-2);
	\end{tikzpicture}}
\,\,\, 
\raisebox{-0.05\height}{\text{$\mathord{\begin{tikzpicture}[scale=0.35]
\node at (14.3,0) [circle,,fill=black,inner sep=0pt,minimum size=0mm,label=center:$ \text{${\scriptstyle \eta = 0}$} $]  {};
	\node at (0,0) [circle,,fill=black,inner sep=0pt,minimum size=0mm,label=above:$ $]  {};
	\node at (1,0) [circle,,fill=black,inner sep=0pt,minimum size=1.5mm,label=above:$\vec{k}_2$]  {};
	\node at (2,0) [circle,,fill=black,inner sep=0pt,minimum size=0mm,label=above:$ $]{};
	\node at (2.5,0) [circle,,fill=black,inner sep=0pt,minimum size=1.5mm,label=above:$\vec{k}_1$]{};
	\node at (4,0) [circle,,fill=black,inner sep=0pt,minimum size=0mm,label=above:$ $]{};
		\node at (6,0) [circle,,fill=black,inner sep=0pt,minimum size=0mm,label=above:$ $]{};
				   \node at (9.5,0) [circle,,fill=black,inner sep=0pt,minimum size=1.5mm,label=above:$\vec{k}_4$]{};
				\node at (10,0) [circle,,fill=black,inner sep=0pt,minimum size=0mm,label=above:$ $]{};
				 \node at (11,0) [circle,,fill=black,inner sep=0pt,minimum size=1.5mm,label=above:$\vec{k}_3$]{};
	\draw[very thick,black] (-1,0) -- (13,0);
		\draw[black] (1,0) -- (4.5,-2);
			\draw[black] (4.5,-2) -- (2.5,0);
			        \draw[black] (7.5,-2) -- (9.5,0);
			        \draw[black] (7.5,-2) -- (11,0);
			        	\draw [fill=white!0] (6,-2) circle (1.5);
			        	\node at (4.5,-2) [circle,,fill=black,inner sep=0pt,minimum size=1.0mm,label=above:$ $]{};
			        	\node at (7.5,-2) [circle,,fill=black,inner sep=0pt,minimum size=1.0mm,label=above:$ $]{};
	\end{tikzpicture}}
$}	
}
$}
\caption{Evaluating the worldsheet formula for the perfect matching on the left gives the tree-level 6-point Witten diagram in the middle. Pasting together the $\pm$ legs using the procedure described in the text then gives the 1-loop 4-point Witten diagram on the right.}\label{6pt}
\end{figure}

Let us illustrate how this works at four points. First we consider the 6-point graph on the left-hand-side of Figure \ref{6pt}, which corredponds to a term in the deformed tree-level correlator in \eqref{deformedtree}, where we fix the positions $\left\{ 1,+,- \right\} $ and remove the rows and columns associated with legs  $\left\{ 1,-\right\} $ from the Pfaffian.
After performing the deformation in \eqref{deformation} and summing over permutations, the deformed tree-level correlator is given by
\begin{equation}
\tilde{\Psi}_6=\delta^{d}\left(\vec{k}_{T}\right)\int_{\tilde{S}_{2}\cap \tilde{S}_{3}\cap \tilde{S}_{4}}\frac{ {\rm d}\s_{2} {\rm d}\s_{3} {\rm d}\s_{4}}{\tilde{S}_{2}\tilde{S}_{3}\tilde{S}_{4}}\frac{\left(\s_{1-}\s_{-+}\s_{+1}\right)^{2}}{\s_{12}\s_{23}\s_{34}\s_{4-}\s_{-+}\s_{+1}}\frac{1}{\s_{14}\s_{2+}\s_{3-}}\frac{{\rm Pf}\tilde{A}_{1-}^{1-}}{\sigma_{1-}}\tilde{\mathcal{C}}_{6}+\rm{perms},
\label{psi6}
\end{equation}
where we sum over cyclic permuations of legs $(2,3,4)$, which will ultimately correspond to the sum over the $s,t,u$ channels of the 1-loop Witten diagrams. Note that the first term in \eqref{psi6} is just the deformed version of \eqref{eq:94} with the relabeling $\left\{ 5,6\right\} \rightarrow\left\{ -,+\right\} $. In particular, the deformed contact diagram is defined in \eqref{contactd} and the deformed Pfaffian is given by
\begin{multline}
{\rm Pf}\tilde{A}_{1-}^{1-}= \frac{(2\mathcal{D}_{2}\cdot\mathcal{D}_{+})(2\mathcal{D}_{3}\cdot\mathcal{D}_{4}-m^{2})}{\s_{2+}\s_{34}}-\frac{(2\mathcal{D}_{2}\cdot\mathcal{D}_{4})(2\mathcal{D}_{3}\cdot\mathcal{D}_{+})}{\s_{24}\s_{3+}}\\
-\frac{(2\mathcal{D}_{2}\cdot\mathcal{D}_{3}-m^{2})[(\mathcal{D}_{4}-\mathcal{D}_{+})^{2}+2m^{2}]}{\s_{23}\s_{4+}}.
\end{multline}
Following the same steps of the evaluation of the undeformed tree-level 6-point correlator in section \ref{sec:sixpoint}, we find
\begin{equation}
\tilde{\Psi}_{6}=\delta^{d}(\vec{k}_{T})(\mathcal{D}_{+12}^{2}+m^{2})^{-1}\tilde{\mathcal{C}}_{6}+\rm{perms}.
\label{6ptdeformedtree}
\end{equation}
Note that the 6-point contact diagram has been split in such a way that the $\pm$ legs are now on different vertices connected by a bulk-to-bulk propagator with boundary momentum $\vec{k}_{+12}$, as illustrated in Figure \ref{6pt}. After taking the limit $\vec{k}_{\pm}\rightarrow \pm\vec{\ell}$ and performing the integral over $\omega$ in \eqref{1loopmain}, the bulk-to-boundary propagators corresponding to legs $\pm$ will then be pasted together to form a bulk-to-bulk propagator, giving rise to the 1-loop Witten diagram illustrated on the right-hand-side of Figure \ref{6pt}. To see this, first note that the only $\omega$ dependence in \eqref{6ptdeformedtree} comes from the bulk-to-boundary propagators. The operator $\mathcal{D}_{+12}^{2}$ does not depend on the spectral parameters of legs $\left\{ +,1,2\right\} $, only on their momenta. As a result, the $\omega$ integral will only involve the bulk-to-boundary propagators for the $\pm$ legs and we are left with the following integral:
\begin{align}
-\frac{1}{\pi}\int_{-\infty}^{\infty} \frac{ {\rm d}\omega \, \omega^{2}}{\omega^{2}+\nu^{2}}\mathcal{K}_{i\omega}(\ell,\eta)\mathcal{K}_{-i\omega}(\ell,\eta')&=\frac{(-1)^{d/2}  (\eta\eta')^{d/2}  }{\pi}  \int_{-\infty}^{\infty}\frac{ {\rm d} \omega \,\omega^{2}}{\omega^{2}+\nu^{2}}\frac{K_{i\omega}(-i \ell\eta)K_{-i\omega}(-i \ell\eta')}{\Gamma(1+i\omega)\Gamma(1-i\omega)}  \nonumber \\
&= G_{\nu}^{dS}(\ell,\eta,\eta'),
\end{align}
where in the first line we have used \eqref{kprop}, and the second line is the AdS propagator in \eqref{eq:AdSbuktobulk} with $z\rightarrow -i \eta$. As explained in section \ref{dspropsmomentum}, Wick-rotating the split representation \eqref{eq:AdSbuktobulk} only agrees with the standard dS propagator for masses satisfying $0 \leq m \leq \frac{d}{2}$. For this reason, we expect that the 1-loop proposal in \eqref{1loopmain} will only hold for dS correlators in this mass range. On the other hand, it should hold for any mass $m^2 \geq -d^2/4$ after Wick-rotation back to AdS. 

Putting everything together, we finally obtain the 1-loop Witten diagram illustrated
in Figure \ref{1loop4pt} summed over permutations:
\begin{equation}
\Psi_{4}^{1-{\rm loop}}=\delta^{d}(\vec{k}_{T})\int {\rm d}^{d} \ell\prod_{i=1}^{2}\mathcal{K}_{\nu}(\vec{k}_{i},\eta)G_{\nu}(|\vec{\ell}|,\eta,\eta')G_{\nu}(|\vec{\ell}+\vec{k}_{12}|,\eta,\eta')\prod_{j=3}^{4}\mathcal{K}_{\nu}(\vec{k}_{j},\eta')+\rm{perms}.
\label{4ptresult1}
\end{equation}
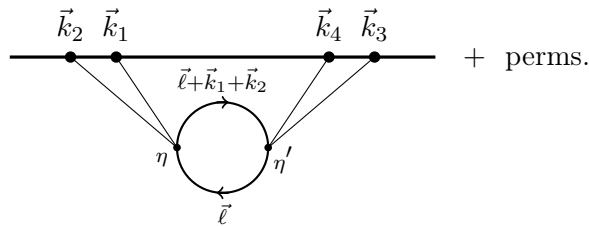
\begin{figure}[h]
\centering
\text{$ 
\mathord{\begin{tikzpicture}[scale=0.4]
	\node at (16,-0.8) [circle,,fill=black,inner sep=0pt,minimum size=0mm,label=above:$\text{$+\,\,$ perms.} $]  {};
	\node at (1,0) [circle,,fill=black,inner sep=0pt,minimum size=1.5mm,label=above:$\vec{k}_2$]  {};
	\node at (2,0) [circle,,fill=black,inner sep=0pt,minimum size=0mm,label=above:$ $]{};
	\node at (2.5,0) [circle,,fill=black,inner sep=0pt,minimum size=1.5mm,label=above:$\vec{k}_1$]{};
	\node at (4,0) [circle,,fill=black,inner sep=0pt,minimum size=0mm,label=above:$ $]{};
				\node at (8,0) [circle,,fill=black,inner sep=0pt,minimum size=0mm,label=above:$ $]{};
				   \node at (9.5,0) [circle,,fill=black,inner sep=0pt,minimum size=1.5mm,label=above:$\vec{k}_4$]{};
				\node at (10,0) [circle,,fill=black,inner sep=0pt,minimum size=0mm,label=above:$ $]{};
				 \node at (11,0) [circle,,fill=black,inner sep=0pt,minimum size=1.5mm,label=above:$\vec{k}_3$]{};
				 \node at (12,0) [circle,,fill=black,inner sep=0pt,minimum size=0mm,label=above:$ $]{};
				 \node at (4,-3.4) [circle,,fill=black,inner sep=0pt,minimum size=0mm,label=center:$ \text{${\scriptstyle \eta}$} $]  {};
				 \node at (8,-3.4) [circle,,fill=black,inner sep=0pt,minimum size=0mm,label=center:$ \text{${\scriptstyle \eta}$}^\prime $]  {};
				  \node at (6,-5.2) [circle,,fill=black,inner sep=0pt,minimum size=0mm,label=center:$ \text{${\scriptstyle \vec{\ell}}$} $]  {};
				   \node at (6,-0.75) [circle,,fill=black,inner sep=0pt,minimum size=0mm,label=center:$ \text{${\scriptstyle \vec{\ell}+\vec{k}_1+\vec{k}_2}$} $]  {};
	\draw[very thick,black] (-1,0) -- (13,0);
		\draw[black,] (1,0) -- (4.5,-3);
			\draw[black] (4.5,-3) -- (2.5,0);
			        \draw[black] (7.5,-3) -- (9.5,0);
			        \draw[black] (7.5,-3) -- (11,0);
			        	\draw [black, thick] (6,-3) circle [radius=1.5];
			        	\draw [<-,thick] (5.8,-4.5) -- (6.2,-4.5);
			        	\draw [->,thick] (5.8,-1.5) -- (6.2,-1.5);
			      	\node at (4.5,-3) [circle,,fill=black,inner sep=0pt,minimum size=1.0mm,label=above:$ $]{};
			      	\node at (7.5,-3) [circle,,fill=black,inner sep=0pt,minimum size=1.0mm,label=above:$ $]{};
	\end{tikzpicture}}
	$}
	\caption{1-loop 4-point Witten diagrams obtained in \eqref{4ptresult1}.}\label{1loop4pt}
\end{figure}
This is just the expected expansion of the 1-loop 4-point cosmological correlator in terms of Witten diagrams. Finally, note that our definition of the deformed $(n+2)$-point tree-level correlator in \eqref{deformedtree} discarded contributions that would give rise to 1-loop corrections to the bulk-to-boundary propagators illustrated in Figure \ref{renorm}.
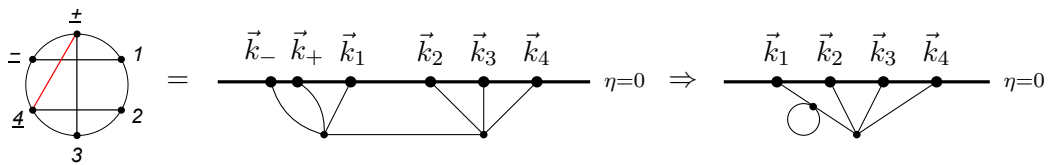
\begin{figure}[h]
\centering
\text{$ 
\raisebox{0.37\height}{\parbox[c]{6.5em}{\includegraphics[scale=0.23]{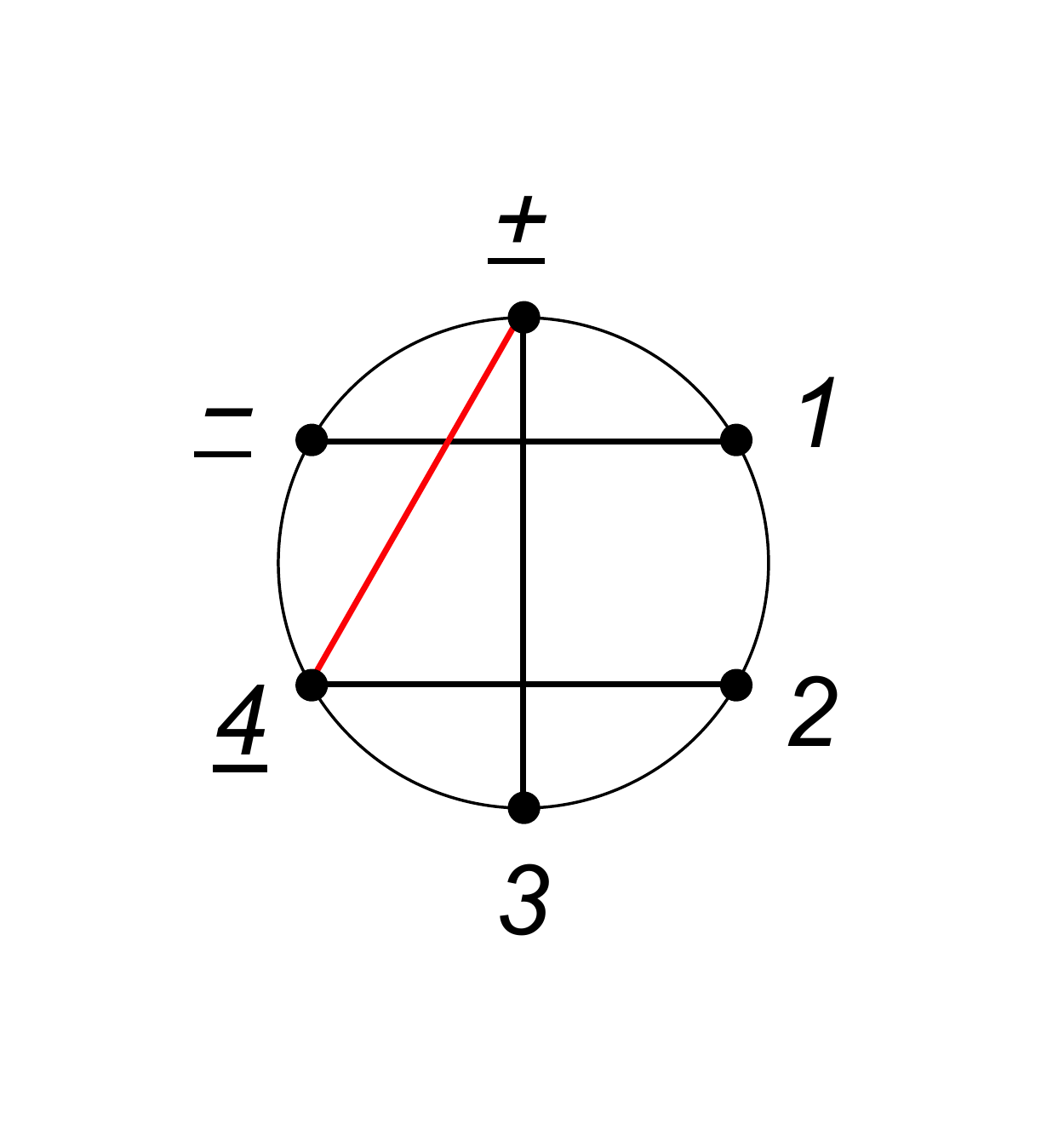}}}
\mathord{\begin{tikzpicture}[scale=0.35]
\node at (-2.5,0) [circle,,fill=black,inner sep=0pt,minimum size=0mm,label=center:$ \text{$=$} $]  {};
\node at (14.3,0) [circle,,fill=black,inner sep=0pt,minimum size=0mm,label=center:$ \text{${\scriptstyle \eta = 0}$} $]  {};
\node at (16.5,0) [circle,,fill=black,inner sep=0pt,minimum size=0mm,label=center:$ \text{$\Rightarrow $} $]  {};
	\node at (0,0) [circle,,fill=black,inner sep=0pt,minimum size=0mm,label=above:$ $]  {};
	\node at (1,0) [circle,,fill=black,inner sep=0pt,minimum size=1.5mm,label=above:$\!\!\!\!\vec{k}_-$]  {};
	\node at (2,0) [circle,,fill=black,inner sep=0pt,minimum size=0mm,label=above:$ $]{};
	\node at (2,0) [circle,,fill=black,inner sep=0pt,minimum size=1.5mm,label=above:$\,\,\, \,\vec{k}_+$]{};
	\node at (4,0) [circle,,fill=black,inner sep=0pt,minimum size=1.5mm,label=above:$\,\vec{k}_1$]{};
		\node at (6,0) [circle,,fill=black,inner sep=0pt,minimum size=0mm,label=above:$ $]{};
			\node at (7,0) [circle,,fill=black,inner sep=0pt,minimum size=1.5mm,label=above:$\vec{k}_2$]{};
				\node at (3,-2) [circle,,fill=black,inner sep=0pt,minimum size=1.0mm,label=above:$ $]{};
				   \node at (9,0) [circle,,fill=black,inner sep=0pt,minimum size=1.5mm,label=above:$\vec{k}_3$]{};
				\node at (10,0) [circle,,fill=black,inner sep=0pt,minimum size=0mm,label=above:$ $]{};
				 \node at (11,0) [circle,,fill=black,inner sep=0pt,minimum size=1.5mm,label=above:$\vec{k}_4$]{};
				 \node at (9,-2) [circle,,fill=black,inner sep=0pt,minimum size=1.0mm,label=above:$ $]{};
	\draw[very thick,black] (-1,0) -- (13,0);
		\draw (1,0) to [bend right]  (3,-2);
		\draw (2,0) to [bend left]  (3,-2);
			   \draw[black] (4,0) -- (3,-2);
			       \draw[black] (9,-2) -- (7,0);
			        \draw[black] (9,-2) -- (9,0);
			        \draw[black] (9,-2) -- (11,0);
			        	\draw[black] (3,-2) -- (9,-2);
	\end{tikzpicture}}
\,\,\, 
\raisebox{-0.0\height}{\text{$\mathord{\begin{tikzpicture}[scale=0.35]
\node at (10.3,0) [circle,,fill=black,inner sep=0pt,minimum size=0mm,label=center:$ \text{${\scriptstyle \eta = 0}$} $]  {};
	\node at (0,0) [circle,,fill=black,inner sep=0pt,minimum size=0mm,label=above:$ $]  {};
	\node at (1,0) [circle,,fill=black,inner sep=0pt,minimum size=1.5mm,label=above:$\vec{k}_1$]  {};
	\node at (2,0) [circle,,fill=black,inner sep=0pt,minimum size=0mm,label=above:$ $]{};
	\node at (3,0) [circle,,fill=black,inner sep=0pt,minimum size=1.5mm,label=above:$\vec{k}_2$]{};
	\node at (4,0) [circle,,fill=black,inner sep=0pt,minimum size=0mm,label=above:$ $]{};
	\node at (5,0) [circle,,fill=black,inner sep=0pt,minimum size=1.5mm,label=above:$\vec{k}_3$]{};
			\node at (7,0) [circle,,fill=black,inner sep=0pt,minimum size=1.5mm,label=above:$\vec{k}_4$]{};
				\node at (4,-2) [circle,,fill=black,inner sep=0pt,minimum size=1.0mm,label=above:$ $]{};
	\draw[very thick,black] (-1,0) -- (9,0);
		\draw[ black] (1,0) -- (4,-2);
			\draw[ black] (4,-2) -- (3,0);
			   \draw[ black] (5,0) -- (4,-2);
			       \draw[black] (4,-2) -- (7,0);
			          	\draw [fill=white!20] (2.0,-1.425) circle (0.6);
			          \node at (2.36,-0.94) [circle,,fill=black,inner sep=0pt,minimum size=1.0mm,label=above:$ $]{};
	\end{tikzpicture}}
$}	
}
$}
\caption{Perfect matching corresponding to a loop correction to an external propagator.}\label{renorm}
\end{figure}
The motivation for discarding such contributions is that in the flat space limit one will get scattering amplitudes obtained by amputating the external legs and putting them on-shell. Alternatively, we can keep such contributions by removing the restriction on perfect matchings in \eqref{deformedtree} and including a factor of 2 for those where the $\pm$ legs are attached to the same vertex.

\section{Conclusion} \label{conclusion}

We have provided further details on the recently proposed worldsheet formula for tree-level correlators of massive $\phi^4$ theory in de Sitter momentum space \cite{Gomez:2021qfd}, which is a toy model for inflationary cosmology. This formula is based on cosmological scattering equations defined in terms of conformal generators acting in the future boundary. Another key ingredient of the formula is a Pfaffian of the conformal generators which appears in the integrand. In principle there could be ambiguities arising from the operatorial nature of the building blocks involved, but they are nontrivially absent in all the examples we have considered. In \cite{Gomez:2021qfd} we verified the proposal up to six points and showed that in the flat space limit it reduces to the CHY formula for $\phi^4$ amplitudes for any number of legs \cite{Cachazo:2014xea}. In this paper, we derived all of these results in detail and extended them up to eight points, using simple graphical rules for evaluating the worldsheet integrals which we derived from a double cover formulation. We also proposed a 1-loop generalization of the worldsheet formula and verified it at 4-points. This relies on a split representation for the bulk-to-bulk propagator in dS obtained by Wick-rotation from AdS and is only valid for masses $0\leq m \leq d/2$, where $d$ is the dimension of the boundary. On the other hand, after Wick-rotating back to AdS it is valid for $m^2\geq -d^2/4$.  It would be interesting to consider generalizations of our formula to arbitrary mass in dS based on the split representations recently found in \cite{Meltzer:2021zin,Sleight:2021plv,Fichet:2021xfn}. 

While we have focused on one of the simplest models of inflationary cosmology, several steps can be taken to make it more realistic. For example, one could consider more general mass deformations which allow particles with different masses to propagate \cite{Naculich:2014naa}. Another interesting direction would be to break some symmetries of the cosmological scattering equations, since cosmological surveys measure correlators of curvature perturbations which become nontrivial when de Sitter boosts are broken \cite{Cheung:2007st,Green:2020ebl,Pajer:2020wxk}. Moreover, it would be of great interest to explore more complicated Pfaffian structures corresponding to more general effective scalar theories, such as the non-linear sigma model, scalar DBI, and special Galileon theories, whose flat space amplitudes have a very elegant description in terms of CHY formulae \cite{Cachazo:2014xea}. When lifting to curved background, new subtleties can arise such as curvature corrections to the effective actions and ordering ambiguities in the worldsheet formulae associated with the operatorial nature of the integrands. It would also be interesting to generalize the construction to spinning correlators in order to make the double copy in (A)dS more systematic \cite{Armstrong:2020woi,Albayrak:2020fyp,Alday:2021odx,Diwakar:2021juk,Farrow:2018yni,Lipstein:2019mpu,Zhou:2021gnu,Sivaramakrishnan:2021srm}, which would in turn streamline the calculation of gravitional correlators in this background. Finally, it would be very desirable to evaluate the worldsheet formula by directly solving the cosmological scattering equations rather than mapping the formula to a sum of Witten diagrams using the global residue theorem. It may be possible to do this analytically using techniques from integrability \cite{Roehrig:2020kck,Eberhardt:2020ewh}, but a numerical approach may ultimately be needed. Indeed, even the scattering equations in flat space require numerical solution above five points for generic kinematics \cite{Cachazo:2013gna,Farrow:2018cqi,DeLaurentis:2019vkf}. In summary, the study of the cosmological scattering equations is still in its infancy, and there are many exciting directions to be explored.  

\begin{center}
\textbf{Acknowledgements}
\end{center}
We thank Charlotte Sleight and David Meltzer for useful discussions. HG and AL are supported by the Royal Society via a PDRA grant and a University Research Fellowship, respectively. RLJ acknowledges the Czech Science Foundation - GA\v{C}R for financial support under the grant 19-06342Y.

\appendix

\section{dS isometries in momentum space} \label{props}

The operator $\mathcal{D}_{k}^{2}$ in equation \eqref{eq:D2def} is the
Casimir of the isometry group of de Sitter space, defined as
\begin{equation}
\mathcal{D}_k^{2}\equiv D^{2}+\tfrac{1}{2}(P_{i}K^{i}+K^{i}P_{i})-\tfrac{1}{2}L_{ij}L^{ij},
\label{dscasimir}
\end{equation}
where the bulk generators for dS$_{d+1}$ can be realized in momentum space as
\begin{eqnarray}
D & = & k^{i}\partial_{i}-\eta\partial_{\eta}+d,\\
P_{i} & = & k_{i},\\
K_{i} & = & k_{i}\partial^{j}\partial_{j}-2k^{j}\partial_{j}\partial_{i}-2d\partial_{i}+2\eta\partial_{\eta}\partial_{i}+\eta^{2}k_{i},\\
L_{ij} & = & k_{i}\partial_{j}-k_{j}\partial_{i},
\end{eqnarray}
where $\partial_{\eta}=\partial/\partial\eta$ and $\partial_{i}=\partial/\partial k^{i}$. Plugging the bulk generators into \eqref{dscasimir} then leads to
\begin{equation}
\mathcal{D}_k^{2} = \eta^{2}\partial_{\eta}^{2}+(1-d)\eta\partial_{\eta}+\eta^{2}k^{2}.
\end{equation}
Moreover they satisfy the standard algebra of the Euclidean conformal group $SO(1,d+1)$:
\begin{equation}
\begin{array}{ccc}
[D,P_{i}]  =  P_{i}, & [D,K_{i}]  = -K_{i}, & [K_{i},P_{j}]  = 2(L_{ij}-\delta_{ij}D).
\end{array}
\end{equation}

When acting on boundary correlators, we can work with the
boundary counterparts of the above generators. The only condition
is that the eigenvalue equation (\ref{eq:eigenstates}) holds at the
future boundary as well, i.e. with $\eta\to 0^-$. Noting that the asymptotic form of the eigenfunctions is $\eta^\Delta$, the generators $P_i$ and $L_{ij}$ are unchanged while $D$ and $K_i$ become 
\begin{subequations}
\begin{eqnarray}
D & = & k_{i}\partial^{i}+(d-\Delta),\\
K_{i} & = & k_{i}\partial^{j}\partial_{j}-2k^{j}\partial_{j}\partial_{i}-2(d-\Delta)\partial_{i}.
\end{eqnarray}
\end{subequations}
The boundary conformal generators are also derived in the next Appendix by Fourier transforming the standard expressions in position space.

\section{Proof of ${\rm SL}(2,\mathbb{C})$} \label{sl2c}

In this Appendix, we will derive the conformal generators in momentum space given by \eqref{eq:CGG-boundary} and prove the ${\rm SL}(2,\mathbb{C})$ symmetry of the CSE.

Let us begin by Fourier transforming a scalar correlator:
\begin{equation}
\left\langle \mathcal{O}\left(\vec{p}_{1}\right)...\mathcal{O}\left(\vec{p}_{n}\right)\right\rangle =\int {\rm d}^{d}x_{1}...{\rm d}^{d}x_{n}e^{i\left(\vec{p}_{1}\cdot\vec{x}_{1}+...+\vec{p}_{n}\cdot\vec{x}_{n}\right)}\left\langle \mathcal{O}\left(\vec{x}_{1}\right)...\mathcal{O}\left(\vec{x}_{n}\right)\right\rangle .\label{fouriertransform}
\end{equation}
Using translational invariance, we can write
\begin{equation}
\left\langle \mathcal{O}\left(\vec{x}_{1}\right)...\mathcal{O}\left(\vec{x}_{n}\right)\right\rangle =F\left(\vec{x}_{2}^{-},...,\vec{x}_{n}^{-}\right),
\end{equation}
where $\vec{x}_{i}^{-}=\vec{x}_{i}-\vec{x}_{1}$, for $i=2,...,n$.
Changing integration variables then gives
\begin{eqnarray}
\left\langle \mathcal{O}\left(\vec{p}_{1}\right)...\mathcal{O}\left(\vec{p}_{n}\right)\right\rangle &=&\int {\rm d}^{d}x_{1}e^{i\vec{x}_{1}\cdot\vec{p}_{T}}\int {\rm d}^{d}x_{2}^{-}...d^{d}x_{n}e^{i\left(\vec{p}_{2}\cdot\vec{x}_{2}^{-}+...+\vec{p}_{n}\cdot\vec{x}_{n}^{-}\right)}F\left(\vec{x}_{2}^{-},...,\vec{x}_{n}^{-}\right)\\
&=&\delta^{d}\left(\vec{p}_{T}\right)\left\langle \left\langle \mathcal{O}\left(\vec{p}_{1}\right)...\mathcal{O}\left(\vec{p}_{n}\right)\right\rangle \right\rangle .\label{deltadecomp}
\end{eqnarray}

Now consider dilatations:
\begin{equation}
\sum_{i=1}^{n}\left(\vec{x}_{i}\cdot\partial_{\vec{x}_{i}}+\Delta\right)\left\langle \mathcal{O}\left(\vec{x}_{1}\right)...\mathcal{O}\left(\vec{x}_{n}\right)\right\rangle =0.\label{dilatation}
\end{equation}
Fourier transforming according to \eqref{fouriertransform} then gives
\begin{equation}
\sum_{i=1}^{n}D_{i}\left\langle \mathcal{O}\left(\vec{p}_{1}\right)...\mathcal{O}\left(\vec{p}_{n}\right)\right\rangle =0,\,\,\,D_{i}=\vec{p}_{i}\cdot\partial_{\vec{p}_{i}}+d-\Delta.
\end{equation}
Moreovoer if we plug in the decomposition in \eqref{deltadecomp} and
integrate deriviatives of the delta function against a test function,
we find
\begin{equation}
\sum_{i=1}^{n}D_{i}\left\langle \left\langle \mathcal{O}\left(\vec{p}_{1}\right)...\mathcal{O}\left(\vec{p}_{n}\right)\right\rangle \right\rangle =d\left\langle \left\langle \mathcal{O}\left(\vec{p}_{1}\right)...\mathcal{O}\left(\vec{p}_{n}\right)\right\rangle \right\rangle .\label{DiWI1}
\end{equation}

Let us derive a slightly different form of the dilation Ward identity.
First note that translational invariance implies that
\begin{equation}
\left\langle \mathcal{O}\left(\vec{x}_{1}+\vec{a}\right)...\mathcal{O}\left(\vec{x}_{n}+\vec{a}\right)\right\rangle =\left\langle \mathcal{O}\left(\vec{x}_{1}\right)...\mathcal{O}\left(\vec{x}_{n}\right)\right\rangle, 
\end{equation}
whose infinitesimal form is
\begin{equation}
\sum_{i=1}^{n}\partial_{\vec{x}_{i}}\left\langle \mathcal{O}\left(\vec{x}_{1}\right)...\mathcal{O}\left(\vec{x}_{n}\right)\right\rangle =0.
\end{equation}
Replacing $\partial_{\vec{x}_{1}}=-\sum_{i=2}^{n}\partial_{\vec{x}_{i}}$
in \eqref{dilatation} then implies 
\begin{equation}
-\sum_{i=2}^{n}\left(\vec{x}_{i}^{-}\cdot\partial_{\vec{x}_{i}^{-}}+\Delta\right)\left\langle \mathcal{O}\left(\vec{x}_{1}\right)...\mathcal{O}\left(\vec{x}_{n}\right)\right\rangle =\Delta\left\langle \mathcal{O}\left(\vec{x}_{1}\right)...\mathcal{O}\left(\vec{x}_{n}\right)\right\rangle .\label{dliatona}
\end{equation}
Fourier transforming this equation then gives
\begin{equation}
\sum_{i=2}^{n}D_{i}\left\langle \left\langle \mathcal{O}\left(\vec{p}_{1}\right)...\mathcal{O}\left(\vec{p}_{n}\right)\right\rangle \right\rangle =\Delta\left\langle \left\langle \mathcal{O}\left(\vec{p}_{1}\right)...\mathcal{O}\left(\vec{p}_{n}\right)\right\rangle \right\rangle .\label{DIWI2}
\end{equation}
Subtracting \eqref{DIWI2} from \eqref{DiWI1} then implies
\begin{equation}
D_{1}\left\langle \left\langle \mathcal{O}\left(\vec{p}_{1}\right)...\mathcal{O}\left(\vec{p}_{n}\right)\right\rangle \right\rangle =\left(d-\Delta\right)\left\langle \left\langle \mathcal{O}\left(\vec{p}_{1}\right)...\mathcal{O}\left(\vec{p}_{n}\right)\right\rangle \right\rangle .\label{dw3}
\end{equation}

Next let's look at conformal boosts:
\begin{equation}
0=\sum_{i=1}^{n}\left(\vec{x}_{i}^{2}\partial_{x_{i}^{\mu}}-2x_{i}^{\mu}\left(\vec{x}_{i}\cdot\partial_{\vec{x}_{i}}+\Delta\right)\right)\left\langle \mathcal{O}\left(\vec{x}_{1}\right)...\mathcal{O}\left(\vec{x}_{n}\right)\right\rangle .\label{boostcwi}
\end{equation}
After Fourier transforming, we find that the boost acts trivially
on the delta function (after integrating against a test function and
using rotational invariance), so the Ward identity reduces to 
\begin{equation}
\sum_{i=1}^{n}K_{i}^{\mu}\left\langle \left\langle \mathcal{O}\left(\vec{p}_{1}\right)...\mathcal{O}\left(\vec{p}_{n}\right)\right\rangle \right\rangle =0,\,\,\,K_{i}^{\mu}=p_{i}^{\mu}\partial_{\vec{p}_{i}}^{2}-2\left(\vec{p}_{i}\cdot\partial_{\vec{p}_{i}}+d-\Delta\right)\partial_{p_{i}^{\mu}}.\label{conformaboota}
\end{equation}
We can obtain another useful form of this Ward identity by substituting $\partial_{\vec{x}_{1}}=-\sum_{i=2}^{n}\partial_{\vec{x}_{i}}$
into \eqref{boostcwi}: 
\begin{equation}
0=\sum_{i=2}^{n}\left(\left(\vec{x}_{i}^{-}\right)^{2}\partial_{x_{i}^{-\mu}}-2x_{i}^{-\mu}\left(\vec{x}_{i}^{-}\cdot\partial_{\vec{x}_{i}^{-}}+\Delta\right)\right)\left\langle \mathcal{O}\left(\vec{x}_{1}\right)...\mathcal{O}\left(\vec{x}_{n}\right)\right\rangle .
\end{equation}
To obtain the above form, we have rotational invariance of the
correlator and \eqref{dliatona}. Fourier transforming \eqref{boostcwi} to momentum space
then implies 
\begin{equation}
\sum_{i=2}^{n}K_{i}^{\mu}\left\langle \left\langle \mathcal{O}\left(\vec{p}_{1}\right)...\mathcal{O}\left(\vec{p}_{n}\right)\right\rangle \right\rangle =0,\label{conformalboostb}
\end{equation}
and subtracting this from \eqref{conformaboota} gives
\begin{equation}
K_{1}^{\mu}\left\langle \left\langle \mathcal{O}\left(\vec{p}_{1}\right)...\mathcal{O}\left(\vec{p}_{n}\right)\right\rangle \right\rangle =0.\label{conformalboostc}
\end{equation}

In summary, the conformal Ward identities in \eqref{DIWI2},\eqref{dw3},\eqref{conformalboostb}, and
\eqref{conformalboostc} follow from using momentum conservation on
support of the momentum delta function to remove all dependence of
$\left\langle \left\langle \mathcal{O}\left(\vec{p}_{1}\right)...\mathcal{O}\left(\vec{p}_{n}\right)\right\rangle \right\rangle $
correlator on $\vec{p}_{1}$. With these identities, it is now straightforward to prove the $SL(2,\mathbb{C})$
symmetry of the CSE. In particular we must show that the following sum vanishes:
\begin{equation}
\sum_{j=1 \atop j\neq i}^n\alpha_{ij}\left\langle \left\langle \mathcal{O}\left(\vec{p}_{1}\right)...\mathcal{O}\left(\vec{p}_{n}\right)\right\rangle \right\rangle =\left(\vec{p}_{i}\cdot\sum_{j\neq i}\vec{K}_{j}+\vec{K}_{i}\cdot\sum_{j\neq i}\vec{p}_{j}+2D_{i}\sum_{j\neq i}D_{j}-2m^{2}\right)\left\langle \left\langle \mathcal{O}\left(\vec{p}_{1}\right)...\mathcal{O}\left(\vec{p}_{n}\right)\right\rangle \right\rangle .
\end{equation}
On support of the momentum delta function, we can remove all dependence
of $\left\langle \left\langle \mathcal{O}\left(\vec{p}_{1}\right)...\mathcal{O}\left(\vec{p}_{n}\right)\right\rangle \right\rangle $
on $\vec{p}_{i}$. The first term on the right-hand-side then vanishes by \eqref{conformalboostb},
the second term vanishes by \eqref{conformalboostc} (applying momentum conservation to the sum over momenta
after taking derivatives), and the third term reduces to 2$\Delta(d-\Delta)=2m^{2}$
by \eqref{DIWI2} and \eqref{dw3}. Hence, the sum indeed vanishes.

\section{Review of Double Cover Formalism} \label{double}

In this Appendix, we will review of the double cover (DC) formalism introduced in the beginning of section \ref{DoubleCover}. Recall that in this approach, the one represents the worldsheet by the quadratic curve
$y_a^2=z_a^2-\Lambda^2$, where $\Lambda\neq0$ is a constant and $(z_a,y_a)  \in \mathbb{C}^2$. One then promotes $\Lambda$, which encodes factorization of the worldsheet, to an integration variable along with the $2n$ variables  $(z_1,\ldots,z_n,y_1,\ldots,y_n)$. Below, we will describe some useful building blocks for constructing the integrand in the double cover formalism and show that it enjoys an ${\rm SL}(2,\mathbb{C})$ along with a scaling symmetry which can be used to fix four of the integration variables.  

In the single-cover formalism, the Cauchy kernel is given by
\begin{eqnarray}
\omega=
\frac{\dif \sigma}{\s-\s_a}\,\,
\left\{
\begin{matrix}
\text {Pole } \sigma=\sigma_a, &  \text{Res. } +1 \\
\text {Pole } \sigma=\infty, &  \text{Res. } -1. \\ 
\end{matrix}
\right. \, 
\end{eqnarray}
The analogue of this in the DC formalism is  
\begin{eqnarray}
\omega= \tau_a(z)\, \dif z\, \,
\left\{
\begin{matrix}
\text {Pole } (z,y)=(z_a,y_a), &  \text{Res. } +1 \\
\text {Pole } (z,y)=(\infty,\infty), &  \text{Res. } -1. \\ 
\end{matrix}
\right. \,  \label{eq:cauchyK}
\end{eqnarray}
where 
\begin{equation}
2\, \tau_a(z)=\frac{1}{z-z_a}\left( \frac{y_a}{y} + 1 \right) + \frac{1}{ y}, \, \qquad \text{with }\,\,\,\,  y^2=z^2-\Lambda^2.
\end{equation}
From \eqref{eq:cauchyK} the following analogue of the CSE in the DC language:
\begin{equation}
S^{\Lambda}_a = \sum_{b=1 \atop b\neq a}^n (2\, {\cal D}_a\cdot {\cal D}_b + \mu_{ab}) \, \tau_{a:b} = \sum_{b=1 \atop b\neq a}^n \a_{ab} \, \tau_{a:b},, \qquad \,
\tau_{a:b}=\frac{1}{2\, z_{ab}}\left( \frac{y_b}{y_a} + 1 \right) + \frac{1}{ 2\,  y_a}. 
\end{equation}
In addition, the Parke-Taylor factor is mapped to
\begin{equation}
{\rm PT}(\mathbb{I}_n)=\frac{1}{\s_{12}\s_{23}\cdots \s_{n1}} \, \, \,  \rightarrow \,\, \, 
{\rm PT}^\Lambda(\mathbb{I}_n) = \tau_{1:2}\tau_{2:3} \cdots \tau_{n:1}\, .
\end{equation}

Unlike the single-cover formalism, the Cauchy kernel is no longer antisymmetric, $\tau_{a:b}\neq -\tau_{b:a}$. Thus, in order to define the $A$-matrix and the reduced Pfaffian  we rewrite $\tau_{a:b}$ in the following way:
\begin{equation}
\tau_{a:b} = \frac{(yz)_a}{y_a} \, T_{ab}, \qquad T_{ab}\equiv \frac{1}{(yz)_a-(yz)_b}, \quad \text{with} \,\, \, (yz)_i\equiv y_i+z_i,
\end{equation}
where $T_{ab}$ is clearly antisymmetric, $T_{ab}=-T_{ba}$.  In terms of $T_{ab}$, the Parke-Taylor factor becomes,
\begin{equation} 
 {\rm PT}^\Lambda(\mathbb{I}_n) = \tau_{1:2}\tau_{2:3} \cdots \tau_{n:1} = \left[\prod_{a=1}^n\frac{(yz)_a}{y_a}\right] T_{12} T_{23} \cdots T_{n1}. \label{eq:Parke-TaylorT}
\end{equation}
Now, we are able to define the $A$-matrix and the reduced Pfaffian,
\begin{align}
A_{rs} & =
\begin{cases} 
\displaystyle  \a_{rs} \, T_{rs},  & r\neq s\\
\displaystyle  ~~ 0 , & r=s
\end{cases}\,\, , \qquad \,
{\rm Pf}^\prime A = \left[\prod_{a=1}^n\frac{(yz)_a}{y_a}\right]  (-1)^{c+d} \, T_{cd} \, {\rm Pf}  A^{cd}_{cd} \, ,
\end{align}
where the prefactor $\prod_a \left[(yz)_a / y_a \right]$ in ${\rm Pf}^\prime A$  must be introduced (as in \eqref{eq:Parke-TaylorT}) to have ${\rm SL}(2,\mathbb{C})$ invariance.

The ${\rm SL}(2,\mathbb{C})$ symmetry of the CSE is generated by the vector fields
\begin{eqnarray}
\left.
\begin{matrix}
& L_{\pm 1} = \L^{\pm 1} \sum_{a=1}^n y_a \, (yz)_a^{\mp 1}\, \partial_{z_a}  \\ 
\\
& \hspace{-2.2cm} L_{0} =  \sum_{a=1}^n y_a \, \partial_{z_a}
\end{matrix}
\right\}
\begin{matrix}
& [L_{\pm 1},L_0]=\pm L_{\pm1}\,  \\
& \!\!\!\!\!\! [L_{1},L_{-1}]=2 L_{0},
\end{matrix}
\, 
\end{eqnarray}
on the support of the curves, $ {\rm C}_i=y_i^2-z_i^2+\Lambda^2=0$. The Faddeev-Popov determinant associated with fixing three of the CSE is then given by 
\begin{eqnarray}
\Delta_{(pqr)} =  \frac{y_p\, y_q\, y_r}{(yz)_p\,(yz)_q\,(yz)_r}\times  \left|
                            \begin{array}{ccc}
                              1 & (yz)_p\, &   (yz)_p^{2}  \\
                              1 &  (yz)_q \,\, &   (yz)_q^{2}\\
                              1  &  (yz)_r\,\,&   (yz)_r^{2} \\
                            \end{array}
                          \right| =  \left(\tau_{p:q} \, \tau_{q:r} \, \tau_{r:p} \right)^{-1}\, .
                       \label{Delta-3}
\end{eqnarray} 

Since the Riemann sphere is embedded in $\mathbb{CP}^2$ there is an extra symmetry inherited from this projective space, which is the dilatation
\begin{equation}
(z_1,\ldots, z_n, \Lambda) \, \,\, \rightarrow \, \, \, \rho \, (z_1,\ldots, z_n, \Lambda).
\end{equation} 
The generator of this symmetry is given by the vector
\begin{equation}
D = \sum_{a=1}^n z_a \, \partial_{z_a} + \Lambda \, \partial_{\Lambda}.
\end{equation}
Using this additional symmetry, we can gauge an extra puncture with the Faddeev-Popov determinant
\begin{eqnarray}
\Delta_{(pqr|m)} =
\Delta_{(pqr)} \, \s_m -  \Delta_{(mpq)} \, \s_r +  \Delta_{(rmp)} \, \s_q -\Delta_{(qrm)} \, \s_p \, ,  \label{Delta-4}
\end{eqnarray} 
where $\Lambda$ now becomes an integration variable.

\section{More on 6 points}\label{sixpoint-16}  

In this Appendix, we will carry out the 6-point calculation in section \ref{sec:sixpoint} using a different choice of Pfaffian, notably
\begin{equation}\label{}
\frac{{\rm Pf}A^{15}_{15}}{ x_{15}}  \,  \rightarrow \,   \frac{ (-1)\, {\rm Pf}A^{56}_{56}}{ x_{56}}.
\end{equation} 
We will show that the final result is independent of this choice, which demonstrates the gauge invariance of the method and illustrates the useful identity in \eqref{pfaffianidentity}.  

\begin{figure}[h]
\centering
\parbox[c]{06.7em}{\includegraphics[scale=0.22]{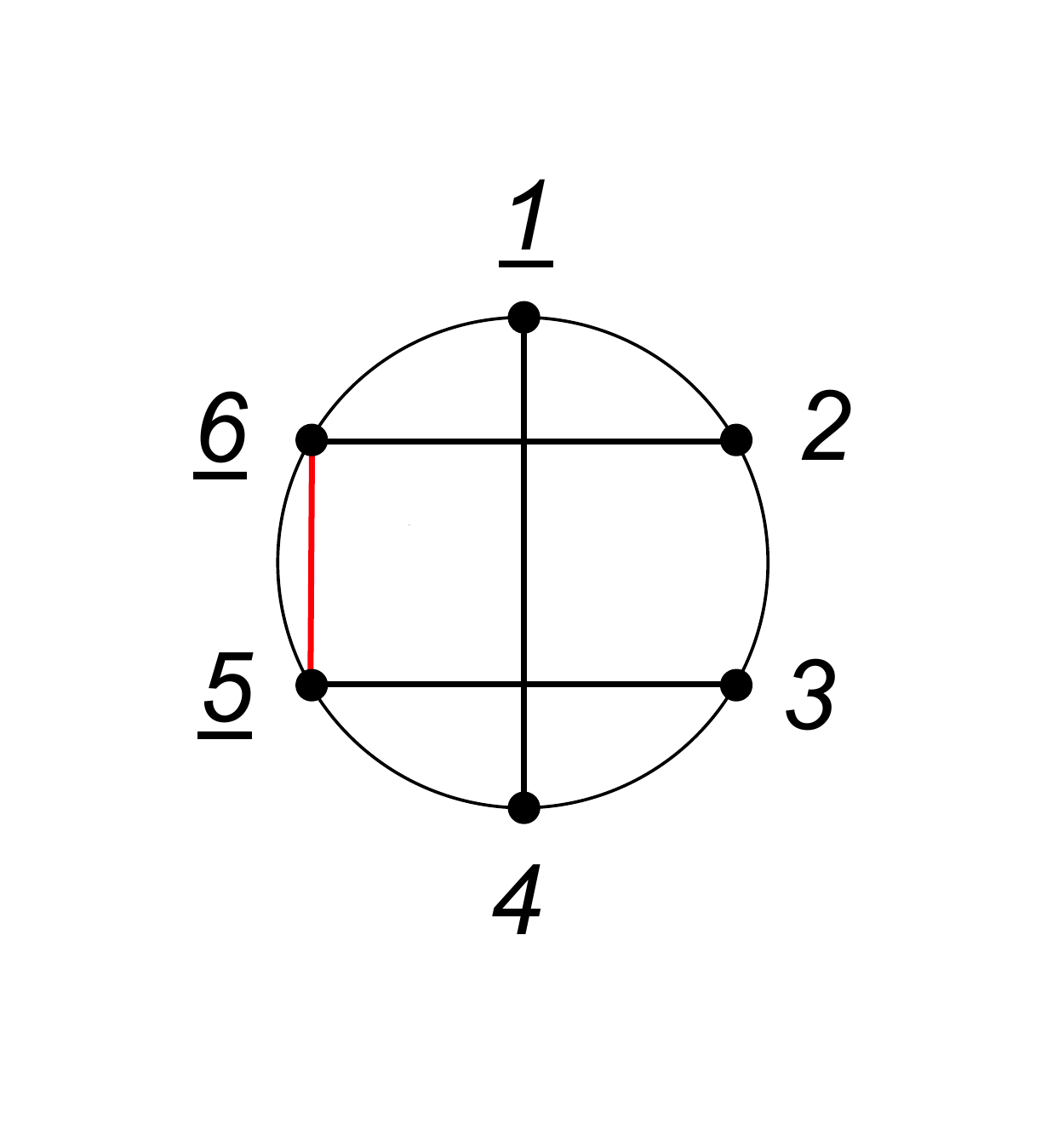}}
$\rightarrow$\,\,
\parbox[c]{06.5em}{\includegraphics[scale=0.22]{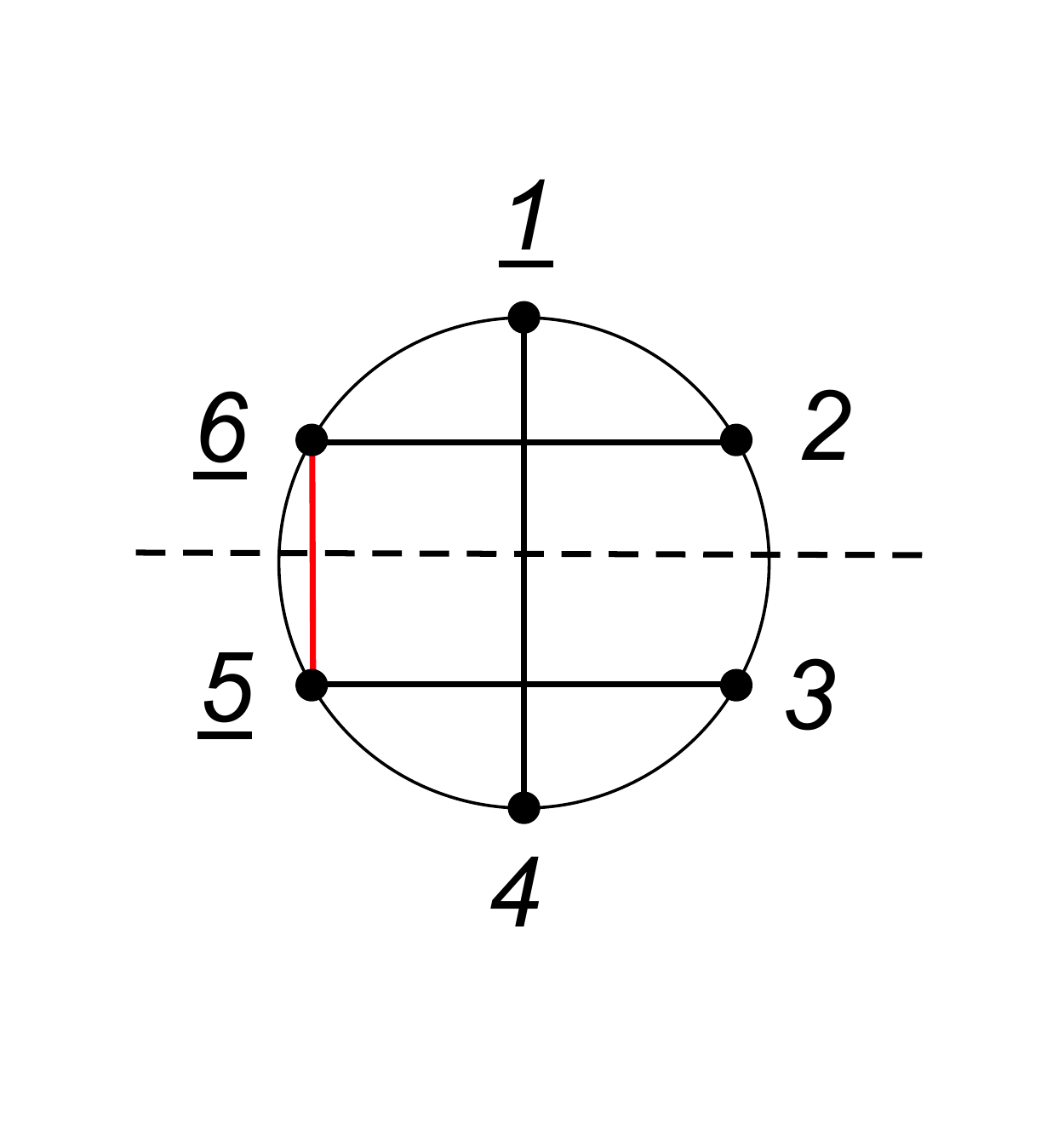}}
\parbox[c]{05.7em}{\includegraphics[scale=0.22]{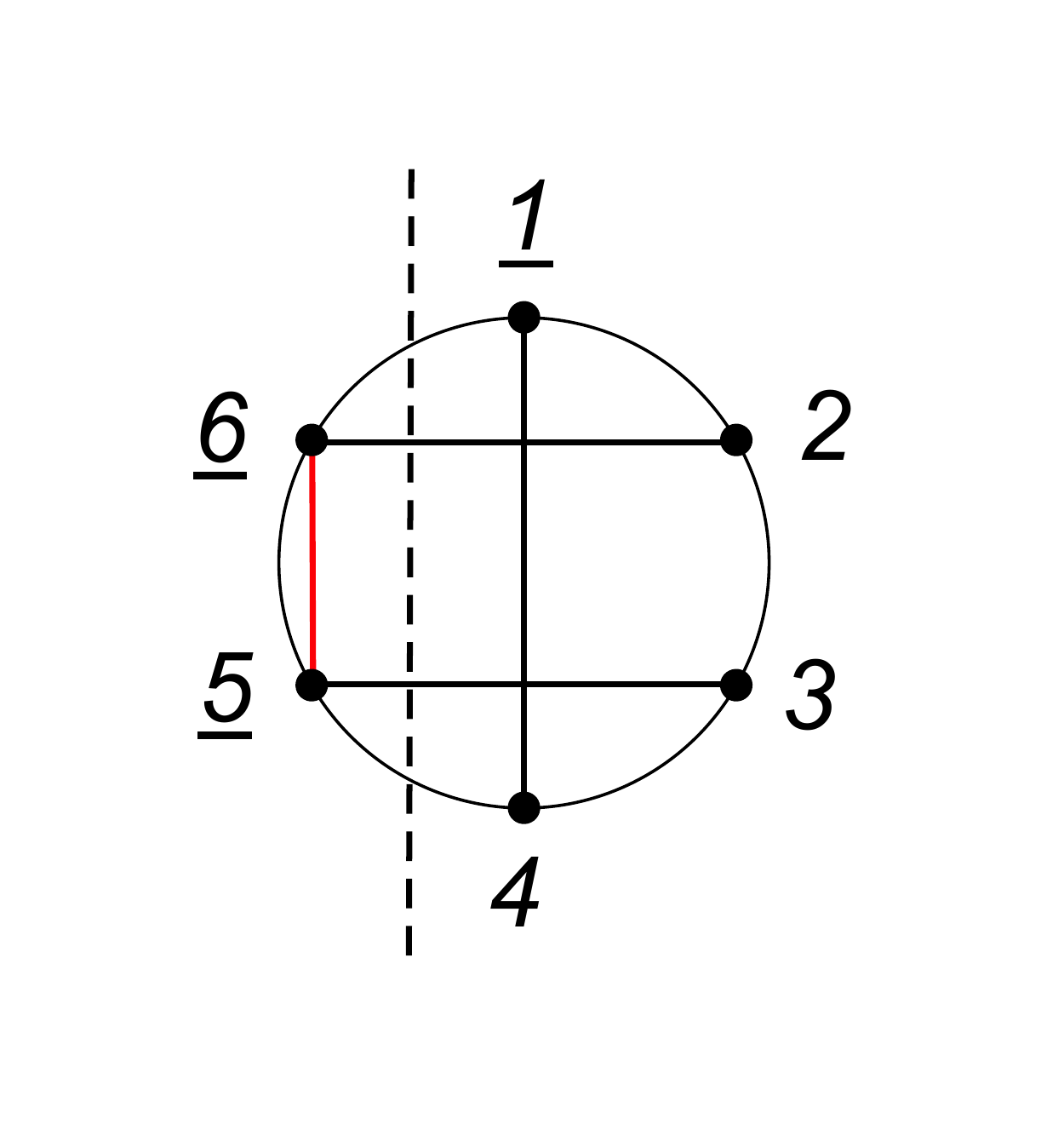}}
\parbox[c]{05.7em}{\includegraphics[scale=0.22]{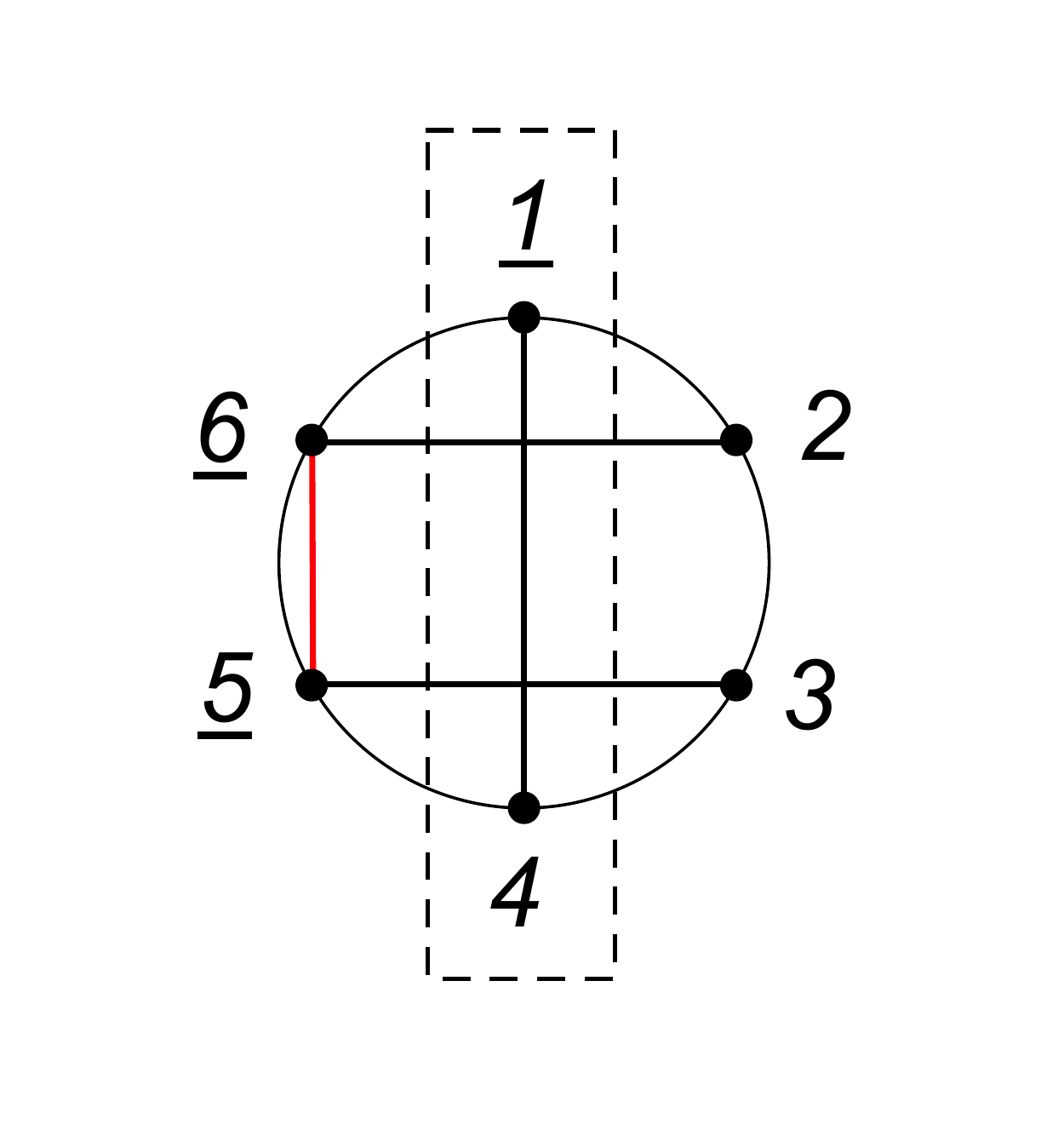}}
\vspace{-0.2cm}
\caption{Another Pfaffian choice for ${\cal A}(\mathbb{I}_6 \!: \! 14,26,35)$ and its factorization contributions.}\label{FigA1} 
\vspace{-0.2cm}
\end{figure}

Let us consider the worldsheet integral, ${\cal A}(\mathbb{I}_6 : 14,26,35 )$, where unlike to expression given in \eqref{eq:94}, 
we remove the rows and columns $\{ 5,6\}$ from the $A$-matrix:
\begin{equation}\label{eq:165}
{\cal A}(\mathbb{I}_6 \!: \! 14,26,35)  =  \! \int_{\gamma}  \dif\sigma_2\dif\sigma_3\dif\sigma_4 ( \sigma_{56} \sigma_{61} \sigma_{15} )^2 (S_2S_3S_4)^{-1}  
{\rm PT}(\mathbb{I}_6) \, \frac{(-1) \, {\rm Pf} A^{56}_{56}}{\sigma_{56} }\,  \frac{ 1 }{ (\sigma_{14} \sigma_{26} \sigma_{35})} .
\end{equation}
By the integration rules, it is simple to see there are three factorization contributions given in Fig. \ref{FigA1}.
We will expand the reduced Pfaffian and compute each term separately. 

\begin{figure}[h]
\centering
\hspace{-1.2cm}
$_{(a)}$\!\!\!\!
\parbox[c]{07.5em}{\includegraphics[scale=0.22]{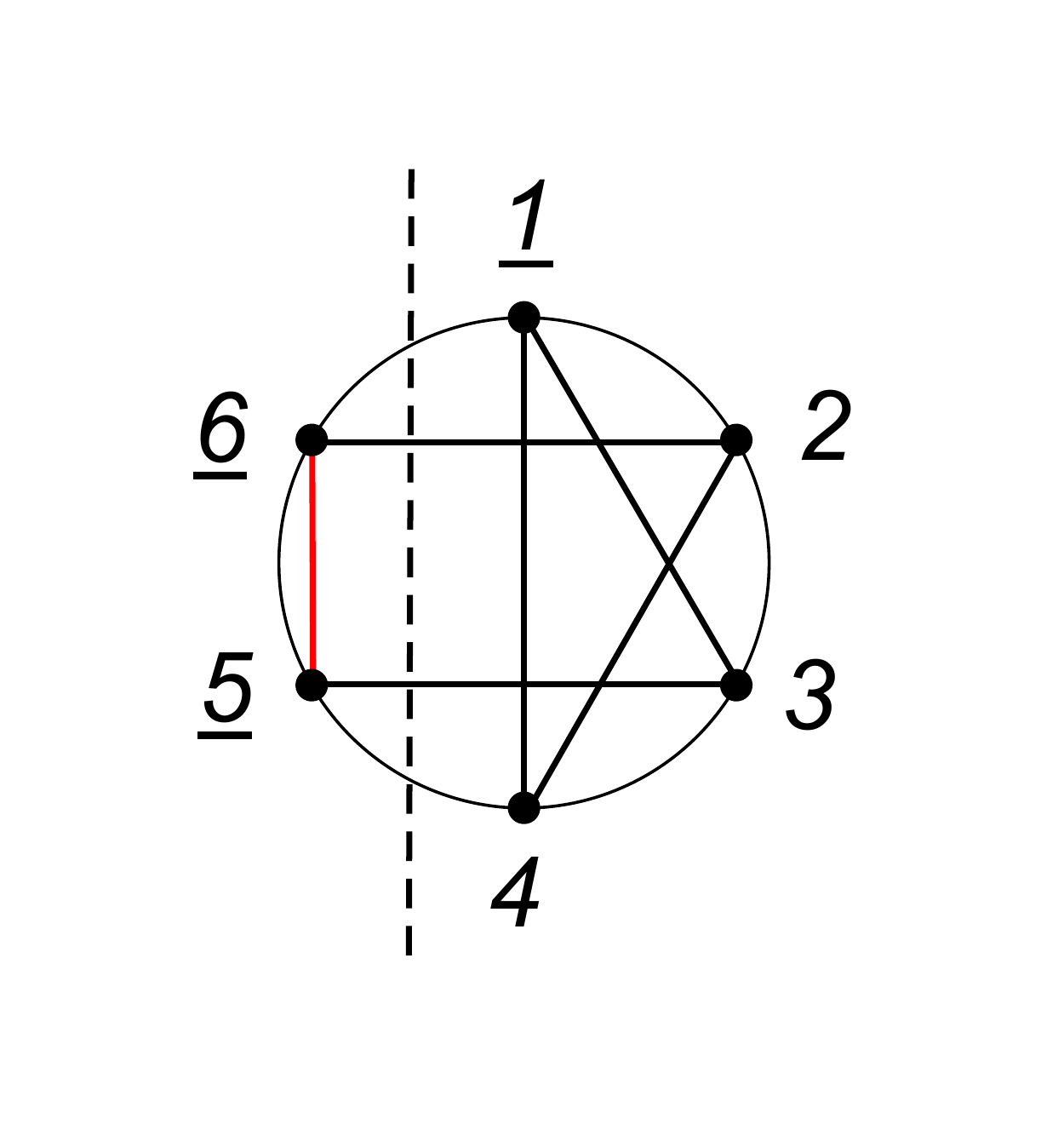}}$_{(b)}$\!\!\!\!
\parbox[c]{05.5em}{\includegraphics[scale=0.22]{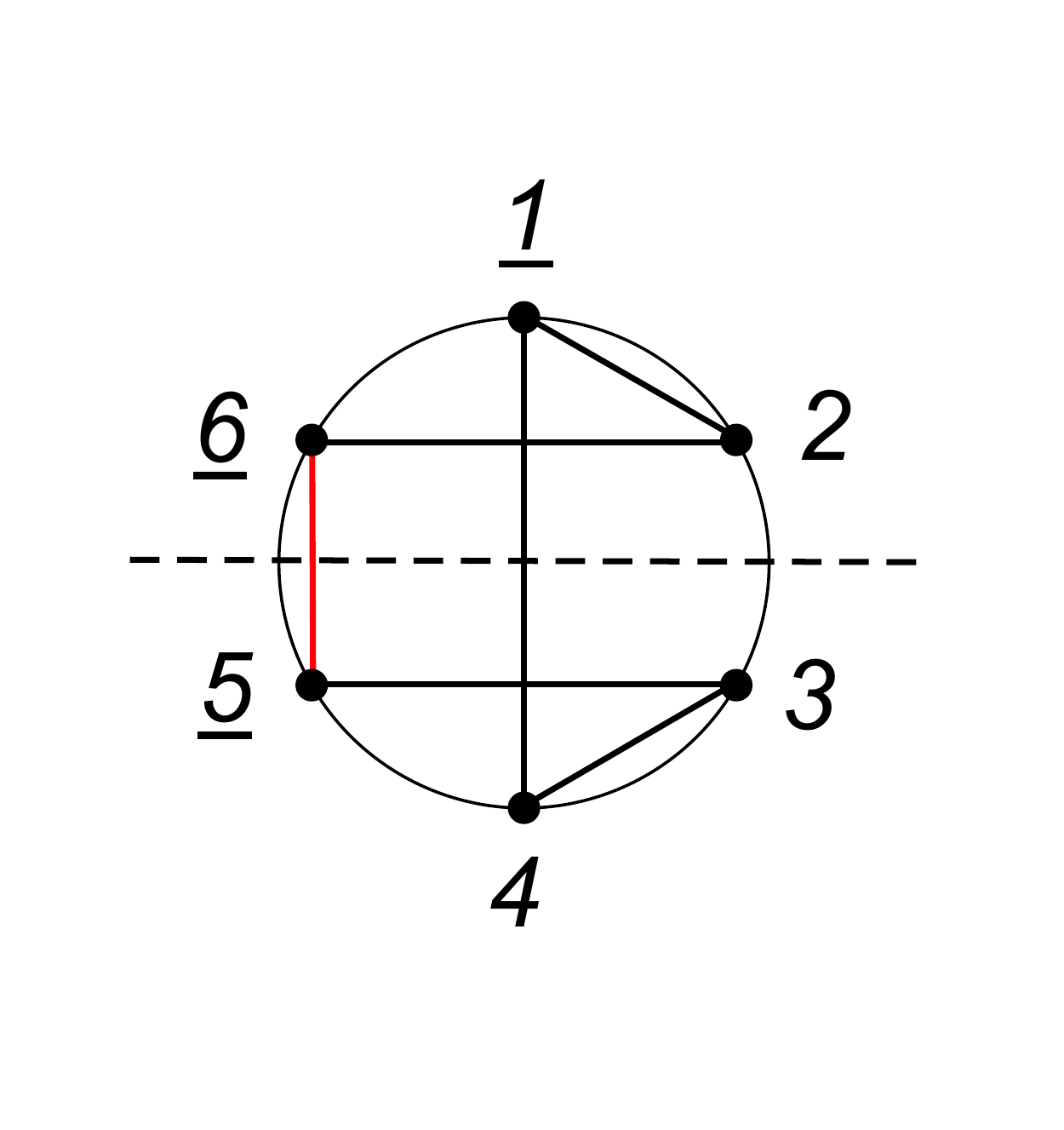}}
\parbox[c]{07.5em}{\includegraphics[scale=0.22]{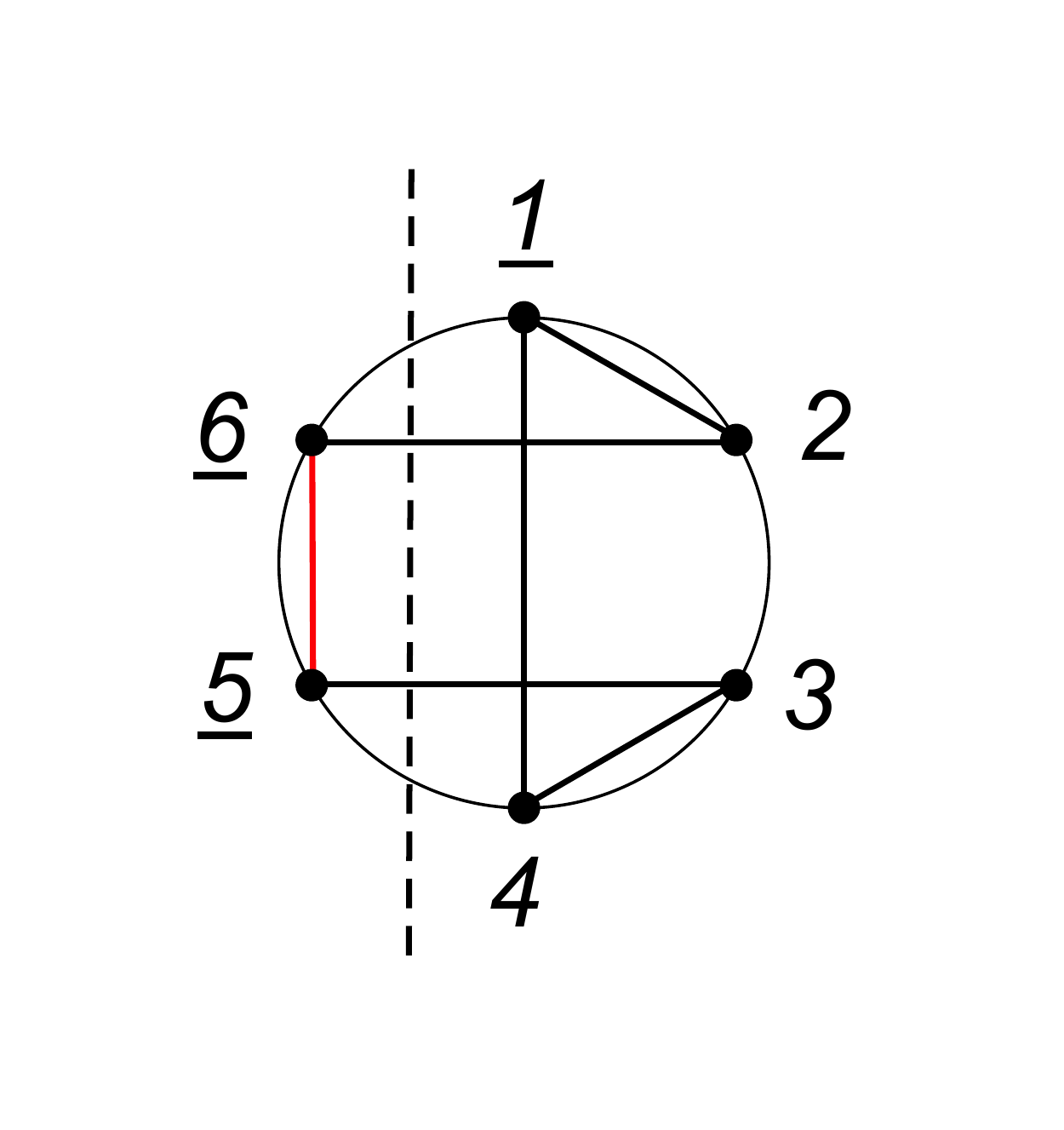}}$_{(c)}$\!\!\!\!
\parbox[c]{05.5em}{\includegraphics[scale=0.22]{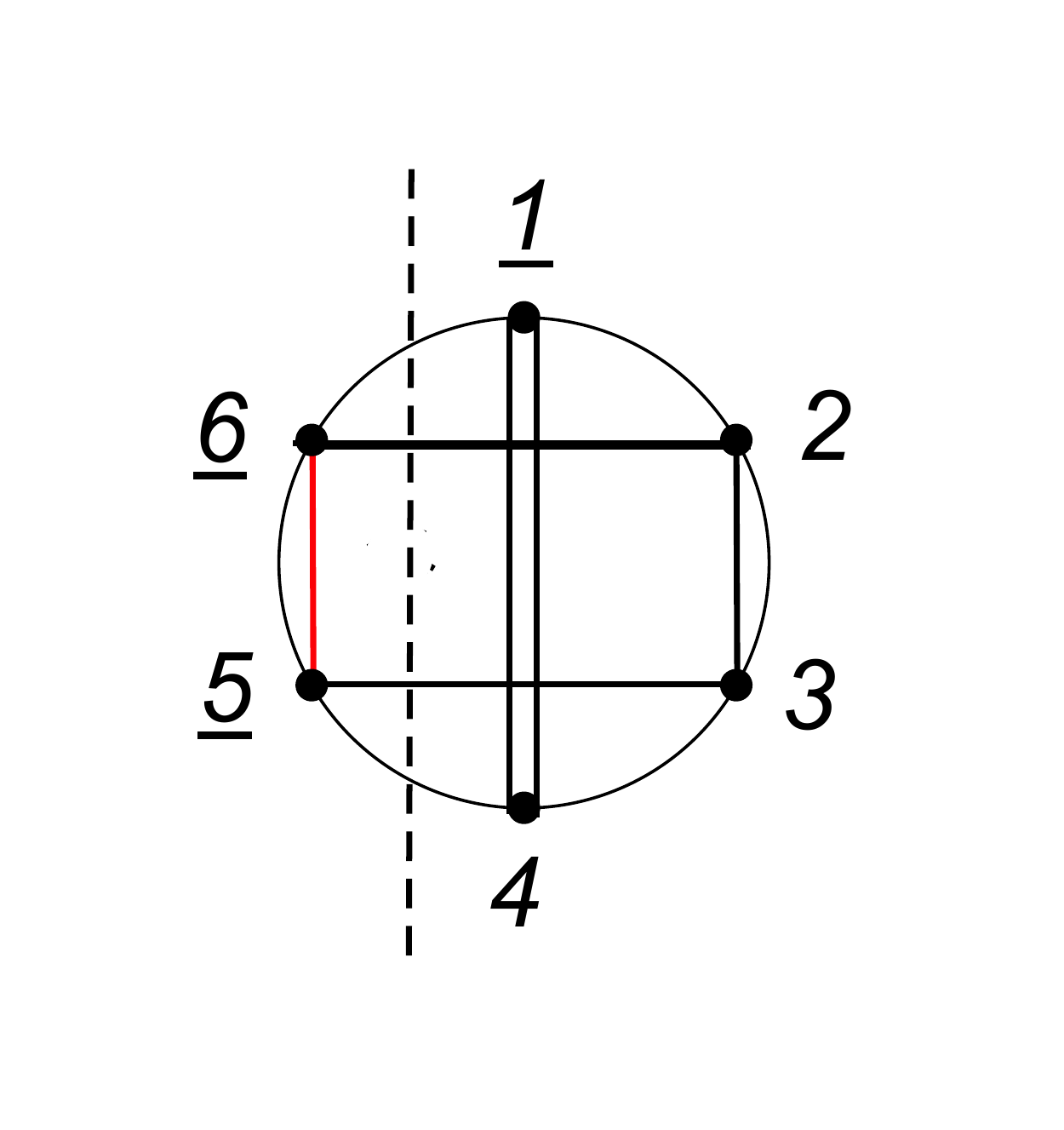}}
\parbox[c]{02.7em}{\includegraphics[scale=0.22]{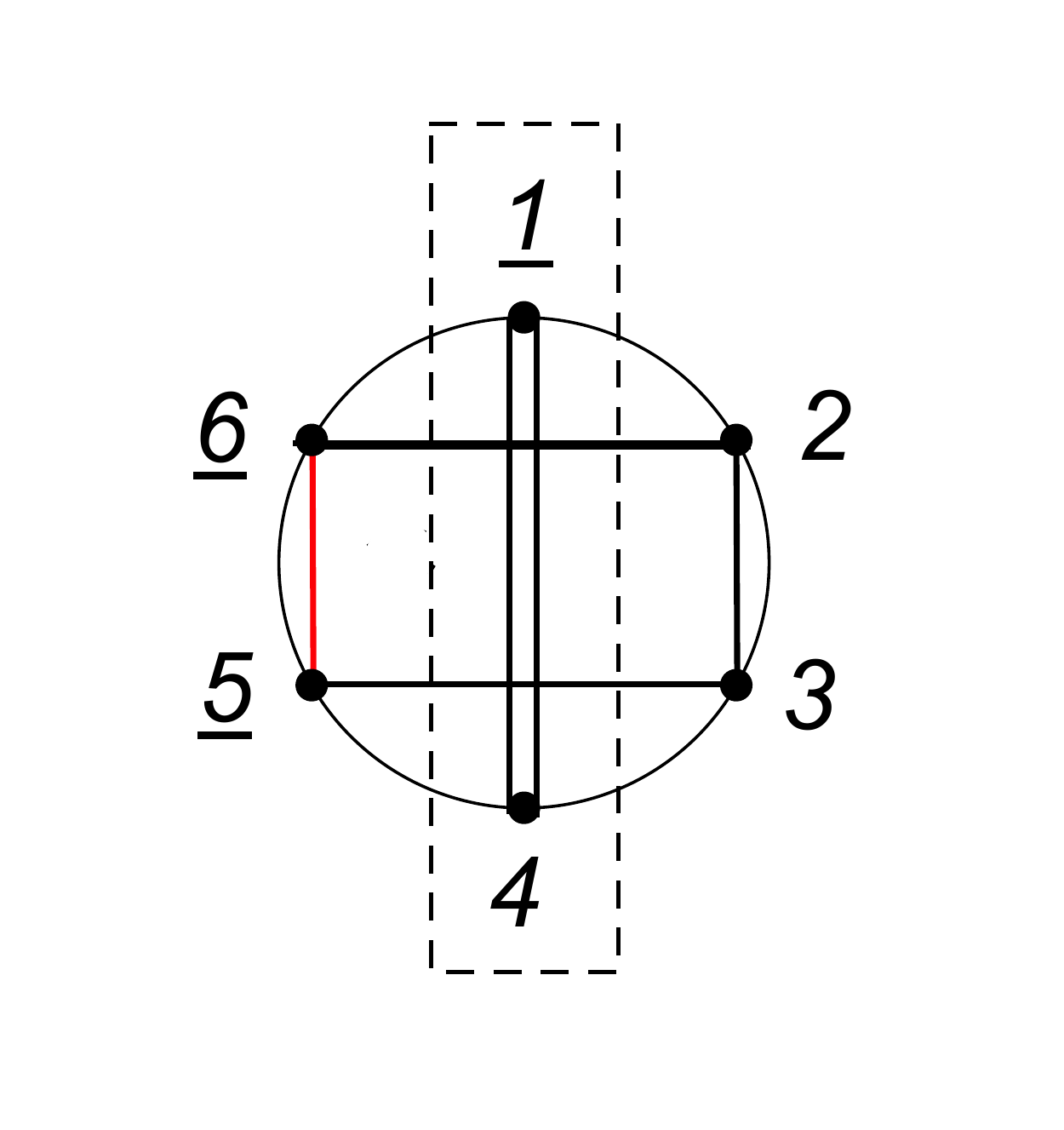}}
\vspace{-0.2cm}
\caption{Graph representation of ${\cal I}^{(1)}$, ${\cal I}^{(2)}$,  and ${\cal I}^{(3)}$, with their factorization contributions.}\label{FigA2} 
\vspace{-0.2cm}
\end{figure}

By definition
\begin{equation}
(-1)\,{\rm Pf} A^{56}_{56} =  \frac{\a_{13}\a_{24}}{\s_{13}\s_{24}} -\frac{\a_{12}\a_{34}}{\s_{12}\s_{34}} -\frac{\a_{14}\a_{23}}{\s_{14}\s_{23}},
\end{equation}
so it is straightforward to see that ${\cal A}(\mathbb{I}_6 \!: \! 14,26,35)  $ becomes
\begin{equation}\label{eq:167}
{\cal A}(\mathbb{I}_6 \!: \! 14,26,35)  =  \! \int_{\gamma} \!  \dif\sigma_2\dif\sigma_3\dif\sigma_4 ( \sigma_{56} \sigma_{61} \sigma_{15} )^2 (S_2S_3S_4)^{-1}  \!
\left\{    {\cal I}^{(1)} \a_{13}\a_{24}  + {\cal I}^{(2)} \a_{12}\a_{34}  +{\cal I}^{(3)}   \a_{14}\a_{23}                  \right\},
\end{equation}
where
\begin{align}
&{\cal I}^{(1)} = {\rm PT} (\mathbb{I}_6) \, {\rm PT}(1,3,5,6,2,4) , \quad  {\cal I}^{(2)} = {\rm PT} (\mathbb{I}_6) \, {\rm PT}(1,2,6,5,3,4), \nonumber \\
& {\cal I}^{(3)} = - {\rm PT} (\mathbb{I}_6) \, {\rm PT}(2,3,5,6)\, {\rm PT}(1,4).
\end{align}
In Fig. \ref{FigA2} we have drawn the diagrams corresponding to 
${\cal I}^{(1)}$, ${\cal I}^{(2)}$  and ${\cal I}^{(3)}$
by including their factorization contributions (notice that they are consistent with Fig. \ref{FigA1}). 

\begin{figure}[h]
\centering
\parbox[c]{06.9em}{\includegraphics[scale=0.22]{6pts-pf1-c1.pdf}}
$\Rightarrow$
\parbox[c]{06.0em}{\includegraphics[scale=0.22]{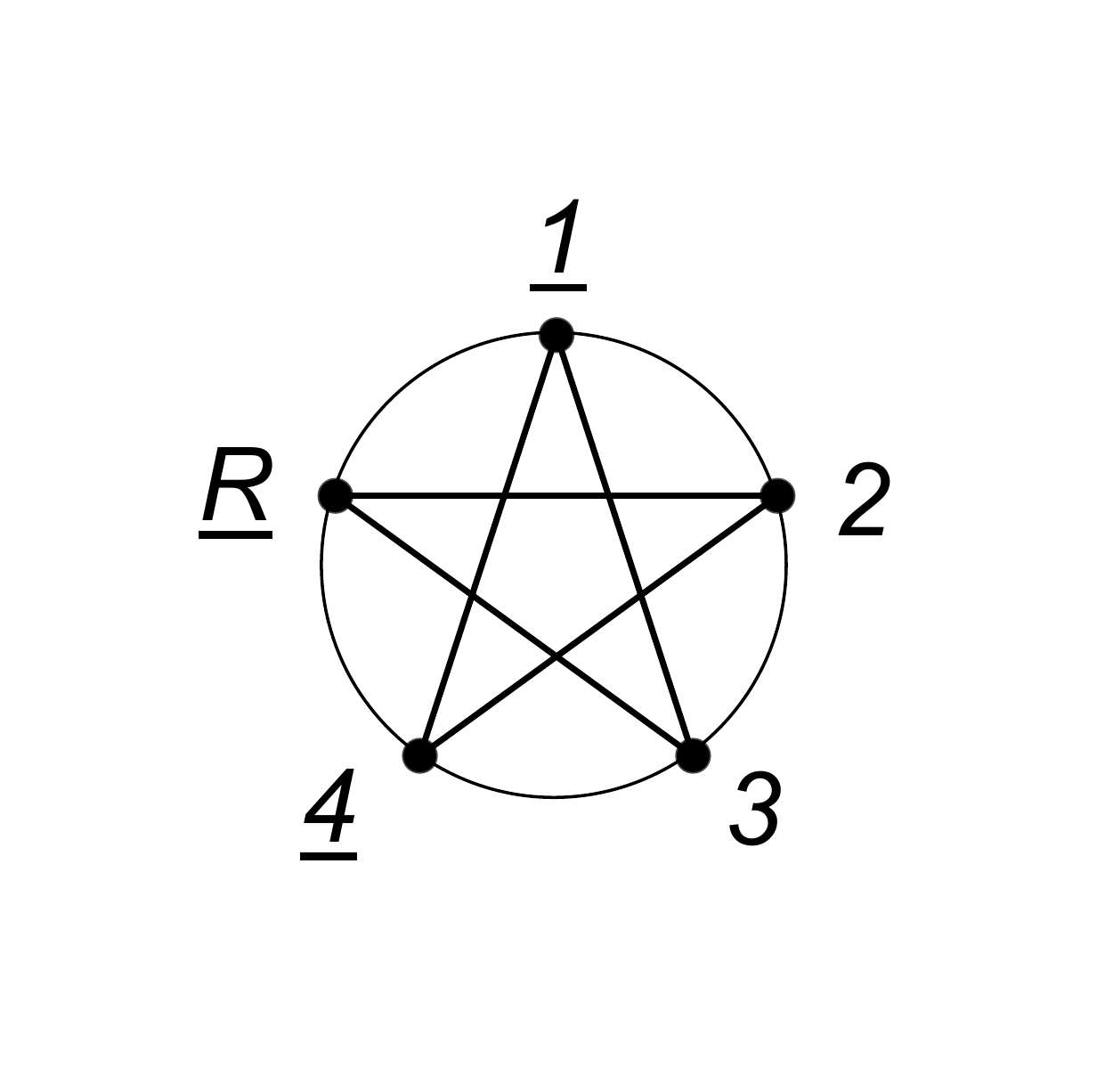}} \, =\,\, 0 
\vspace{-0.2cm}
\caption{Resulting graph from the factorization contribution of ${\cal I}^{(1)}$. The five-point diagram vanishes.}\label{FigA3} 
\vspace{-0.2cm}
\end{figure}

First, it is straightforward to check that the graph in Fig. \ref{FigA2}$(a)$ vanishes. 
Given the  factorization cut, $\s_4\rightarrow \s_3 \rightarrow \s_2 \rightarrow \s_1=constant$, we use the parametrization, $\s_a=\epsilon x_a + \s_1$, $a=1,2,3,4$ and $x_{1}=0$, $x_4=constant$ and $x_R=\infty$. The integral over ${\cal I}^{(1)}$ then becomes
\begin{align}\label{eq:169}
& \int_{\gamma} \!  \dif\sigma_2\dif\sigma_3\dif\sigma_4 ( \sigma_{56} \sigma_{61} \sigma_{15} )^2 (S_2S_3S_4)^{-1}      {\rm PT} (\mathbb{I}_6) \, {\rm PT}(1,3,5,6,2,4)  = [({\cal D}_5+{\cal D}_6)^2+m^2]^{-1} \nonumber\\
& 
\int_{\hat\gamma} \!  \dif x_2\dif x_3 ( x_{R1} x_{14} x_{4R} )^2 (\hat S_2 \hat S_3)^{-1}      {\rm PT} (\mathbb{I}_5) \, {\rm PT}(1,3,R,2,4), 
\end{align}
where $\hat \gamma=\gamma_{\hat S_{2}} \cap \gamma_{\hat S_{3}} $, 
\begin{align}\label{eq:170}
\hat S_2= \frac{\a_{21}}{x_{21}} + \frac{\a_{23}}{x_{23}} +\frac{\a_{24}}{x_{24}} + \frac{\a_{2R}}{x_{2R}} , \qquad
\hat S_3= \frac{\a_{31}}{x_{31}} + \frac{\a_{32}}{x_{32}} +\frac{\a_{34}}{x_{34}} + \frac{\a_{3R}}{x_{3R}},
\end{align}
and $\a_{aR} = \a_{a5}+\a_{a6}$, $a=2,3$. The diagram associated to the integral in \eqref{eq:169} is shown in Fig. \ref{FigA3},  which vanishes after applying the integration rules.

\begin{figure}[h]
\centering
\parbox[c]{06.9em}{\includegraphics[scale=0.22]{6pts-pf2-c2.pdf}}
$\Rightarrow$
\parbox[c]{06.9em}{\includegraphics[scale=0.22]{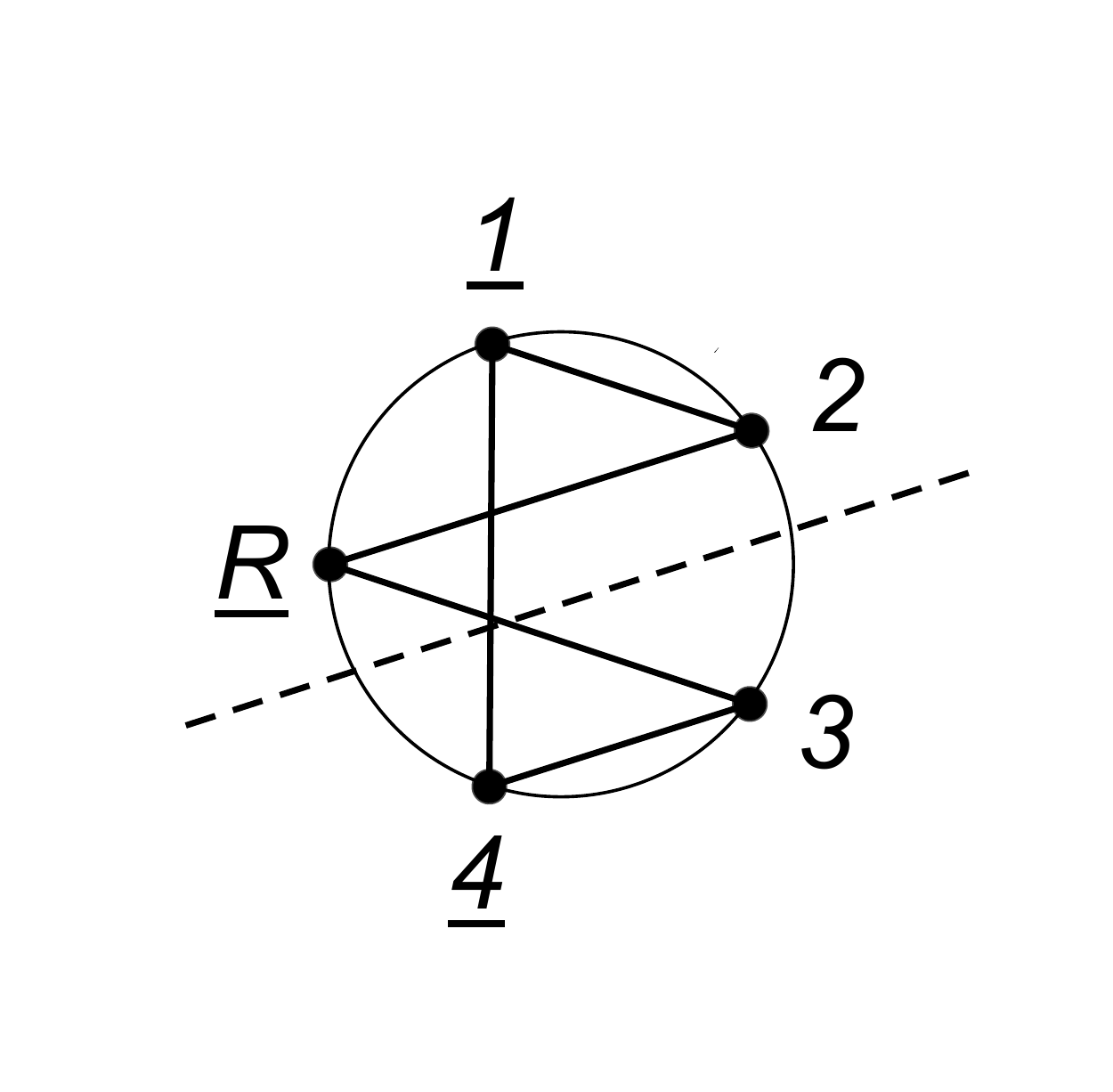}} 
$\Rightarrow$\!\!
\parbox[c]{06.0em}{\includegraphics[scale=0.22]{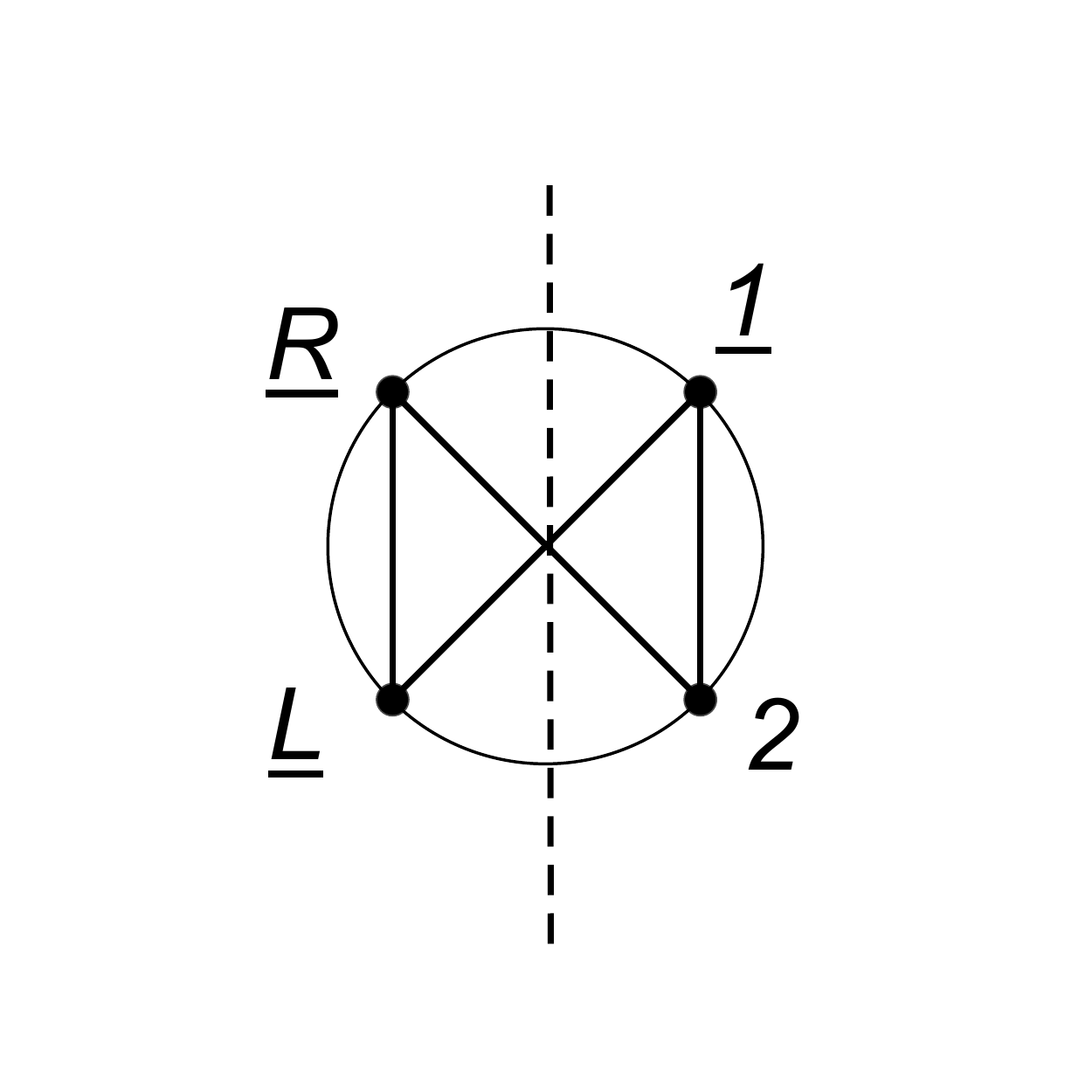}} 
\vspace{-0.2cm}
\caption{Five and four-point integrands obtained from the second factorization contribution of ${\cal I}^{(2)}$.}\label{FigA4} 
\vspace{-0.2cm}
\end{figure}

The second diagram in Fig. \ref{FigA2} has two factorization contributions. The first contribution is very similar to \eqref{eq:94}. Thus, it is straightforward to show that
\begin{align}\label{eq:171}
& \left[ \int_{\gamma} \!  \dif\sigma_2\dif\sigma_3\dif\sigma_4 ( \sigma_{56} \sigma_{61} \sigma_{15} )^2 (S_2S_3S_4)^{-1} \,     {\cal I}^{(2)} \, \a_{12} \a_{34}\right]\Big|_{\s_3\rightarrow \s_4 \rightarrow \s_5}  = [({\cal D}_3+{\cal D}_4 +{\cal D}_5)^2+m^2]^{-1}. 
\end{align}
To compute the second factorization contribution, where $\s_4\rightarrow \s_3 \rightarrow \s_2\rightarrow \s_1=constant$,
we use the usual parametrization, $\s_a=\epsilon x_a + \s_1$, $a=1,2,3,4$ and $x_{1}=0$, $x_4=constant$ and $x_R=\infty$. Following the same procedure used to evaluate \eqref{eq:94} we arrive at
\begin{align}\label{eq:172}
& \left[ \int_{\gamma} \!  \dif\sigma_2\dif\sigma_3\dif\sigma_4 ( \sigma_{56} \sigma_{61} \sigma_{15} )^2 (S_2S_3S_4)^{-1} \,     {\cal I}^{(2)} \, \a_{12} \a_{34}\right]\Big|_{\s_4\rightarrow \s_3 \rightarrow \s_2\rightarrow \s_1}  = [({\cal D}_5+{\cal D}_6 )^2+m^2]^{-1}  \nonumber \\
& 
\left[
\int_{\hat\gamma} \!  \dif x_2\dif x_3 ( x_{R1} x_{14} x_{4R} )^2 (\hat S_2 \hat S_3)^{-1}      {\rm PT} (\mathbb{I}_5) \, {\rm PT}(1,2,R,3,4) \right]\a_{12} \a_{34},
\end{align}
where $\hat \gamma=\gamma_{\hat S_{2}} \cap \gamma_{\hat S_{3}} $, and $\hat S_2$ and $\hat S_3$ are the same ones given in \eqref{eq:170}. In Fig. \ref{FigA4}, we depict the five-point diagram corresponding to the integral in \eqref{eq:172}. 
Using the contour $\gamma_{\hat S_3}$ to perform the GRT, it is simple to see that there is only one factorization contribution from the five-point diagram, $x_3\rightarrow x_4=constant\equiv x_L$, as depicted in Fig. \ref{FigA4}. Using the parametrization $x_a=\epsilon y_a + x_L$, $a=3,4$, $y_3=constant$ and $y_4=0$, the five-point diagram then reduces to
\begin{align}\label{eq:173}
& \left[
\int_{\hat\gamma} \!  \dif x_2\dif x_3 ( x_{R1} x_{14} x_{4R} )^2 (\hat S_2 \hat S_3)^{-1}      {\rm PT} (\mathbb{I}_5) \, {\rm PT}(1,2,R,3,4) \right]\Big|_{x_3\rightarrow x_4 }  = [\a_{34}]^{-1}  
\nonumber \\
& 
\left[
\int_{\gamma_{\tilde S_2}} \!  \dif x_2 ( x_{R1} x_{1L} x_{LR} )^2 (\tilde S_2)^{-1}      {\rm PT} (\mathbb{I}_4) \, {\rm PT}(1,2,R,L) \right],
\end{align}
where
\begin{equation}
\tilde S_2 =\frac{\a_{21}}{x_{21}}+ \frac{\a_{2R}}{x_{2R}} +\frac{\a_{2L}}{x_{2L}}, \qquad \a_{2L} = \a_{23} + \a_{24}.
\end{equation}
The resulting four-point diagram in Fig. \ref{FigA4} is trivial to compute: 
\begin{align}\label{eq:175}
\left[
\int_{\gamma_{\tilde S_2}} \!  \dif x_2 ( x_{R1} x_{1L} x_{LR} )^2 (\tilde S_2)^{-1}      {\rm PT} (\mathbb{I}_4) \, {\rm PT}(1,2,R,L) \right]\Big|_{x_2\rightarrow x_1 } = [\a_{12}]^{-1},
\end{align}
Combining \eqref{eq:171}, \eqref{eq:172}, \eqref{eq:173} and \eqref{eq:175}, the integral over ${\cal I}^{(2)}$ becomes

\begin{align}\label{eq:176}
& \left[ \int_{\gamma} \!  \dif\sigma_2\dif\sigma_3\dif\sigma_4 ( \sigma_{56} \sigma_{61} \sigma_{15} )^2 (S_2S_3S_4)^{-1} \,     {\cal I}^{(2)} \, \a_{12} \a_{34}\right] = [({\cal D}_3+{\cal D}_4 +{\cal D}_5)^2+m^2]^{-1} +
[\a_{56}]^{-1}.
\end{align}

The two factorization contributions of the integral  over ${\cal I}^{(3)}$ can be similarly computed to give
\begin{align}\label{eq:177}
& \left[ \int_{\gamma} \!  \dif\sigma_2\dif\sigma_3\dif\sigma_4 ( \sigma_{56} \sigma_{61} \sigma_{15} )^2 (S_2S_3S_4)^{-1} \,     {\cal I}^{(3)} \, \a_{12} \a_{34}\right] = -
[\a_{56}]^{-1}.
\end{align}
Summing the three diagrams we finally obtain
\begin{align}\label{}
{\cal A}(\mathbb{I}_6 \!: \! 14,26,35)  &=  \! \int_{\gamma} \!  \dif\sigma_2\dif\sigma_3\dif\sigma_4 ( \sigma_{56} \sigma_{61} \sigma_{15} )^2 (S_2S_3S_4)^{-1}  \!
\left\{    {\cal I}^{(1)} \a_{13}\a_{24}  + {\cal I}^{(2)} \a_{12}\a_{34}  +{\cal I}^{(3)}   \a_{14}\a_{23}      \right\} \nonumber \\
&= [({\cal D}_3+{\cal D}_4 +{\cal D}_5)^2+m^2]^{-1} ,
\end{align}
which agrees with \eqref{eq:17}.


\end{document}